\documentclass{aa}  

\usepackage{graphicx}
\usepackage{txfonts}
\usepackage{relsize}
\usepackage{booktabs}
\usepackage[thinc]{esdiff}
\usepackage{nicefrac}
\usepackage[justification=centering]{subcaption}
\usepackage{multirow}
\usepackage[colorlinks=true, allcolors=blue]{hyperref}
\usepackage[noabbrev, capitalise]{cleveref}

\makeatletter
\renewcommand*\aa@pageof{, page \thepage{} of \pageref*{LastPage}}
\makeatother

\def\dif{\text{d}}

\begin{document} 

   \title{A Bayesian approach to the halo galaxy -- supermassive black hole connection through cosmic time}

   \author{C. Boettner
          \inst{1}
          \and
          M. Trebitsch
          \inst{1}
          \and
          P. Dayal
          \inst{1}
          \fnmsep
          }

   \institute{Kapteyn Astronomical Institute, University of Groningen,
              Landleven 12 (Kapteynborg, 5419) 9747 AD Groningen\\
              \email{boettner@astro.rug.nl}
             }

   \date{Received xxx; accepted xxx}
 
  \abstract
   {}
   {The evolution of dark matter halos, galaxies, and supermassive black holes are deeply interdependent. We study whether this co-evolution can be qualitatively understood by connecting the evolution of a dark matter structure with simple empirical prescriptions for baryonic processes.}
        {We established expressions for the (star-forming) galaxy stellar mass function, galaxy UV luminosity function, active black hole mass function, and quasar bolometric luminosity function by assuming a direct and physically motivated relationship between the properties of galaxies and supermassive black holes, and the mass of their host halo. We calibrated the baryonic prescriptions using a fully Bayesian approach to reproduce observed population statistics. The derived parameterisations were then utilised to investigate the connection between galaxy and black hole characteristics and how these characteristics change with redshift.}
   {The galaxy stellar mass -- UV luminosity relation, black hole mass -- stellar mass relation, black hole mass -- AGN luminosity relation, and redshift evolution of these quantities obtained from the model are qualitatively consistent with observations. Based on these results, we present upper limits on the expected number of sources for $z=5$ up to $z=15$ for scheduled JWST and \textit{Euclid} surveys, thus showcasing that empirical models can offer qualitative predictions at a high redshift in a fast, easy, and flexible manner that complements more computationally expensive approaches.}
   {}


   \maketitle
%

\section{Introduction}
The distribution of dark matter in the Universe serves as the framework upon which galaxies form and evolve. It is widely accepted that galaxies originate and develop within gravitationally bound dark matter halos, resulting in a strong correlation between the characteristics of galaxies and the dark matter halos that host them \citep[e.g.][]{Fall1980, Efstathiou1983, Blumenthal1984, Wechsler2018}.
    
In standard Lambda cold dark matter ($\Lambda$CDM) cosmology, the growth of dark matter halos is thought to occur hierarchically \citep{Peebles1965, Silk1968, White1978}, with steady accretion of intergalactic matter and merging of gravitationally bound halos playing significant roles \citep[e.g.][]{Toomre1972, White1978, Barnes1988}. Since the behaviour of dark matter on large scales is governed solely by gravity, dark matter structure formation is generally considered a well-understood process supported by analytical and numerical models \citep{Press1974, Sheth2001, Despali2015}. In contrast, the evolution of galaxies within these halos is a more intricate process involving a diverse range of baryonic processes across all scales that regulate the growth of galaxies. The interplay between these baryonic processes and the underlying dark matter distribution has been extensively explored through abundance matching techniques, which connect observed galaxy properties to the statistics of dark matter halos \citep[e.g.][]{Kravtsov2004, Vale2004, Shankar2006}. The gravitational evolution of galaxies is dominated by the much more massive halos they inhabit; however, simultaneously, non-gravitational interactions such as radiative cooling, stellar evolution, and feedback from stars and active galactic nuclei (AGN) add an additional layer of complexity. The evolution of galaxies is mainly influenced by the gravitational effects of the massive halos surrounding them. However, non-gravitational interactions, including radiative cooling, stellar evolution, and feedback from stars and AGN, add another level of complexity to the process, shaping the galaxies' characteristics and behaviours \citep[e.g.][]{Dekel1986,White1991,Naab2017}. An area of particular interest is the connection between galaxies and their central supermassive black holes (SMBHs), as research has revealed they have properties that are closely related \citep{Gebhardt2000, Ferrarese2000, Davis2017}, suggesting their evolution to be tightly interconnected. The observed near-linear relationship between the rate of star formation and the stellar mass of galaxies termed the galaxy main sequence, implies that the baryonic processes of star formation and stellar mass growth are closely intertwined and play a significant role in regulating the overall growth and evolution of galaxies \citep{Brinchmann2004, Whitaker2014, Tomczak2014, Popesso2019, Sherman2021, Lilly2013}. To gain a comprehensive understanding of galaxy evolution, it is therefore necessary to study the evolution of dark matter, galaxies, and black holes in conjunction.
    
Star formation and AGN powered by accretion onto the central SMBH are principal contributors to gas heating and outflows in galaxies. Stars inject a considerable amount of energy and momentum into the interstellar medium through stellar winds, electromagnetic radiation, and supernovae. These processes heat or directly eject gas from the galaxy, thereby depleting it of fuel for further star formation \citep[e.g.][]{Larson1974b, Larson1974a, Dekel1986, Hopkins2012}. Abundance matching studies have demonstrated the significant impact of AGN feedback in shaping the stellar mass function and suppressing star formation in massive galaxies \citep{Shankar2006}. At the same time, the accretion of matter onto SMBHs in the centre of galaxies releases vast quantities of energy and momentum into their surroundings. This energy and momentum heats and expels nearby gas, which subsequently similarly influences star formation \citep{Silk1998, Croton2006}.The effectiveness of stellar and AGN feedback in removing gas from galaxies depends on the mass of the galaxy. \citet{Dekel1986} showed that stellar feedback is more effective at removing gas from low-mass galaxies because the gas is less tightly bound to the galaxy. In contrast, \citet{Silk1998} argued that AGN feedback is more effective at removing gas from massive galaxies, while its role in dwarf galaxies remains uncertain \citep{Dashyan2018, Koudmani2019, Koudmani2021, Koudmani2022, Trebitsch2018, Sharma2020}. These arguments are largely in agreement with observational studies on the stellar mass -- halo mass relation \citep[e.g.][]{Guo2010, Moster2010, Behroozi2010, Reddick2013, Moster2013} and galaxy stellar mass function \citep[e.g.][]{Ilbert2013, Duncan2014, Davidzon2017}. In addition, \citet{Bower2017} propose that a shutdown of stellar-driven outflows in massive galaxies leads to enhanced accretion rates onto SMBHs and, consequently, an increase in the effectiveness of AGN feedback. It is crucial to simultaneously account for the impact of stellar feedback and AGN feedback when modelling the evolution of galaxies and SMBHs.
    
Because galaxy evolution involves a wide range of physical scales and processes, no single model can fully encapsulate all of its characteristics. To address this, various modelling techniques have been created, each with its own balance between complexity and comprehensiveness. The most commonly used methods are semi-analytical and semi-numerical models that combine numerical simulations of dark matter with analytical prescriptions for baryonic physics \citep{White1991, Kauffmann1993, Cole1994, Somerville1999, Benson2002, Lacey2016, Poole2016} and full hydrodynamical simulations that jointly track the assembly of dark and baryonic matter \citep{Navarro1994, Vogelsberger2014, Schaye2015, Dubois2016, Nelson2019}. While these models have revolutionised our understanding of galaxies, their high computational costs and lengthy runtimes make it challenging to thoroughly explore the available parameter space. Moreover, galaxy evolution involves many processes that are still not fully resolvable in simulations. These processes must be represented by parameterisations derived from physical or empirical considerations. In these models, these prescriptions are usually physics-based, meaning that they are derived directly from fundamental physical processes. These processes result in specific characteristics of galaxy populations that can be compared with observational data.
    
Empirical models, on the other hand, rely on observational relations and conceptual
arguments alone to infer physical constraints \citep{White1991, Rodriguez2015,
Sharma2019}. These models can either be completely analytical or merger tree-based. They can include various physical processes or be simplified to contain only key components. This allows for the study of specific questions. Examples of such models include   
\textsc{EMERGE} \citep{Moster2018a}, \textsc{UniverseMachine} \citep{Behroozi2019}, and
\textsc{Trinity} \citep{Zhang2022}. These are comprehensive empirical models
containing around 50 free parameters and datasets of ten observational constraints,
aimed to study the evolution of galaxies, SMBHs, and their connection to halos from
$z=0$ -- $10$. These models produce detailed results, but they also come with a high
computational cost and can be challenging to interpret the impact of individual
parameters. On the other hand, the simple model proposed by \citet{Salcido2020}, aimed
at connection halo and stellar population statistics, has few and easily
interpretable parameters but does not include SMBHs and does not account for evolution
in baryonic processes.
    
In this study, we aim to bridge this gap in empirical models of the co-evolution of halos, galaxies, and AGN. We developed a model that connects halos to the properties of galaxies and AGN using simple analytical relations, which were calibrated using observational data on the galaxy stellar mass function (GSMF, $z=0-10$), galaxy UV luminosity function (UVLF, $z=0-10$), active black hole mass function (active BHMF, $z=0-5$), and quasar bolometric luminosity function (QLF, $z=0-7$). This allowed us to study the co-evolution of these four quantities over a redshift range of $z=0-10$. We especially aim to assess if the simple idea of connecting physical observables directly to the halo mass statistics yields sensible results for the relation between these different observables.
    
The simplicity of our model makes it easy to interpret the parameters involved and their evolution, and its reduced computational complexity allowed us to perform a full Bayesian exploration of parameter space, providing a comprehensive understanding of
the scope and limitations of the model while utilising all the information in the observational data. We validated our model using independent observational datasets on the relationships between the observables, specifically the galaxy stellar mass -- UV luminosity relation, SMBH mass -- stellar mass relation, and SMBH mass -- AGN bolometric luminosity relation, and we make qualitative predictions on the expected number densities of galaxies at $z>10$, which in good agreement with the JWST Early Data Release results \citep{Donnan2022, Harikane2022}. This model, being easy to interpret and computationally efficient, can be used to gain a qualitative understanding of galaxy evolution at high redshift and to inform more complex and computationally expensive models in a straightforward manner, as well as make quantitative predictions for upcoming instruments such as \textit{Euclid}.
    
The paper is organised as follows: We begin by presenting the theoretical framework of our model, including the assumptions we used to develop analytical relations for the observables in \cref{sec:modeldescription}. In \cref{sec:methods}, we describe the datasets used to calibrate the model, as well as the statistical method employed to match our model to these observations. Validation of the model is covered in \cref{sec:validation} where we compare the model output against independent datasets on the interrelationship between the observables, and we also examine the limitations of the assumptions made. In \cref{sec:redshiftevolution}, we explore the evolution of our model's parameters with redshift and provide predictions for an as-of-yet unobserved redshift. Finally, in \cref{sec:conclusion}, we summarise our findings and provide an overview of the implications and potential applications of our model.

\section{Model description}
\label{sec:modeldescription}
\begin{figure}
        \centering
        \includegraphics[width=\columnwidth]{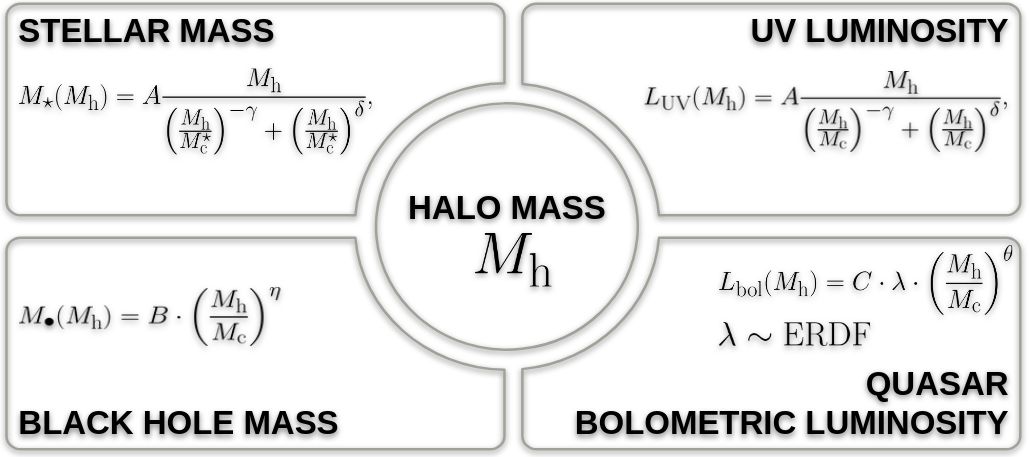}
        \caption{\textbf{Model schematic.} The properties of galaxies and supermassive black holes are assumed to be directly related to the halo mass through simple analytical prescriptions. Using a halo mass function, such as the one devised by \citet{Press1974}, the number densities of the derived quantities can be calculated and compared to observations.}
        \label{fig:model_schematic}
\end{figure}
In this study, we present a model that establishes connections between observable baryonic structures, such as galaxies and supermassive black holes (SMBHs), and their host halos through empirical relations. We attribute properties to galaxies and SMBHs based on the mass of their host halos. This enables us to replicate average relations and monitor how observed quantities change as redshift increases. Moreover, this approach allows us to investigate the relationships between observable quantities, which are detailed in this section. A summary of the model and its parameters can be found in \cref{tab:summary}.

\subsection{Connecting observables to halo statistics}
\label{subsec:modelbasics}
 The halo mass function (HMF) describes the number density of halos at a given redshift. This function can be derived analytically using knowledge of the matter density power spectrum \citep{Press1974} and closely corresponds to the results obtained from simulations of dark matter assembly. In its analytical form, the HMF is given by
 \begin{equation}
        \phi (M_\mathrm{h}) = \diff{n}{\log M_\mathrm{h}} (M_\mathrm{h}, z) =  \frac{\overline{\rho}}{M_\mathrm{h}} f\left(\nu (M_\mathrm{h}, z)\right) \left|\diff{\log \nu(M_\mathrm{h}, z) }{\log M_\mathrm{h}}\right|,
        \label{eq:HMF}
\end{equation}
where $\bar{\rho}$ is the mean matter density, $\nu$ is the mass variance at a given mass scale. $f$ is called the multiplicity function which depends on the details of the dark matter collapse model (see \cref{ApA:HMF} for details). For this work, we use the Sheth--Tormen HMF \citep{Sheth2001} for ellipsoidal collapse.

To construct our model, we make two simplifying assumptions:
\begin{enumerate}
        \item In a given cosmic volume, the total quantity of halos, central galaxies, and supermassive black holes $n$ are equal. This implies that each halo harbours a single galaxy and a single supermassive black hole (i.e. the occupation fractions $\equiv 1$ across all mass ranges). \footnote{This approach has certain limitations. It disregards halo and galaxy substructure, including satellites, as well as mergers. Additionally, it assumes that all halos host (active) SMBHs, when this is not necessarily the case for low-mass halos. Studies suggest that the occupation fraction of SMBHs deviates meaningfully from unity for $M_\mathrm{h} < 10^{11} M_\odot$ at $z=0$ \citep{Volonteri2016}. However, our focus is primarily on halos in a higher-mass range, as these are more easily detectable at high redshift. Several studies \citep[e.g.][]{Stefanon2021} have applied abundance matching with success in this regime, suggesting that this simplification is a valid assumption for these high-mass halos.}
        \item An observable $q$ of the galaxy is completely determined by an invertible function of the halo mass (but may evolve with redshift), that is,
        \begin{align}
        & &q = \mathcal{Q} (M_\mathrm{h}; z).
        \label{eq:qhmrel}
        \end{align}
\end{enumerate}
The second assumption will not hold for individual galaxies, which are subject to a wide range of physical mechanisms but can be understood in a statistical sense when averaged over a large number of objects. For this reason, we expect the model to be able to reproduce average relations between halo mass and observables, but not the scatter in these relations.

Given these two assumptions, the number density of the observables is directly linked to the HMF. For example, given a stellar mass -- halo mass relation $M_\star = \mathcal{Q} (M_\mathrm{h})$, the stellar mass function is given by
\begin{equation}
        \phi (M_\star) =
        \diff{n}{\log M_\star} \left( \mathcal{Q}(M_\mathrm{h})\right) =
        \diff{n}{\log M_\mathrm{h}} (M_\mathrm{h}) \cdot
        \diff{\log M_\mathrm{h}}{\log M_\star} \left(\mathcal{Q}(M_\mathrm{h})\right)
        \label{eq:qNDF}
\end{equation}
($\log = \log_{10}$). Thus, the form of \cref{eq:qhmrel} fully determines the population statistics of the observables, and by matching this number density to observations we can constrain the $q$ -- halo mass relation. Further, thanks to the invertibility of \cref{eq:qhmrel} we can directly link various observable quantities (see \cref{subsec:obsinterrelation}).

 Differences in the shape of the HMF and number density of observable properties are attributed to the influence of baryonic processes, including the effects of stellar and AGN feedback, on the formation and evolution of galaxies. We distinguish three feedback regimes \citep[see also][]{Salcido2020}:
\begin{enumerate}
        \item \textbf{Stellar Feedback Regime:} In low-mass halos, the star formation injects sufficient amounts of energy into the galaxy and efficiently drives gas outflows. This regulates the gas budget of the galaxy and prevents gas build-up in the galactic centre, inhibiting star formation and black hole growth.
        \item \textbf{Turnover Regime:} In this regime, the mass approaches a critical value at which gravity overcomes the stellar-driven outflows, leading to a build-up of gas in the galaxy, leading to rapid star formation and black hole growth.
        \item \textbf{AGN Feedback Regime:} In massive halos, black holes grow large enough for AGN to drive effective gas outflows, again regulating gas content and slowing star formation and black hole growth.
\end{enumerate}
 Depending on halo mass, galaxies will be dominated by different feedback mechanisms. To connect the observed population statistics to the HMF, we need to account for these effects. In the following, we derive relations based on these physical ideas.
\subsection{Star-forming galaxies}
\begin{figure}
        \centering
        \includegraphics[width=\columnwidth]{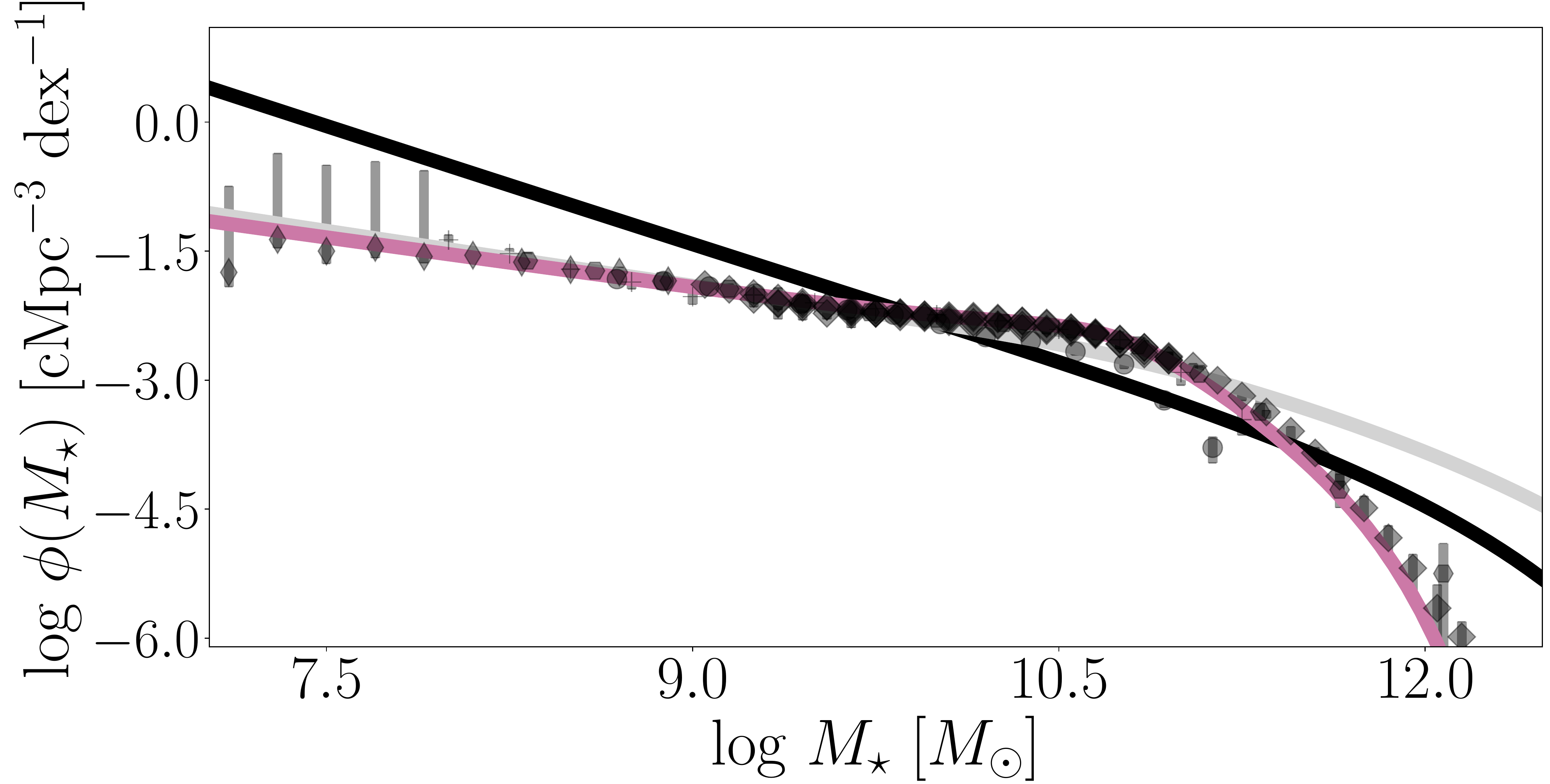}
        \caption{\textbf{Demonstration of feedback effects on GSMF:} Least squares regression of \cref{eq:qNDF} to the observed galaxy stellar mass function at $z=0$ (black dots, see text and \cref{fig:mstar_ndf_intervals} for details). The black curve shows the maximum likelihood estimate for a model without feedback ($\gamma_\star = \delta_\star = 0$), i.e. a simple scaling of the HMF; the light grey curve is a stellar feedback-only model ($\gamma_\star>0$, $\delta_\star=0$), while the pink curve includes stellar and AGN feedback ($\gamma_\star, \delta_\star>0$).}
        \label{fig:mstar_ndf_best_fit}
\end{figure}
The interaction between stellar and AGN feedback across various halo mass scales results in a distinctive relationship in the stellar mass – halo mass relation. Research has demonstrated that the galaxy stellar mass function exhibits a steeper slope at both the low and high-mass ends compared to the halo mass function. This observation prompted \citet{Moster2010} to propose a double power law relationship between halo mass and stellar mass. This parameterisation has been found to closely match the observed relation at $z=0$, which is obtained from abundance matching, clustering analysis and empirical modelling. The turnover halo mass of the power-law is  $M_\mathrm{c} \approx 10^{12}  M_\odot$ \citep{Wechsler2018}.

Since feedback processes regulate the rate of star formation, it is reasonable to anticipate a similar relation for the population of newly formed stars. Notably, these stars are the primary contributors to the UV luminosity of galaxies actively undergoing star formation. We can therefore parametrise \cref{eq:qhmrel} for the total stellar mass and UV luminosity as
\begin{align}
        {\mathcal{Q}}(M_\mathrm{h}) = A\frac{M_\mathrm{h}}{\left( \frac{M_\mathrm{h}}{M_\mathrm{c}}\right)^{-\gamma} + \left(\frac{M_\mathrm{h}}{M_\mathrm{c}}\right)^\delta},
        \label{eq:galaxyhalorelation}
\end{align}
where $\mathcal{Q}$ is the stellar mass $M_\star$, or UV luminosity $L_\mathrm{UV}$, and  $A$, $\gamma$, $\delta > 0$. In this parameterisation, the critical mass $M_\mathrm{c}$ indicates the mass scale at which the two feedback processes are of equal strength. 

While the galaxy main sequence establishes a direct relationship between stellar mass and star-formation rate (SFR; and by proxy UV luminosity) up to at least $z=2.5$ \citep{Whitaker2014}, a well-known bi-modality exists between star-forming and quenched galaxies \citep[e.g.][]{Bell2004, Cano-Diaz2016}, with more massive galaxies being increasingly likely to have low SFRs. To ensure that the assumption of a monotonic relation between halo mass and UV luminosity is valid, we therefore restrict ourselves to modelling star-forming galaxies only, leaving quenching for future work.

\cref{eq:galaxyhalorelation} is motivated by the fact that in the low halo mass limit (stellar feedback regime), this function behaves as a power law $q \propto M_\mathrm{h}^{1+\gamma}$, while in the high-mass limit (AGN feedback regime) we get $q \propto M_\mathrm{h}^{1-\delta}$, which resulting in the expected alteration of the halo mass function. For $\delta>1$, the function becomes non-invertible with a maximum at $M_\mathrm{max} = M_\mathrm{c} \left(\frac{\gamma +1}{\delta -1}\right)^\frac{1}{\gamma+\delta}$, meaning parameter space is restricted to $0\leq\delta<1$. To construct the GSMF and UVLF of the observable quantities (stellar mass and UV luminosity) specified by \cref{eq:qNDF}, we need the derivative of this function given by
\begin{equation}
        \diff{\log q}{\log M_\mathrm{h}} (M_\mathrm{h}) = 1- \left[ \ \frac{-\gamma \left( \frac{M_\mathrm{h}}{M_\mathrm{c}}\right)^{-\gamma}+ \delta \left( \frac{M_\mathrm{h}}{M_\mathrm{c}}\right)^\delta}{\left( \frac{M_\mathrm{h}}{M_\mathrm{c}}\right)^{-\gamma} + \left( \frac{M_\mathrm{h}}{M_\mathrm{c}}\right)^\delta}\right].
        \label{eq:galaxyhalorelationderivative}
\end{equation}
In the low-mass limit, the relation, which inversely contributes to the expression for the number density, takes the form $1+\gamma$. In contrast, at the high-mass end, it assumes the form $1-\delta$. The resulting number density is therefore suppressed at the low and high-mass end compared to the HMF. In $\Lambda$CDM, the HMF is given by a Schechter -- like function: a power law at the low-mass end and an exponential drop-off after some critical value (although it is more complicated in reality, see \cref{ApA:HMF}). If the power law slope of the HMF is denoted by $-\alpha_\mathrm{HMF}$, the corresponding observable quantity will exhibit a low-mass slope given by $-\alpha = -\frac{\alpha_\mathrm{HMF}}{1+\gamma}$.

\cref{eq:galaxyhalorelationderivative} having an inflection point at
\begin{equation}
        M_\mathrm{in} = M_\mathrm{c} \left(\frac{\gamma + \delta -1}{\gamma + \delta +1}\right)^\frac{1}{\gamma+\delta},
        \label{eq:inflectionpoint}
\end{equation}
results in the number density function having distinct slopes on either end of the mass scale $M_\mathrm{h} \approx M_\mathrm{c}$, as long as $\gamma+\delta>1$. This finding is consistent with observational evidence suggesting that galaxy stellar mass distribution and UV luminosity function are more accurately represented by a double Schechter function, unlike the halo mass function, which is described by a single Schechter function \citep{Tomczak2014, Weigel2016, McLeod2021}. 

Finally, note that for $\gamma=\delta=0$ (a model without feedback), the relation turns into $q \propto M_\mathrm{h}$ and $\diff{\log q}{\log M_\mathrm{h}} (M_\mathrm{h}) = 1$. The resulting number density is therefore identical in shape to the HMF, but shifted by $A/2$. If $\gamma \neq 0$ and $\delta=0$, we get a model without AGN feedback where the number density is suppressed at the low-mass end but traces the HMF at high masses. In \cref{fig:mstar_ndf_best_fit}, we show model galaxy stellar mass functions for all three cases imposed on observations at $z=0$. It is evident that both feedback mechanisms are needed to reproduce the shape of the observed GSMF.

\subsection{Supermassive black holes}
Despite the challenges associated with estimating the masses and characteristics of SMBHs, especially for those not actively accreting, studies have consistently found robust correlations between the properties of SMBHs and their host galaxies at low redshift. These correlations indicate a co-evolutionary relationship between SMBHs and their host galaxies. This is primarily demonstrated by the stellar mass -- velocity dispersion ($M-\sigma$) relation \citep{Gebhardt2000, Ferrarese2000, Davis2017}, and to a lesser extent, the black hole mass-bulge mass \citep{Kormendy2013} and black hole mass-stellar mass \citep{Reines2015} relations. 

Given that stellar properties are closely linked to halo properties, we can connect the observed SMBH number density directly to the halo mass function in a similar fashion as for the galaxy properties. In this work, we will specifically focus on the black hole mass function (BHMF) and the bolometric luminosity function of active black holes, also known as the Quasar luminosity function (QLF).
\subsubsection{Black hole mass function}
\label{subsubsec:BHMF}
In accordance with the three regimes of galaxy formation we have discussed earlier, we model black hole growth to occur in three distinct phases: a slow growth phase in the stellar feedback regime, rapid growth in the turnover phase and slower growth again in the AGN feedback regime. This picture is supported by the empirical discovery that SMBH growth appears to commence anti-hierarchically, with more massive SMBH forming first \citep{Kelly2013}. A simple parameterisation for this idea is given by
\begin{equation}
        M_\bullet = B \cdot
        \begin{cases}
        \left(\frac{M_\mathrm{h}}{M_\mathrm{c}^\bullet}\right)^{\eta_1} & \text{for }
                M_\mathrm{h} \ll M_\mathrm{c}^\bullet,\\
        \left(\frac{M_\mathrm{h}}{M_\mathrm{c}^\bullet}\right)^{\eta_2} & \text{for }
                M_\mathrm{h} \approx M_\mathrm{c}^\bullet,\\
        \left(\frac{M_\mathrm{h}}{M_\mathrm{c}^\bullet}\right)^{\eta_3} & \text{for }
                M_\mathrm{h} \gg M_\mathrm{c}^\bullet,\\
        \end{cases}
        \label{eq:bhmasshalomassrelation_full}
\end{equation}
where we expect $\eta_2 > \eta_1 > \eta_3$. However, as will become clear in the following sections, the observational datasets we use are restricted to actively accreting black holes, suggesting the available data to only constrain the rapid growth phase. We will therefore restrict our model to a single power law,
\begin{equation}
        M_\bullet (M_\mathrm{h}) = B \cdot \left(\frac{M_\mathrm{h}}{M_\mathrm{c}^\bullet}\right)^\eta.
        \label{eq:bhmasshalomassrelation}
\end{equation}
The derivative of this equation is
\begin{equation}
        \diff{\log M_\bullet}{\log M_\mathrm{h}} = \eta,
        \label{eq:mbhderivative}
\end{equation}
which is similar to the low-mass behaviour of \cref{eq:galaxyhalorelation} and similarly flattens the slope of the power law part of the number density function.

It is important to note that this single power law relation will break down at very low black hole masses, however for the available BHMF data, this prescription yields equal or better results compared to using multiple power laws. Note that the parameters $B$ and $\eta$ are strongly anti-correlated when fitting the black hole mass function, since $B$ shifts the HMF horizontally while $\eta$ shifts it vertically.

Besides the black hole mass -- halo mass relation, we further have to account for the fact that the number statistics on black holes and their masses are not complete. In fact, the dataset used in this work is limited to the black hole mass function of Type 1 AGN (see \cref{subsubsec:ABHMFdata}). To account for this, to connect this dataset to the halo mass function, we have to include a prescription of the \textit{duty cycle}, i.e. the fraction of actively accreting black holes, into \cref{eq:qNDF}. In this work, we use a simple, halo mass- and redshift-independent duty cycle, $f_\mathrm{d}$,
\begin{equation}
        \phi (M_\bullet) =
        f_\mathrm{d} \cdot \diff{n}{\log M_\bullet} \left( M_\bullet(M_\mathrm{h})\right).
        \label{eq:BHMF_duty_cycle}
\end{equation}

For a given observed AGN black hole mass function, a lower duty cycle therefore scales the intrinsic black hole mass function by a larger factor, with the effect that a black hole with a given fixed mass will be mapped to a lower-mass halo. 

\subsubsection{Quasar luminosity function}
\label{subsubsec:QLF}
Actively accreting supermassive black holes are among the brightest objects in the Universe, resulting in the QLF being well-sampled up to $z \sim 3$ and partially constrained up to $z \sim 7$. This makes it the primary tool for understanding AGN evolution. The bolometric luminosity in particular is tightly connected to the accretion rate of the black holes. Unlike the previous quantities, the number statistics of AGN luminosities differ qualitatively from that of the halo masses. The GSMF, UVLF, and BHMF are all well-described by Schechter functions similar to the HMF. However, the QLF is better characterised by a broken power law \citep[e.g.][]{Hasinger2005,Schneider2010,Ueda2014}. 

Replicating this form using the formalism introduced up till now is difficult because it is predicated upon the HMF, which exhibits asymptotic exponential behaviour. This difference may be associated with the much weaker correlation of the bolometric luminosity to the black hole mass (and therefore, according to our BHMF model, to the halo mass) compared to the other quantities. Theoretically, two SMBHs with equal masses residing in similar mass halos can exhibit significant variations in their bolometric luminosities. The range spans tens of orders of magnitude, from virtually inactive black holes with undetectable emissions to luminosities approaching or surpassing the Eddington limit. In the literature, it is common to describe the relation between the black hole mass and bolometric luminosity using the Eddington luminosity relation
\begin{equation}
        L_\mathrm{bol} (\lambda) = \lambda \cdot 10^{38.1} \cdot \frac{M_\bullet}{M_\odot} \quad\mathrm{erg \, s^{-1}},
        \label{eq:eddingtonrelation}
\end{equation}
where $\lambda$ is the Eddington ratio and $L_\mathrm{bol}(\lambda=1)$ is the Eddington luminosity.

The Eddington ratio for a given supermassive black hole relies on various factors apart from the black hole mass, and the precise relationships are still not fully understood. In modelling approaches, it is often assumed that the Eddington ratio behaves like a random variable that follows an Eddington ratio distribution function (ERDF). This function, denoted as $\xi(\lambda)$, might depend on black hole mass, halo mass, redshift, and other parameters like the central gas density and temperature.

In our model, we can incorporate this intrinsic spread. We can do this by assuming that the observed bolometric luminosity function represents the expectation value over all possible values of $\lambda$. This expectation value is weighted by the probability of a given $\lambda$, as determined by the ERDF,
\begin{equation}
        \phi(L_\mathrm{bol}) = \int_0^\infty \phi_\lambda(L_\mathrm{bol},\lambda) \xi(\lambda) d\lambda.
        \label{eq:obsQLF}
\end{equation}

Various shapes and functional dependencies of the ERDF have been proposed in the literature \citep[see][]{Shankar2013}. \citet{Caplar2015} and \citet{Weigel2017} have shown that a $M_\mathrm{h}$-independent ERDF can reproduce the bright end behaviour of the QLF, if it behaves as a power law in the limit $\lambda\rightarrow\infty$. 

\citet{Caplar2015} have calculated the ERDF with an unbroken as well as a broken power law form and found that in this approach the faint end of the QLF is only weakly affected by the chosen ERDF, while the bright end is dominated the $\lambda\rightarrow\infty$ behaviour of the ERDF and insensitive to the $\lambda\rightarrow 0$ end. Following this approach, we construct the QLF model using the following ingredients:
\begin{equation}
        L_\mathrm{bol} (M_\mathrm{h}) = C \cdot \lambda \cdot \left(\frac{M_\mathrm{h}}{M_\mathrm{c}^\mathrm{bol}}\right)^\theta,
        \label{eq:lbolmhrelation}
\end{equation}
\begin{equation}
        \diff{\log L_\mathrm{bol}}{\log M_\mathrm{h}} = \theta
        \label{eq:lbolderivative}
\end{equation}
\begin{equation}
        \xi(\lambda) \dif \lambda = \frac{ \frac{1}{1+\left(\frac{\lambda}{\lambda_\mathrm{c}}\right)^\rho}}
        {\mathlarger{\int_{-\infty}^\infty} \frac{1}{1+\left(\frac{\lambda}{\lambda_\mathrm{c}}\right)^\rho} \dif \lambda} \dif \lambda,
        \label{eq:ERDF}
\end{equation}
\begin{equation}
        \phi(L_\mathrm{bol}) = \int_0^\infty \phi_\mathrm{\lambda}(L_\mathrm{bol},\lambda) \xi(\lambda) d\lambda,
        \label{eq:bolLF}
\end{equation}
where \cref{eq:lbolmhrelation} is the relation between halo mass and bolometric AGN luminosity with ($C,\theta$) being free parameters. The ERDF is given by \cref{eq:ERDF}    with ($\lambda_c, \rho$) being the free parameters.

\begin{figure}
        \centering
        \includegraphics[width=\columnwidth]{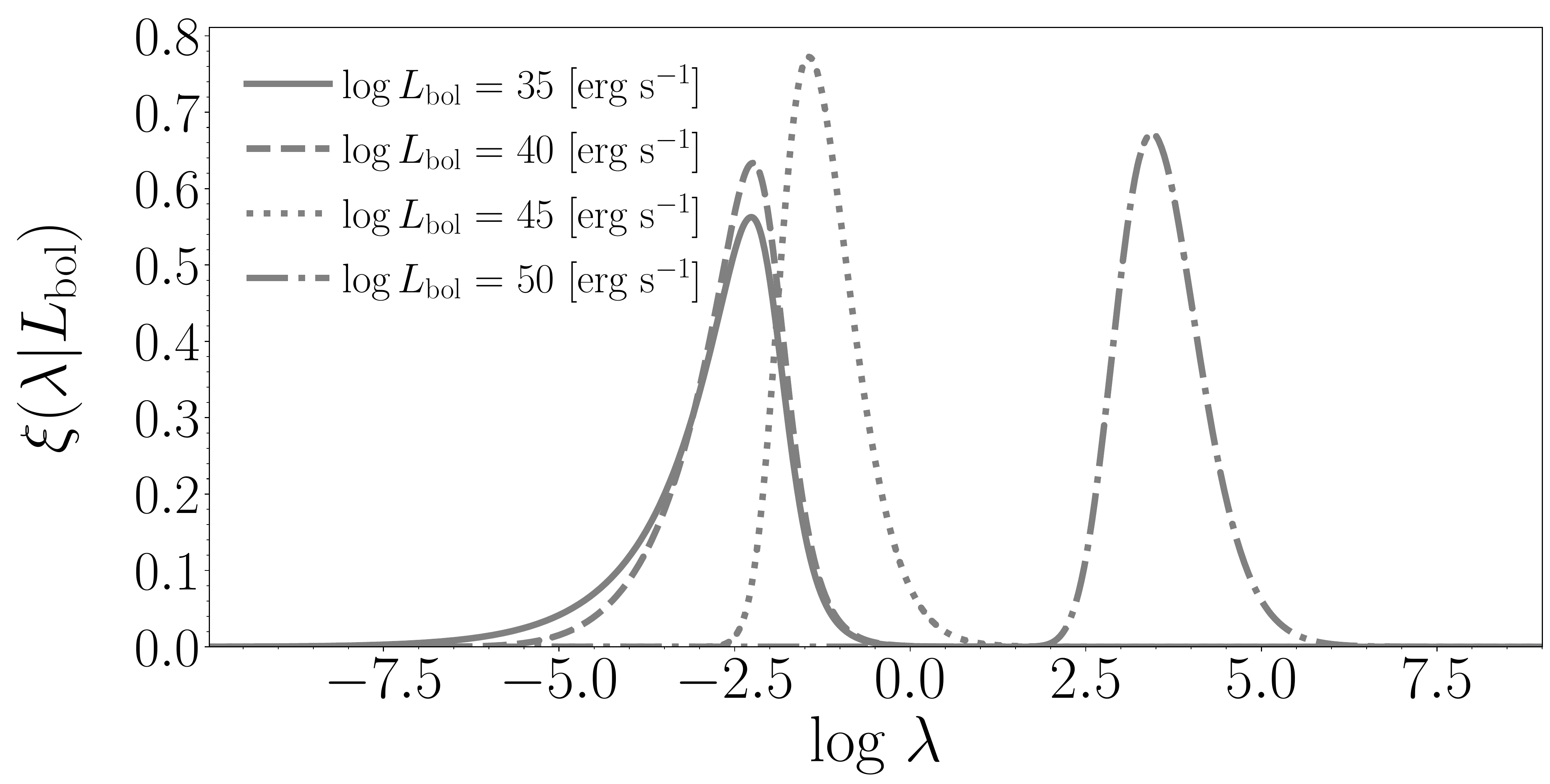}
        \caption{\textbf{Luminosity dependence of the conditional ERDF.} While our ERDF model given by \cref{eq:ERDF} is independent of halo and black hole properties, the conditional ERDF defined by \cref{eq:condERDF} varies as a function of the bolometric luminosity: higher luminosities are associated with larger Eddington ratios due to the lower number density of high-mass black holes. Shown are the conditional ERDFs for a set of luminosities at $z=0$ and $(C, \theta, \lambda_\mathrm{c}, \rho) = (40, 2, -2, 2)$.}
        \label{fig:Lbol_conditional_ERDF}
\end{figure}

 The model for the observed QLF is calculated using \cref{eq:bolLF}, where $\phi_\mathrm{\lambda}$ is calculated using \cref{eq:qNDF} and \cref{eq:lbolderivative}. The value under the integral in \cref{eq:bolLF} is the contribution to the bolometric luminosity function for a given value of $\lambda$, and if normalised to unity, represents the conditional probability of observing a given Eddington ratio for a fixed bolometric luminosity, i.e.
\begin{equation}
        \xi(\lambda|L_\mathrm{bol}) =    
                \frac{\phi_\lambda(L_\mathrm{bol},\lambda) \xi(\lambda)}
                {\int \phi_\lambda(L_\mathrm{bol},\lambda) \xi(\lambda)}.
                \dif \lambda,
        \label{eq:condERDF}
\end{equation}
It is worth noting that this function can vary for different bolometric luminosities even though the original ERDF was assumed mass-independent, due to the varying number densities of halos for different masses. An example of this is shown in \cref{fig:Lbol_conditional_ERDF}. 

The conditional distribution has a maximum approximately when $\phi_\mathrm{bol,\lambda}(L_\mathrm{bol},\lambda) \approx \xi(\lambda)$. In the limit $\lambda \rightarrow 0$ it is dominated by the exponential decay of the HMF and asymptotically has the same exponential behaviour, while in the limit $\lambda \rightarrow \infty$ the function decays as a power law under the assumption that the HMF has a power law-like behaviour in the low-mass limit. 

The asymptotic slope of $\xi(\lambda|L_\mathrm{bol})$ is given by $-\left(\rho-\frac{\alpha_\mathrm{HMF}}{\theta}\right)$. Note that we require $\rho>1$ for the ERDF to be normalisable and $\rho-\frac{\alpha_\mathrm{HMF}}{\theta}>1$ for the $\phi_\mathrm{bol}$ integral to converge. The cumulative distribution function for \cref{eq:ERDF} is given by $F(\lambda) = \lambda \cdot {}_2F_1\left(1, \frac{1}{\rho}; 1 + \frac{1}{\rho}, \left(\frac{\lambda}{\lambda_\mathrm{c}}\right)^\rho\right)$, where ${}_2F_1(\lambda)$ is the hypergeometric function. If the previous conditions are met, the resulting observed bolometric luminosity function has the shape of a broken power law with an asymptotic faint end slope $\frac{\alpha_\mathrm{HMF}}{\theta}$ and an asymptotic bright end slope $-\rho$.

\subsection{Relating observable properties}
\label{subsec:obsinterrelation}
Since in our model, all observable quantities are dependent solely on the halo mass (and redshift) through invertible relations, we can express any observable quantity $q_1$ as a function of any other quantity $q_2$ by $q_1(M_\mathrm{h}) = q_1 \left(M_\mathrm{h}(q_2)\right) = q_1(q_2)$. This allows us to study the relations between these quantities directly from their observed number densities. In particular, we are able to relate galaxy and black hole properties.

We choose to study the performance of our model (\cref{sec:validation}) using three relations:
\begin{enumerate}
        \item \textbf{The galaxy stellar mass -- UV luminosity relation:} A proxy for the galaxy main sequence which has been robustly observed up to at least $z=6$ \citep{Santini2017a}.
        \item \textbf{The SMBH mass -- stellar mass relation:} One of the main clues that SMBHs and galaxies might co-evolve and well established at low redshift, albeit with a larger scatter than the $M$ -- $\sigma$ relation.
        \item \textbf{The SMBH mass -- AGN bolometric luminosity relation:} One of the main avenues to study AGN and black hole accretion.
\end{enumerate}
The asymptotic limits of the $M_\bullet$ -- $M_\star$ and $L_\mathrm{UV}$ -- $M_\star$ relations can be readily calculated since their parameterisations are all asymptotically power laws. For example, combining \cref{eq:galaxyhalorelation} and \cref{eq:bhmasshalomassrelation} yields $M_\bullet \sim M_\star^{\nicefrac{\eta}{(1+\gamma_\star)}}$ for $M_\mathrm{h} \ll M_\mathrm{c}$. The power law slopes for these two relations and different limiting cases of $M_\mathrm{h}$ can be found in \cref{tab:q1q2relations}.

The situation is a little more complicated for the $M_\bullet$ -- $L_\mathrm{bol}$ relation, due to the adapted approach for modelling the QLF. Combing the relations for black hole mass and the bolometric luminosity given by \cref{eq:bhmasshalomassrelation,eq:lbolmhrelation} yields
$ L_\mathrm{bol} \sim \lambda \cdot M_\bullet^{\nicefrac{\eta}{\theta}}$, with behaviour of this relation depending on the distribution of $\lambda$: if the AGN sample selection is based on the host halo mass, the form of \cref{eq:ERDF} guarantees that the expectation value $\langle \lambda \rangle = \int \lambda \xi(\lambda) \dif \lambda$ is $M_\mathrm{h}$-independent and $\langle L_\mathrm{bol} \rangle \sim M_\bullet^{\nicefrac{\eta}{\theta}}$ behaves as a power law. In practice, AGN samples are however primarily selected on luminosity, so we have to use
\begin{equation}
\langle \lambda |L_\mathrm{bol} \rangle
        = \int \lambda \xi(\lambda|L_\mathrm{bol}) \dif \lambda
        = \frac{\int \lambda \phi_\lambda(L_\mathrm{bol},\lambda) \xi(\lambda)}
        {\int           \phi_\lambda(L_\mathrm{bol},\lambda) \xi(\lambda)}
        \dif \lambda,
\end{equation}
which will differ for varying values of $L_\mathrm{bol}$, altering the functional form of the relation.
\begin{table}
        \centering
        \caption{Power law slopes for black hole mass $M_\bullet$ -- stellar mass $M_\star$ relation and UV luminosity $L_\mathrm{UV}$ -- stellar mass $M_\star$ relation for different halo mass limits.}
        \begin{tabular}{llll}
        Relation & $M_\mathrm{h} \rightarrow 0$ & $M_\mathrm{h} \approx M_\mathrm{c}$ & $M_\mathrm{h} \rightarrow \infty$ \\ \hline \hline
         
        $M_\bullet$ -- $M_\star$ & $\nicefrac{\eta}{(1+\gamma_\star)}$ & $\eta$ & $\nicefrac{\eta}{(1-\delta_\star)}$ \\
         
        $L_\mathrm{UV}$ -- $M_\star$ & $\nicefrac{(1+\gamma_\mathrm{UV})}{(1+\gamma_\star)}$ & $1$ & $\nicefrac{(1-\delta_\mathrm{UV})}{(1-\delta_\star)}$ \\
        \end{tabular}
        \label{tab:q1q2relations}
\end{table}

\section{Observational datasets and model calibration}
\label{sec:methods}
\begin{figure*}
     \centering
     \includegraphics[width=\textwidth]{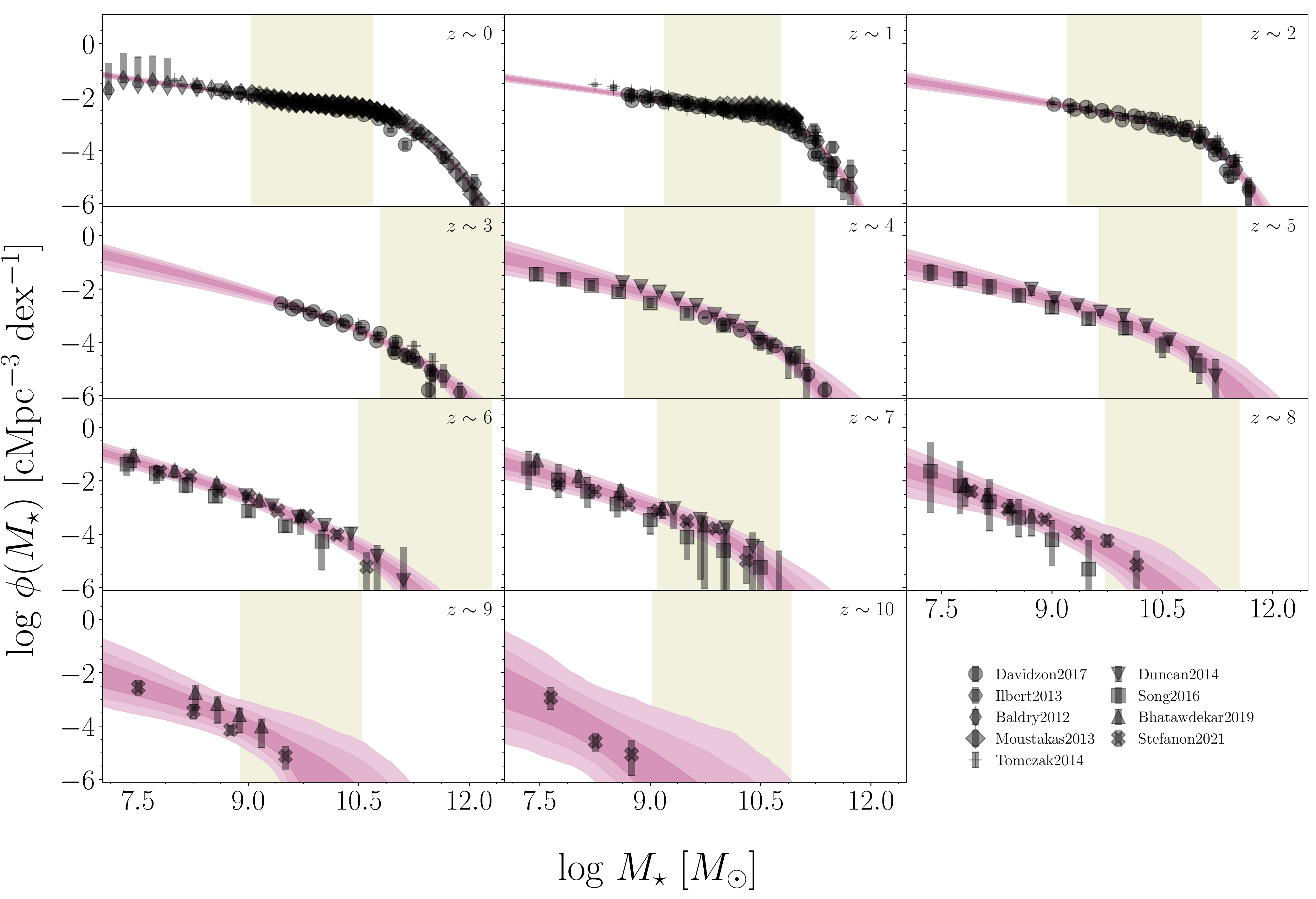}
     \caption{\textbf{Credible regions of modelled galaxy stellar mass functions:} The model number densities as calculated from \cref{eq:qNDF} matched to observational data. The credible regions cover (in decreasing order of colour saturation) 68\%, 95\% and 99.7\% of the posterior distributions. The shaded area in the background marks the regime where the two feedback mechanisms contribute about equally to the stellar mass -- halo mass relation given by \cref{eq:galaxyhalorelation}, the white area on the low-mass side is dominated by stellar feedback, $(\left(\nicefrac{M_\mathrm{h}}{M_\mathrm{c}^\star}\right)^{\delta_\star+\gamma_\star} < 0.1)$, while the area on the high-mass side is AGN feedback-dominated, $\left(\nicefrac{M_\mathrm{h}}{M_\mathrm{c}^\star}\right)^{\delta_\star+\gamma_\star} >10$. The regime borders are calculated as the median value from a sample of GSMFs drawn from the posterior.}
     \label{fig:mstar_ndf_intervals}
\end{figure*}
\begin{figure*}
     \centering
     \includegraphics[width=\textwidth]{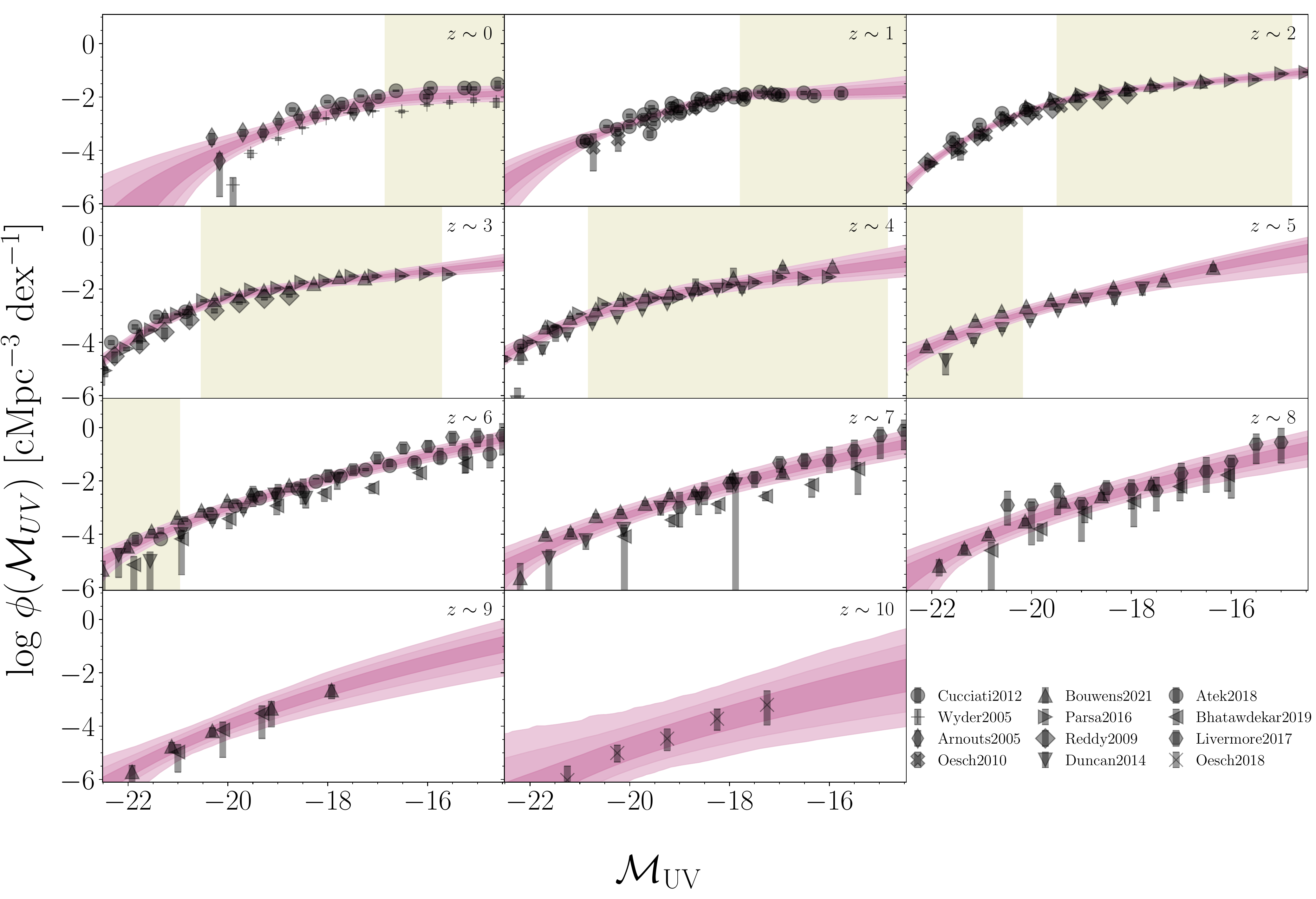}
      \caption{\textbf{Credible regions of modelled galaxy UV luminosity functions:} Similar to \cref{fig:mstar_ndf_intervals}. Here, the white region on the bright side is the AGN feedback-dominated regime, while the area on the faint end is stellar feedback-dominated.}
     \label{fig:Muv_ndf_intervals}
\end{figure*}
\begin{figure}
     \centering
     \includegraphics[width=\columnwidth]{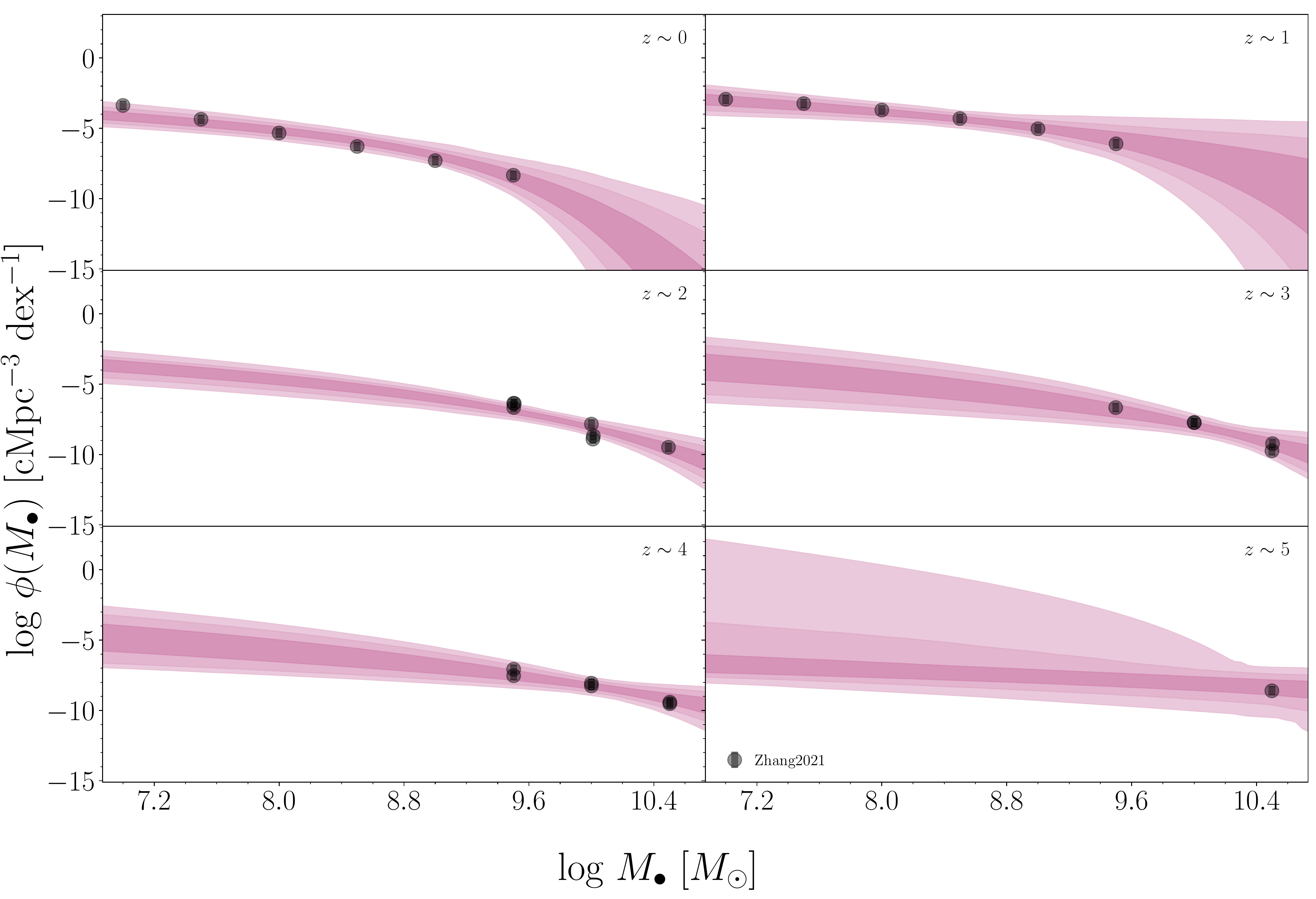}
      \caption{\textbf{Credible regions of modelled black hole mass functions.}  Note that the available data does not constitute the full BHMF, but the mass function of Type 1 AGN (\cref{subsubsec:ABHMFdata}).}
     \label{fig:mbh_ndf_intervals}
\end{figure}
\begin{figure*}
     \centering
     \includegraphics[width=\textwidth]{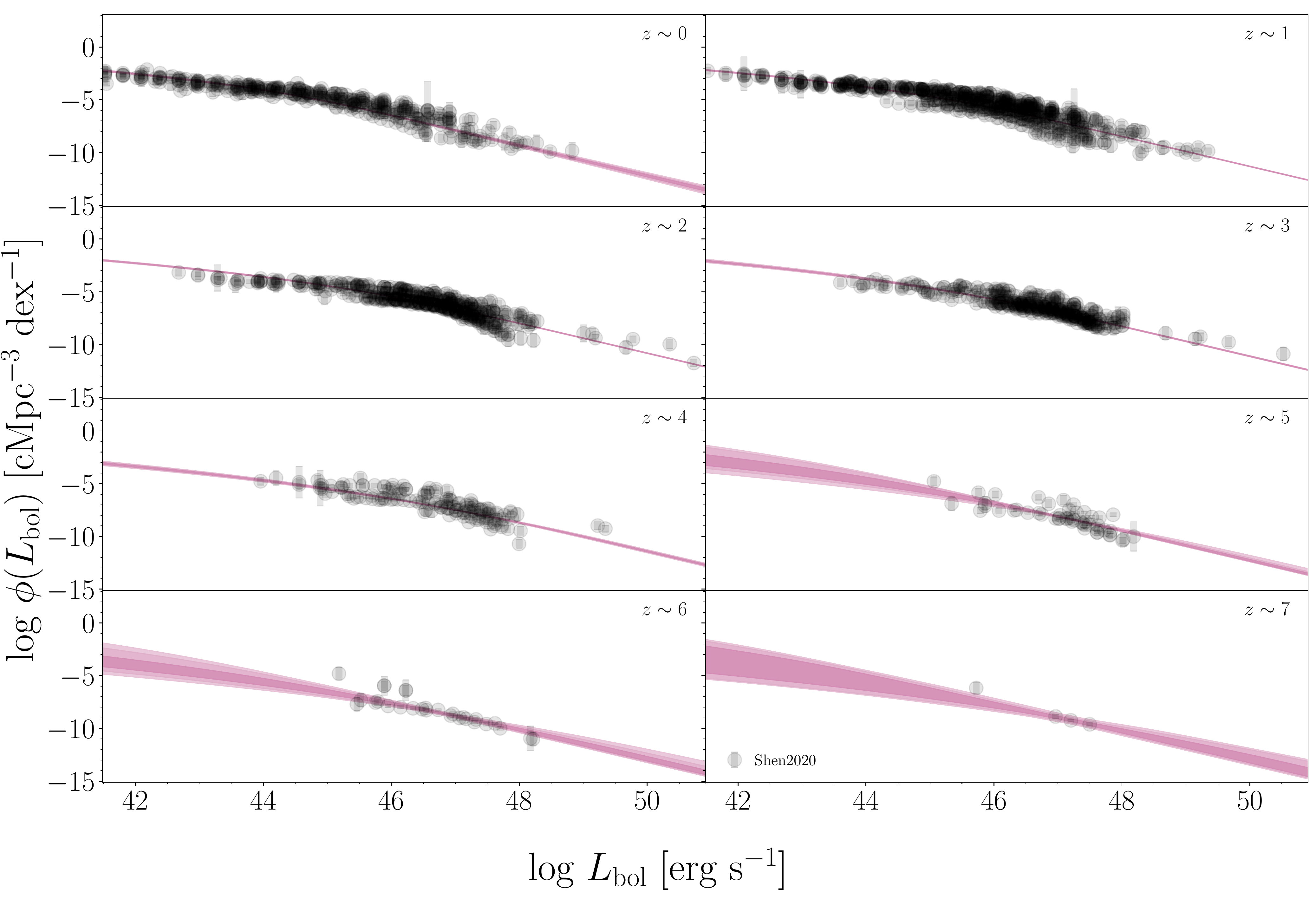}
      \caption{\textbf{Credible regions of modelled quasar luminosity functions.}  The ERDF parameters $(\lambda_\mathrm{c}, \rho)$ are free at $z=0$, but fixed to the $z=0$ MAP estimates $(\lambda_\mathrm{c}, \rho) \approx (-2.2,  1.4)$ for $z>0$.}
     \label{fig:Lbol_ndf_intervals}
\end{figure*}
The focus of our empirical model is not on making exact numerical predictions, but on gaining a qualitative comprehension of how the physical mechanisms involved change and how important they are. For this reason, we turn to a probabilistic approach for calibrating our model: rather than focusing on expectation values or best-fit parameters, we study the evolution of the probability distributions for the parameter as a whole. This approach capitalises on the model's computational simplicity, which is particularly beneficial when dealing with sparsely sampled data (such as those at high redshifts). This is also advantageous when making predictions beyond the current observational limitations. In this section, we describe the datasets used to calibrate the model, the statistical framework and the assumption made for the statistical inference to match the observed number density functions.

\subsection{The datasets}
\label{subsec:datasets}
To ensure comparability, we have collected observational datasets from various sources and standardised them. However, there may still be discrepancies in the data processing methods used by the original authors. We attempt to address this issue by incorporating additional uncertainty when utilising datasets from multiple authors (see \cref{subsec:likelihoodfunction} for details).
\subsubsection{GSMF data}
We collect data on the GSMF from a variety of sources \citep{Baldry2012b, Moustakas2013, Ilbert2013, Duncan2014, Tomczak2014, Song2016, Davidzon2017, Bhatawdekar2018, Stefanon2021} spanning $z=0$ -- $10$. We correct the data for the assumed IMF, employing a \citet{Chabrier2003} IMF with a mass range of $0.1$--$100 M_\star$, when necessary and re-binning into integer redshift bins (e.g. mapping $0.5 \leq z \leq 1.5$ to $z \sim 1$). The stellar masses have primarily been constructed by SED fitting to available photometry, and have been corrected for dust extinction. The GSMF is best described by a double Schechter function for $z < 3$ \citep{Davidzon2017}, while at higher redshift a single Schechter function suffices. The low-mass slope of the GSMF has been robustly found to decrease towards lower redshift, while the overall number density increases.
\subsubsection{Galaxy UVLF data}
The galaxy UV luminosity (rest frame wavelength centred around $\lambda \approx 1500$-$1600$ \r{A}) data also spans a redshift range from
$z=0$ -- $10$, collected from multiple sources
\citep{Wyder2005,Parsa2016,Cucciati2012,Duncan2014,Atek2018,Arnouts2005,Livermore2017,Bhatawdekar2018,Bouwens2021}. The provided data is corrected for dust extinction, and when necessary is adjusted for different IMFs. The UVLF is best constrained for $z\geq2$ for which the rest frame UV is redshifted to the optical and IR bands. The UVLF peaks across most luminosities around $z=2$ -- $3$, which is known to be the peak of star formation and drops towards higher and lower redshift. Similar to the GSMF, the UVLF slope increases robustly between $z=2$ and $z=10$, with some debate on the evolution on the very faint end \citep{Bowler2015,Bowler2017,Harikane2022}. The evolution for $z<2$ is more strongly contested but seemingly consistent with a constant slope \citep{Cucciati2012}.
\subsubsection{Data on the active BHMF of type 1 AGN}
\label{subsubsec:ABHMFdata}
Current constraints on the SMBH masses from direct kinematic modelling are limited to a small sample of local galaxies \citep{Kormendy2013} from which no reliable number statistics can be discerned. Larger and more distant samples can be inferred using indirect methods in AGN, predominantly by estimating velocity dispersion (and consequently virial mass) of the accretion disk from the broad line emission width for Type 1 (unobscured) AGN \citep{Vestergaard2006}. Lately, broad line - narrow line correlations have been used to extend the estimation to Type 2 (obscured) AGN as well \citep{Baron2019}. These types of studies can be used to constrain the BHMF up to $z=5$ \citep{Kelly2013}, but with the caveat that these are limited to the sub-population of active black holes rather than the total black hole population. We employ the active black hole mass function of Type 1 AGN as collected by \citet{Zhang2021} based on the work of \citet{Schulze2010}, \citet{Kelly2013} and \citet{Schulze2015}. These active BHMFs follow the known evolution of the AGN population: number densities peak between $z=2$ and $3$, which corresponds to the era of peak star formation and black hole growth (and thus increased black hole activity). The number densities at high masses drop faster towards low redshift compared to lower-mass black holes (cosmic downsizing). It needs to be stressed that the relation between the number densities of Type 1 AGN and the total SMBH population is not trivial and that we have not included a mechanism in our model that accounts for this selection of the sub-population. Interpreting the results of the redshift evolution of the BHMF and relating the black hole masses to other quantities in the model therefore needs to be done in light of this limitation (see \cref{sec:validation}).
\subsubsection{QLF data}
We employ the quasar bolometric luminosity functions constructed by \citet{Shen2020} covering $z=0$-$7$ \footnote{We refer to their original publication by \citet{Shen2020} for a detailed list of datasets used.}. These QLFs are based on observations in the IR, optical, UV and X-ray bands from which the bolometric luminosities have been constructed using a template quasar SED constructed by the authors, where the bolometric luminosity is defined to cover from the range 30$\mu$m (far IR) up to 500keV (ultra-hard X-ray). The QLFs show a considerable redshift evolution in normalisation as well as slope, with similar signs of cosmic downsizing \citep{Cowie1996, Hasinger2005}. For example, the number density of AGN with $ \log L_\mathrm{bol} \approx 46$ erg s$^{-1}$ appears to peak at $z \approx 2.4$ \citep{Shen2020}.

\begin{figure*}
     \centering
     \begin{subfigure}[b]{\columnwidth}
         \centering
         \includegraphics[width=\textwidth]{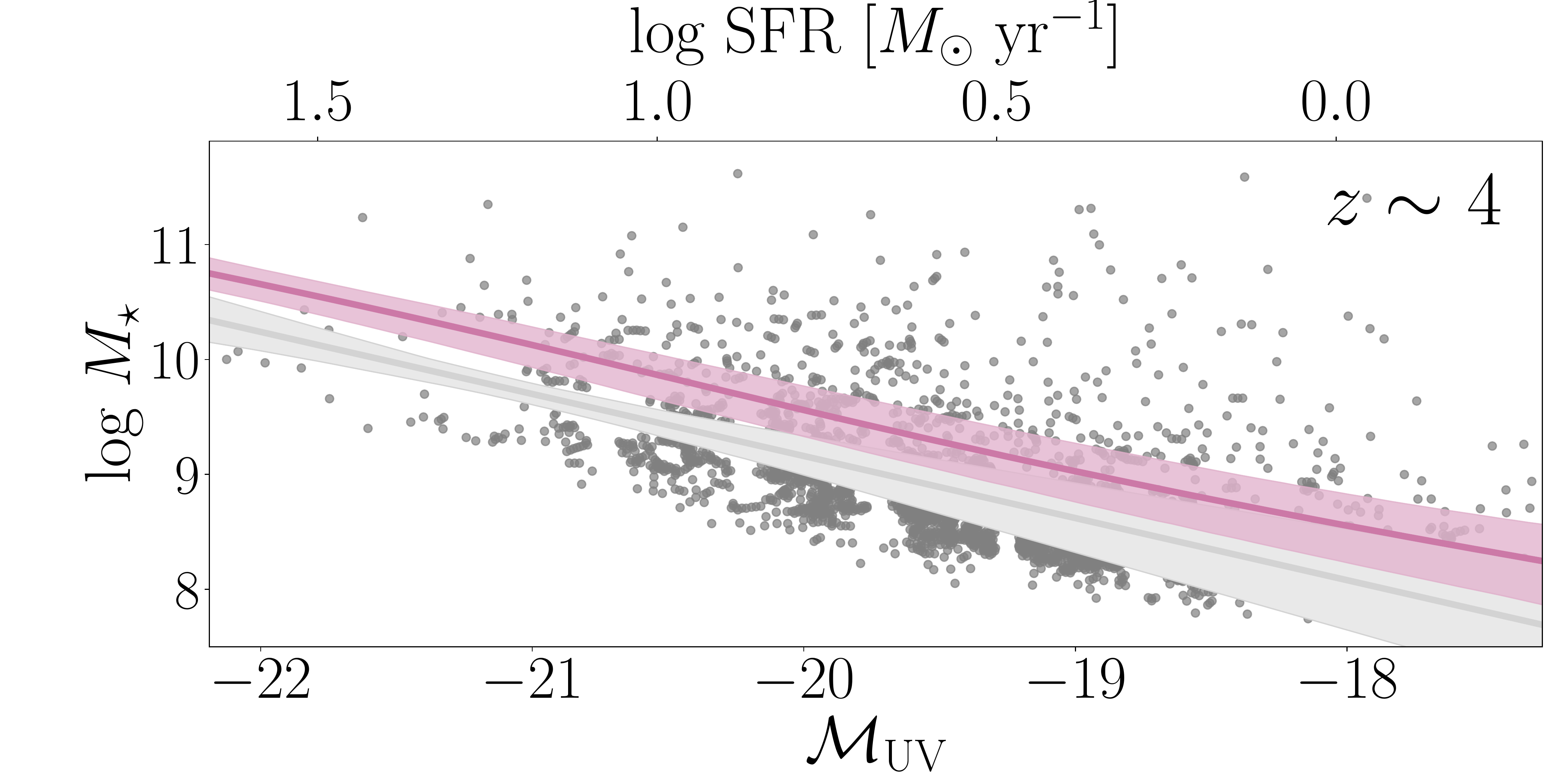}
     \end{subfigure}
     \hfill
     \begin{subfigure}[b]{\columnwidth}
         \centering
         \includegraphics[width=\textwidth]{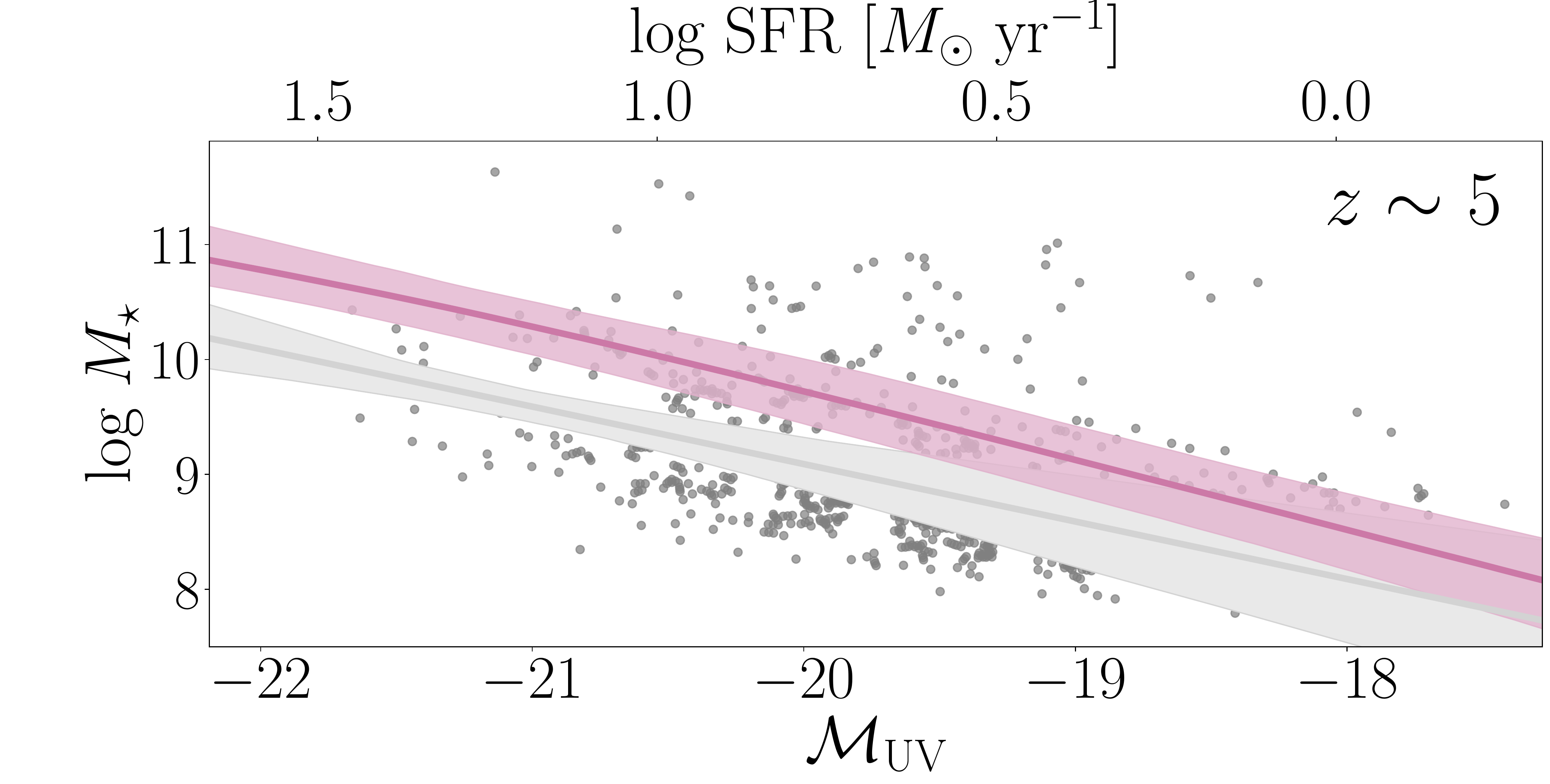}
     \end{subfigure}
     
    \begin{subfigure}[b]{\columnwidth}
         \centering
         \includegraphics[width=\textwidth]{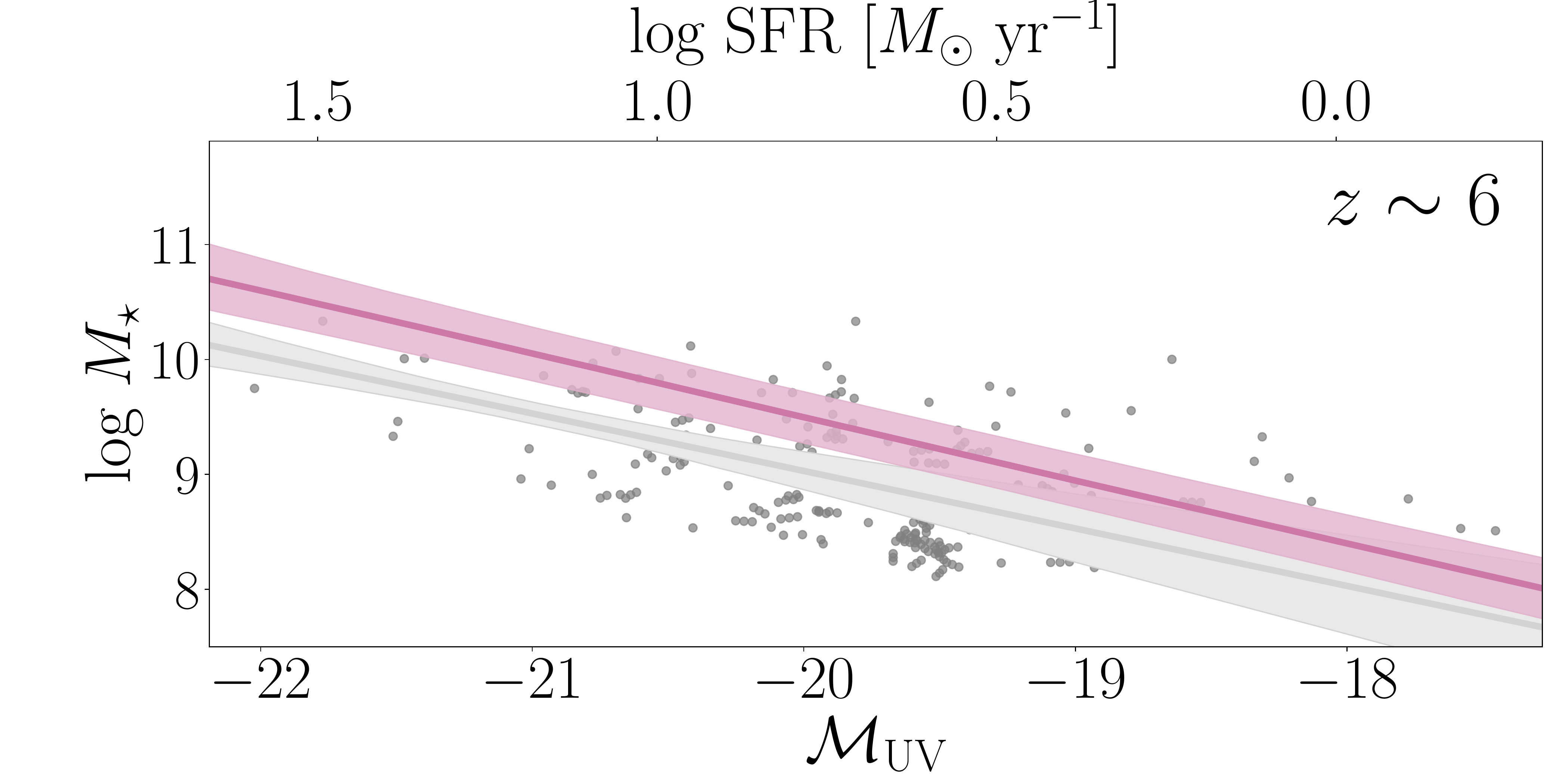}
     \end{subfigure}
     \hfill
     \begin{subfigure}[b]{\columnwidth}
         \centering
         \includegraphics[width=\textwidth]{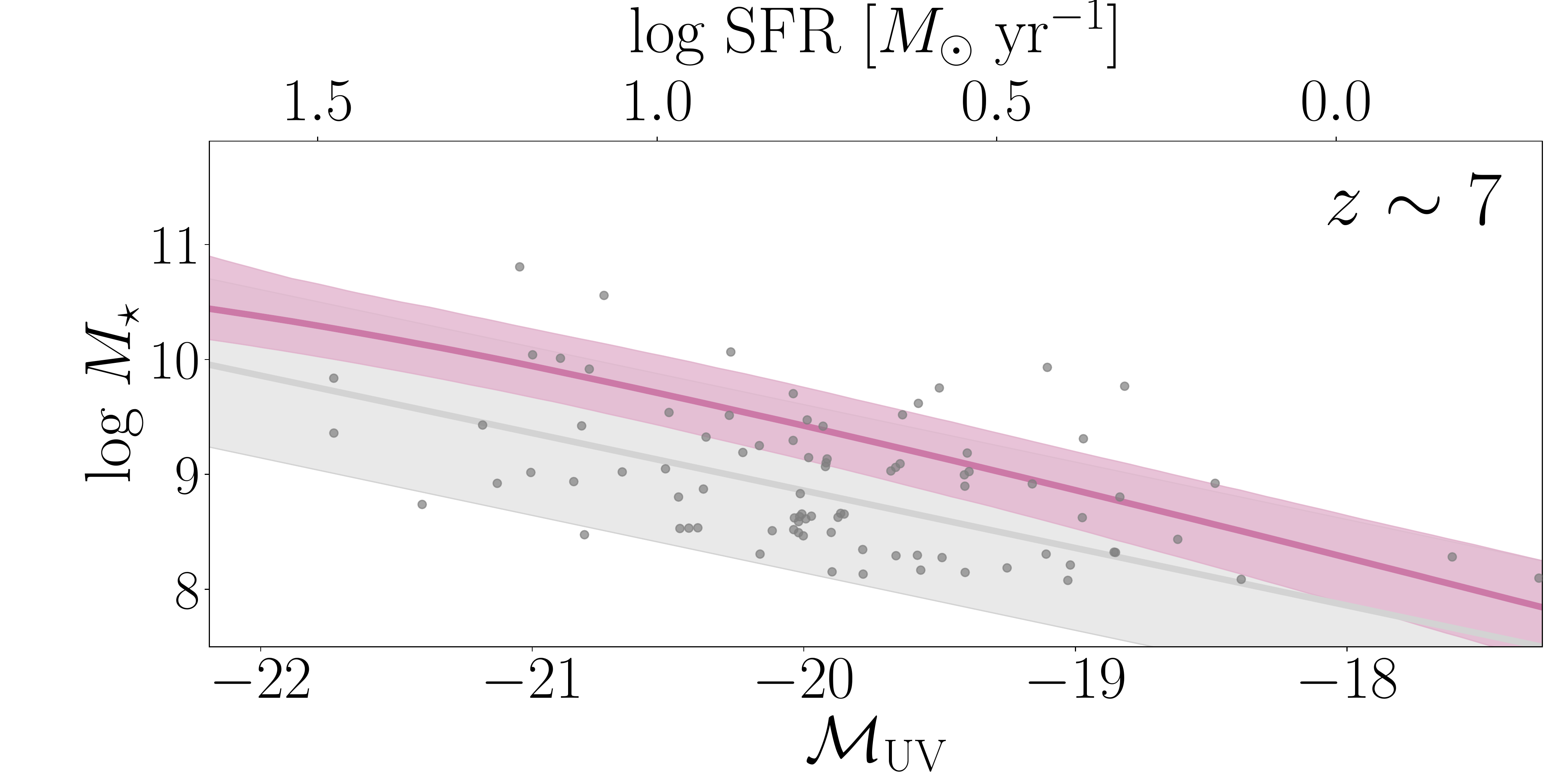}
     \end{subfigure}
     \caption{\textbf{Galaxy UV luminosity -- stellar mass relation at} $\mathbf{z=4-7.}$ \textbf{} The pink line shows the model median, with the shaded area being the 95\% credible region. The reference data is obtained from \citet[grey points]{Song2016}, while the grey lines are their reported best-fit $\log (M_\star)$ -- $\mathcal{M}_\mathrm{UV}$ relations. Also shown is the SFR corresponding to a given UV luminosity, following \citet{Madau2014}.
     }
     \label{fig:Muv_mstar_relation}
\end{figure*}
\subsection{Likelihood function and priors}
\label{subsec:likelihoodfunction}
We construct the parameter probability distributions using a least squares approach and sample the posterior distributions using the MCMC algorithm \textsc{emcee} \citep{ForemanMackey2013}. We assume a Gaussian likelihood function
\begin{equation}
        \ln \mathcal{L}(z; p) = - \frac{1}{2} \mathlarger{\sum_i}\left[
        \frac{\mathrm{R}_i(z;p)^2}{s_i(z;p)^2} + \ln \left(2 \pi s_i(z;p)^2\right)\right],
\label{eq:likelihoodfunction}
\end{equation}
where we calculate the residuals
$ R_i(z;p) = \log \phi_\mathrm{obs} \left(q_i, z\right)
                -\log \phi_\mathrm{model} \left(q_i(M_\mathrm{h}), z ; p\right)$
in log space. For the BHMF and QLF, which were collected from a single source each, we use the relative uncertainties to weight the residuals,  $s_i(z;p) = r_i(q_i)$. We use therelative uncertainties since we work in log space. 

For the GSMF and UVLF, we consider two contributions to the variance $s_i(z;p) = r_i(q_i) +
\sigma\left(R_i(z;p)\right)$. In addition to the reported uncertainties, combining the data from multiple groups introduces an additional systematic uncertainty since every
group performs the data reduction in their own ways using varying assumptions and
methods (such as dust correction prescription, SED templates, etc.). It is prohibitively difficult to account for these discrepancies in detail, so for simplicity, we assume a Gaussian distribution for these effects and use the variance of the residuals $\sigma^2\left(R_i(z;p)\right)$ as an estimate for the spread of this distribution.

To establish a basic redshift-independent prior, we can employ a set of independent uniform distributions for each parameter within sensible bounds $a_i$. The prior probability for the parameter $p_i$ is represented as $P_z(p_i) = \mathcal{U}_{a_i,b_i}$, where $a_i$ and $b_i$ denote the lower and upper bounds, respectively. Consequently, the overall prior probability is calculated as $P(p) = \prod_i P(p_i)$. Under this approach, the likelihood of all possible parameter combinations is considered equal at the beginning, and only the information obtained from the redshift bin modifies the probability of the parameters.

As an alternative, we adopt a successive prior approach. This implies that, for the distribution at redshift $z+1$, we use the posterior distribution at redshift $z$  as the prior. The posterior is given by $P_z(p) =
\frac{\mathcal{L}(z;p) P_{z-1}(p)}{\int \mathcal{L}(z;p) P_{z-1}(p) \dif p}$, where $\mathcal{L}$ is the likelihood function defined in \cref{eq:likelihoodfunction}. We assume a uniform prior at $z=0$. Essentially, this approach assumes that any evolution in the parameter will be gradual and smooth across redshift, resulting in distributions that do not differ significantly between different redshift bins. Working from this assumption, we can use more stringent constraints at low redshift (where we have more data), and then progress to the less constrained redshift regime. However, this method sacrifices the independence of the distributions between redshift bins.

\subsection{Recreating the population statistics}
We constrain the free parameters of the model by matching the observed number density functions (\cref{subsec:datasets}) to our model (\cref{sec:modeldescription}) using the previously described probabilistic approach (\cref{subsec:likelihoodfunction}). Using this approach, we can derive probability distributions for the parameter. These distributions can then be employed to construct sample number density functions. A summary of the parameter distributions can be found in \cref{ApC:parametertable}. The resulting number densities are shown in \cref{fig:mstar_ndf_intervals,fig:Muv_ndf_intervals,fig:mbh_ndf_intervals,fig:Lbol_ndf_intervals}. We mark the 68\%, 95\% and 99.7\% credible regions for the average number densities at every redshift. 

For the parameter distributions to be well-behaved, especially at high redshift where data is sparse, we have to make several additional assumptions:
\begin{enumerate}
        \item For the GSMF and UVLF, we treat the critical mass $M_\mathrm{c}$ and AGN feedback parameter $\delta$ as free parameters up to redshift $z=2$ and $4$, respectively. At higher redshift, the two parameters are marginalised over when needed, since the AGN-dominated regime is not sampled (see \cref{fig:mstar_ndf_intervals,fig:Muv_ndf_intervals}).
        \item For the GSMF, we enforce an upper limit on the normalisation parameter A so that ${M_\star}/{M_\mathrm{h}}$ peaks at the cosmic baryon fraction $\approx 0.2$.
        \item For the BHMF and QLF, we set the critical mass, $M_\mathrm{c}$, to the most probable parameter estimates provided by the maximum a posteriori (MAP) estimator of the GSMF, i.e. $M_\mathrm{c}^\mathrm{bol} = M_\mathrm{c}^\star$ and $M_\mathrm{c}^\mathrm{UV} = M_\mathrm{c}^\star$, motivated by the connection between stellar mass and black hole mass growth.
        \item For the QLF, we fix the parameter of the ERDF ($\lambda_\mathrm{c}$, $\rho$) for $z>0$ to the MAP estimates at $z=0$, i.e. assume an un-evolving ERDF.
\end{enumerate}
The scarcity of data in the high-mass end of the GSMF and the bright end of the UVLF due to low number density hinders detailed study of the evolution of $M_\mathrm{c}$ and $\delta$ parameters at high redshift. However, the number densities can still be closely matched when treated as nuisance parameters and some information about their evolution can be inferred (see \cref{sec:redshiftevolution}). 

There is no strong reason to assume the ERDF is un-evolving, however, the observations are still reasonably well reproduced at higher redshift justifying this assumption. Note that the distributions for the bright end of the QLF are extremely localised for $z=1$-$3$; this exemplifies that the bright end is completely determined by the shape of the ERDF (while the faint end is strongly constrained by the amounts of available observations).

\subsection{The black hole duty cycle}
\label{subsec:dutycycle}
As discussed in \cref{subsubsec:BHMF}, to connect the halo mass function with the Type 1 AGN BHMF we need to choose a value for the duty cycle $f_\mathrm{d}$.

Based on our results so far, we can already make a rough estimate of this value. \citet{Hatziminaoglou2009} find that the ratio between Type 2-to-Type 1 AGN is approximately 2-2.5, based on their low-luminosity, low-redshift AGN sample. To estimate the ratio of active to non-active SMBHs, we can use the ERDF obtained by modelling the QLF (\cref{subsubsec:QLF}), and argue that SMBH below a cutoff Eddington ratio can be considered inactive. For this cutoff, we choose an Eddington ratio of $\lambda_\mathrm{cutoff}=0.01$, since this is the regime where AGN switch between Jet mode and radiative mode \citep{Heckman2014} and the because the sample collected by \citet{Baron2019} primarily contains supermassive black holes with $\lambda>0.01$.  The same cutoff value was used by \citet{Aird2018}. For the MAP parameter estimate of the ERDF at $z=0$, the probability of a randomly selected black hole having an Eddington ratio below 0.01 is ~0.90. Since we assume our ERDF to be mass-independent this means we expect 1 in 10 black holes to be active. Combining these estimates, we therefore find a duty cycle of $f_\mathrm{d} \approx 3$\%.

As a comparison, we also use values of the duty cycle found by \citet{Aird2018}, based on a sample of X-ray selected AGN. They find that the duty cycle changes as a function of redshift, stellar mass of the host galaxy, and SFR. However, to keep it simple in this work, we simply use their minimum value ($\sim 0.15$\%) and maximum value  ($\sim 20$\%) to establish bounds for our relations.

\section{Assessing the relation between observables}
\label{sec:validation}
In the preceding section, we showed that our model can generate number densities that closely align with available observations. Nevertheless, utilising observed number densities to constrain the model parameters does not adequately reflect the model's performance. It is essential to validate the model's performance by comparing its output to a dataset that was independent of the data used for calibration. Since a key feature of the model is the ability to relate different observable quantities (see \cref{subsec:obsinterrelation}), it is insightful to compare these interrelations to available datasets.
\begin{figure}
    \centering
    \includegraphics[width=\columnwidth]{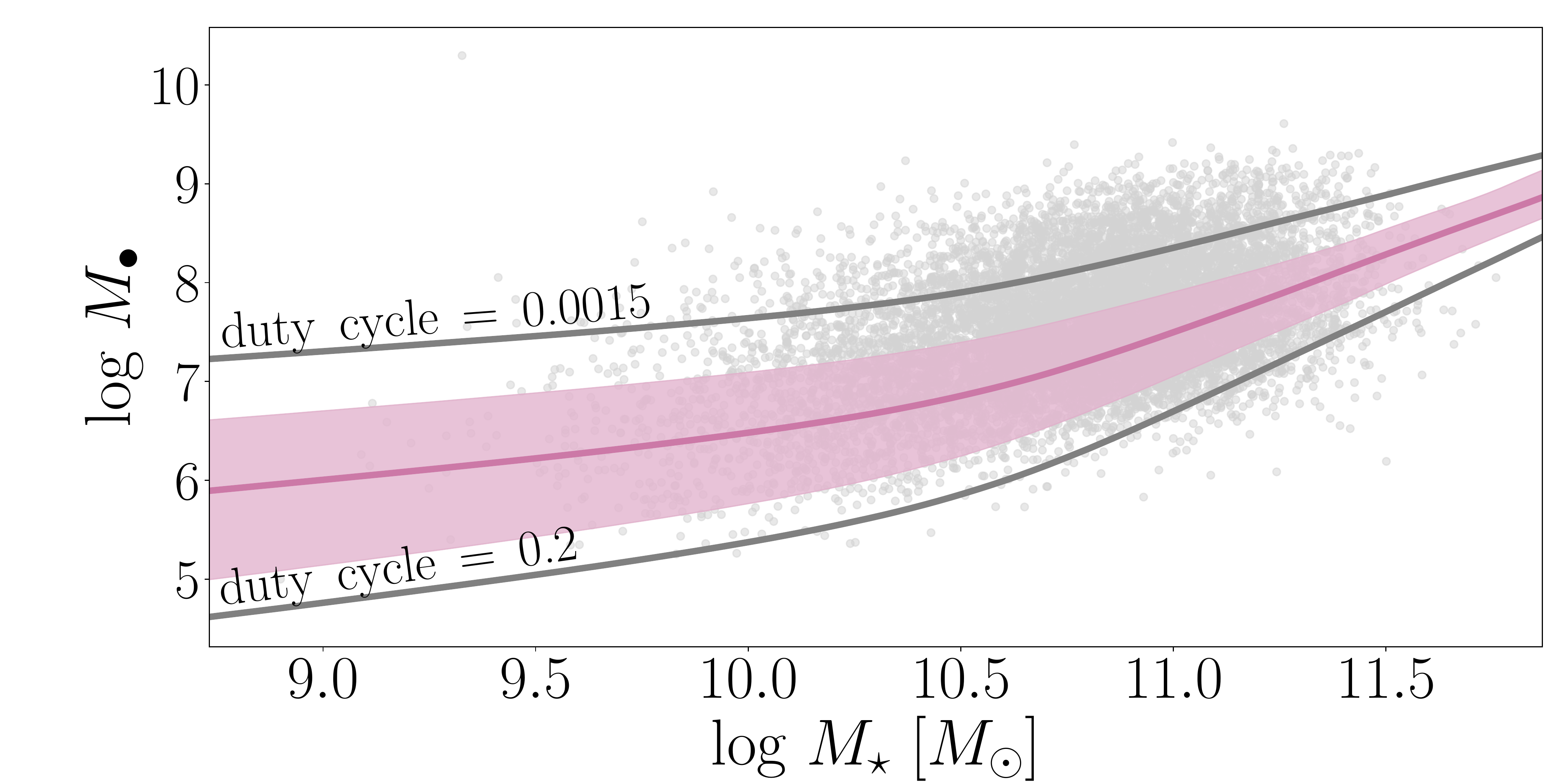}
    \caption{\textbf{SMBH mass -- galaxy stellar mass relation at} $\mathbf{z=0}$\textbf{.} The reference observations are obtained from \citet{Baron2019} for a sample of Type 1 and Type 2 AGN at $z<0.3$. The pink line shows the model median and the 95\% credible region, using a duty cycle $f_\mathrm{d} = 3$\%. The grey and black lines show the relation using MAP estimates for the parameter of the GSMF and BHMF models with $f_\mathrm{d}=0.15$\% and  $f_\mathrm{d}=20$\%, covering the range of values found by \citet{Aird2018}.}
    \label{fig:mstar_mbh_relation}
\end{figure}
\subsection{Galaxy stellar mass -- UV luminosity relation}
\label{subsec:mstarmuvrel}
In star-forming galaxies, the relationship between stellar mass and SFR is known as the galaxy's main sequence. This near-linear relation is well-studied and has been reported in the literature \citep{Brinchmann2004, Whitaker2014, Sherman2021}. Since the intrinsic UV luminosity of a galaxy is an indicator of the instantaneous SFR, the stellar mass-UV luminosity relation can serve as a proxy to study the evolution of the galaxy's main sequence. We collect data on this relation from \citet{Song2016}, which covers a redshift range of $z=4$-$7$. This range is ideal for studying our model output because the GSMF and UVLF are well sampled within this range.

\cref{fig:Muv_mstar_relation} shows the anticipated relationship derived from the model in comparison to observational data. Although the model aligns with the range of observations, it consistently overestimates stellar masses for a given UV luminosity when compared to the $\log M_\star$ -- $\mathcal{M}_\mathrm{UV}$ relation as inferred by \citet{Song2016}.

The data themselves hold a hint for a potential resolution to this discrepancy. When considering fixed luminosity bins, the stellar mass distribution is asymmetric, with a skew towards higher masses. This asymmetry can be explained by the data likely comprising two distinct galaxy populations: one population comprises galaxies that actively form stars and adhere to the main sequence, resulting in a tight correlation in the $\log  M_\star$ -- $\mathcal{M}_\mathrm{UV}$ plane. The other population consists of dusty star-forming or inactive galaxies, hinted at by recent JWST observations \citep{Naidu2022}.

 As the observed quiescent fraction increases towards lower redshift, it is important to note that the asymmetry also increases. The modelled relation is situated approximately halfway between the minimum and maximum observed stellar masses per luminosity bin, implying that the source of the discrepancy lies in our model's assumption of a one-to-one correspondence between stellar mass/UV luminosity and halo mass, as defined by \cref{eq:galaxyhalorelation}. This is a simplifying assumption, as one in reality would expect a distribution of stellar masses and UV luminosities for a given halo mass. Indeed, recent ALMA observations are beginning to show galaxies that lie significantly above the main sequence \citep{Algera2023}. 

If the distribution were symmetric and reasonably localised, it would minimally influence the relation between stellar mass/UV luminosity and halo mass, primarily creating a spread around the relation. However, since the distributions are skewed (with an identical expectation value but a large portion concentrated around values lower or higher than the expectation value), it may result in an overestimation of the modelled mean values. This is because lower-mass halos contribute more significantly due to their higher number densities. In principle, this effect can be taken into account in the model by including this asymmetric distribution using the machinery described in \cref{ApB:scatter}, but in practice the distribution of stellar mass/UV luminosity for a given halo mass is hard to constrain observationally. A proof-of-concept example is shown in \cref{fig:mstar_Muv_scatter_distribution}. At this stage, the model reproduces the stellar mass -- UV luminosity relation reasonably well given the simplicity of the model and scatter in the observed relation, but it is good to keep this systematic bias in mind if the model is used in practice.

\subsection{SMBH mass -- galaxy stellar mass relation}
\label{subsec:mbhmstarrel}
Our model not only enables us to establish connections between various stellar properties of galaxies but also allows us to link stellar and AGN properties. In the local universe, the relationship between galaxy stellar mass and the mass of the central SMBH is relatively well-defined, making it suitable for testing our model's accuracy.

To investigate this, we employ the dataset compiled by \citet{Baron2019} for Type 1 and Type 2 AGN at $z < 0.3$. \cref{fig:mstar_mbh_relation} depicts the relationship between stellar and SMBH masses at $z=0$ as derived from our model. When assuming a duty cycle $f_\mathrm{d} = 3\%$, as discussed in \cref{subsec:dutycycle}, we find a good agreement between observations and model prediction. Furthermore, using the extreme values of $f_\mathrm{d}=0.15$\% and $f_\mathrm{d}=20$\% yields sensible bounds that are also consistent. Similar results have been found by studies on the AGN fraction, for example, by \citet{LaMarca2024}. We can see that a lower duty cycle leads to black holes of a given mass being mapped to a lower halo mass, which corresponds to a lower stellar mass. 

Of course, the actual duty cycle relation is a lot more complicated than this simple scaling. \citet{Gilli2007, Hatziminaoglou2009} and \citet{U2022} suggest that the Type 2-to-Type 1 ratio evolves with black hole luminosity (with is in our model directly linked to black hole mass), and the ratio between active and non-active black holes (commonly described through the occupation fraction and duty cycle) is expected to evolve with black hole mass and redshift \citep{Shankar2013, Volonteri2017, Heckman2014} which is related to the cosmic downsizing discussed earlier. In particular, the fact that our way of estimating the active to non-active ratio using the ERDF, which is constructed using the quasar luminosity function (i.e. only active BHs) and is only weakly constrained on the low-Eddington ratio end (see \cref{subsubsec:QLF}), leads to such a good match to the available data may be in part a coincidence but seems to suggest the correct order of magnitude increase needed to reconcile the model with the available data. More sophisticated models for the duty cycle are given e.g. by \citet{Aird2018} and \citet{Georgakakis2019}. \citet{Georgakakis2019} find in their model that X-ray luminous AGN are primarily found in haloes with masses around $10^{11}$ -- $10^{12} M_\odot$, matching the turnover mass found in our model. \citet{Aird2018} find that the ERDF for AGN in star-forming galaxies evolves between $z=0.5$ and $z=2$. The average accretion rate increases over this range for low and high-mass galaxies, with the distributions also becoming steeper. This may explain why our model slightly underestimates the number densities of bright AGN ($\log L_\mathrm{bol} > 47$ erg/s) at $z=2$ and $z=3$.

\begin{figure}
     \centering
     \includegraphics[width=\columnwidth]{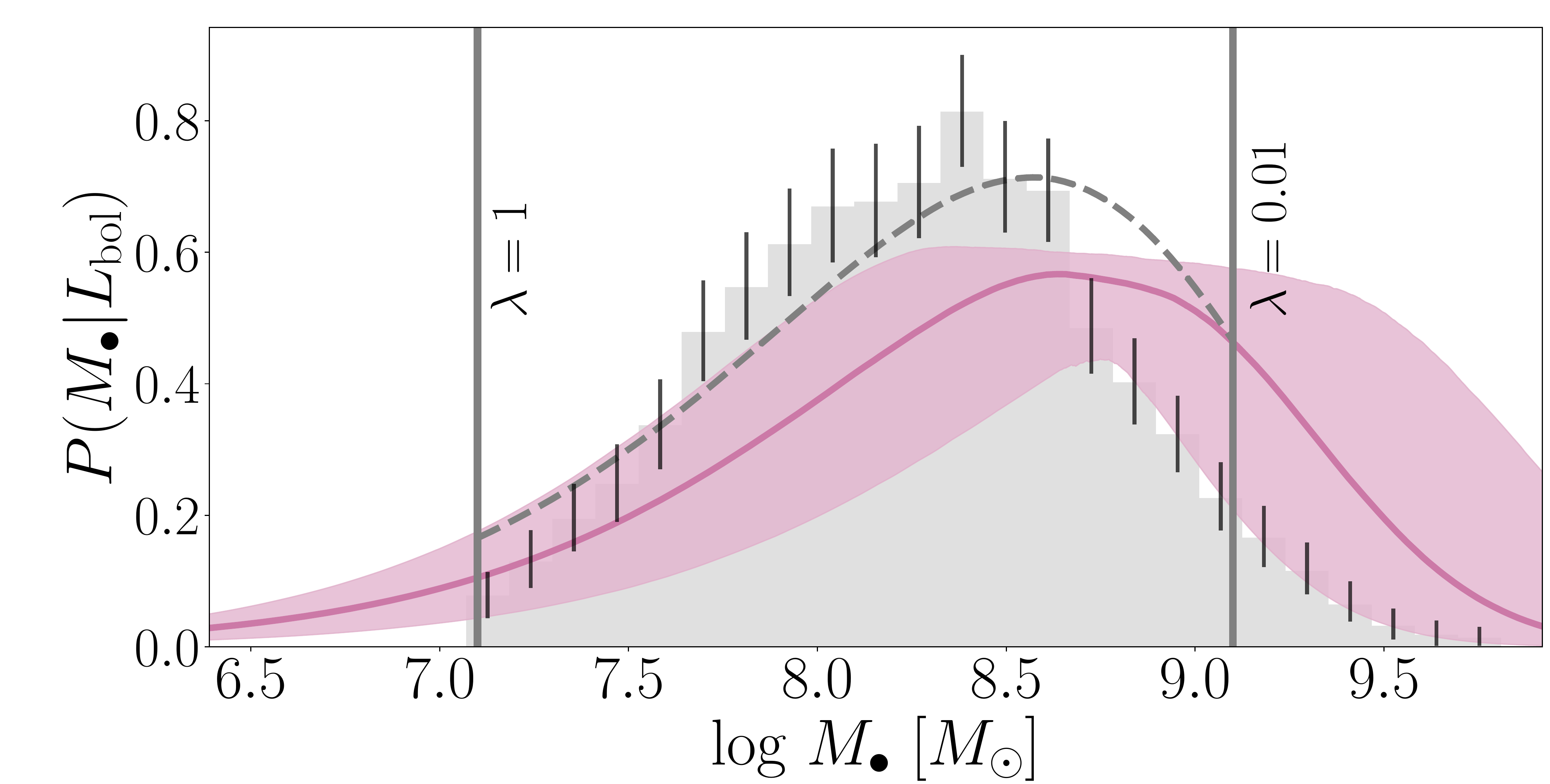}
     \caption{\textbf{Probability density of black hole masses for} $\log L_\mathrm{bol} = 45.2$ [erg s$^{-1}$] \textbf{at} $\mathbf{z=0}$\textbf{.} The pink line shows the model median distribution with the 95\% credible region, while the histogram shows the observational data on Type 1 AGN gathered from \citet{Baron2019}. The grey vertical lines show where the Eddington ratio reaches 0.01 and 1, which marks the boundaries where we expect additional effects to play a role that are not included in the model. The grey dotted line is the expected distribution calculated from the MAP parameter estimates when invoking a hard cutoff outside of these boundaries. Doing so decreases the mismatch between the distribution means from 0.5 dex to 0.3 dex compared to the observational sample. The black lines mark the 95\% credible region for the observational bins, based on the uncertainties in the black hole masses.}
     \label{fig:Lbol_bh_mass_distribution}
\end{figure}

\subsection{SMBH mass -- AGN bolometric luminosity relation}
\label{subsec:mbhlbolrel}
Finally, we can connect the properties of active SMBH. As described in
\cref{subsec:obsinterrelation}, the relation between black hole mass and AGN luminosity is
a power law with a slope given by $\eta / \theta$. Since both parameters are assumed
to be positive, the model predicts that average AGN luminosity increases with black hole
mass. As argued before however, the number statistics of black hole masses plays a crucial role in observed black hole quantities. To include this effect, we can estimate the black
hole mass distribution from the conditional ERDF (see \cref{subsubsec:QLF}), and calculate the mean black hole mass for every luminosity this way. 

This approach not only considers the intrinsic AGN luminosities but also incorporates the number statistics of black hole masses. A notable advantage of this approach is that it solely relies on information from the bolometric luminosity function, thereby avoiding the uncertainties associated with the BHMF.

 At low luminosities, the number densities play no significant role and the produced relation is similar to the direct calculation (and is within model uncertainties and the proposed scaling of the BHMF). For luminosities $>10^{45}$ erg s$^{-1}$ (at $z=0$), the exponential drop in the BHMF becomes dominant and the expected black hole mass stops to grow with luminosity. Simply put, at $z=0$ active black holes with $M_\bullet>10^{8.5} M_\odot$ are expected to be so rare that high luminosities are much more likely caused by large Eddington ratios rather than very massive black holes. The difference between the two predictions is in essence a selection effect: if AGN are selected based on their luminosity (as is done in observational surveys), the flattening relation is to be expected. If the AGN were selected based on their mass (so that the number statistics of the BHMF play no role), we'd expect the simple power law relation.

In \cref{fig:Lbol_bh_mass_distribution}, we show the empirical black hole mass distribution from a luminosity-selected sample of \textasciitilde 2000 Type 1 AGN from \citet{Baron2019}, with mean bolometric luminosity of $10^{45.2}$ erg s$^{-1}$ and a scatter of \textasciitilde 0.2 dex, and the modelled distribution obtained from the conditional ERDF. Overall, the two distributions show good agreement. However, the expected and observed mean black hole mass differ by approximately 0.5 dex. The model also underestimates the probability of high Eddington ratios (and overestimates the probability of low Eddington ratios) compared to the data. Additionally, the model predicts wider tails to the distribution than observed. 

At the bright end, the simple Eddington model of isotropic accretion breaks down for $\lambda \geq 1$. It is therefore expected that our model, which is built on the linear $L_\mathrm{bol}$-$M_\bullet$ relation derived from Eddington theory, would not match observations in this regime. Previous studies have shown that the fraction of AGN drops sharply for accretion rates with $\lambda \geq 1$ \citep{Heckman2004}. At the high-mass end (low Eddington ratios, $\lambda \sim 0.01$), AGN tend to be more likely supported by advection-dominated accretion flows, with geometrically thick and optically thin accretion disks, which affects the black hole mass-luminosity relation as well as the ability to estimate black hole masses using broad line emissions \citep{Narayan2005}. The radiative mode -- Jet mode transition does not occur instantaneously at any specific Eddington ratio, but becomes more likely as the Eddington ratio decreases \citep{Best2012, Russell2013}, which matches the discrepancy in our model for $\lambda \sim 0.01$. If we, for simplicity, invoke hard cutoffs for $\lambda < 0.01$ and $\lambda > 1$ (and normalise the distribution accordingly), the mismatch between expected and observed mean black hole mass decreases to \textasciitilde 0.3 dex, as can be seen from the dotted grey line in \cref{fig:Lbol_bh_mass_distribution}.

\section{Evolution of the population statistics over cosmic time}
\label{sec:redshiftevolution}
\begin{figure*}
     \centering
     \begin{subfigure}[b]{\columnwidth}
         \centering
         \includegraphics[width=\textwidth]{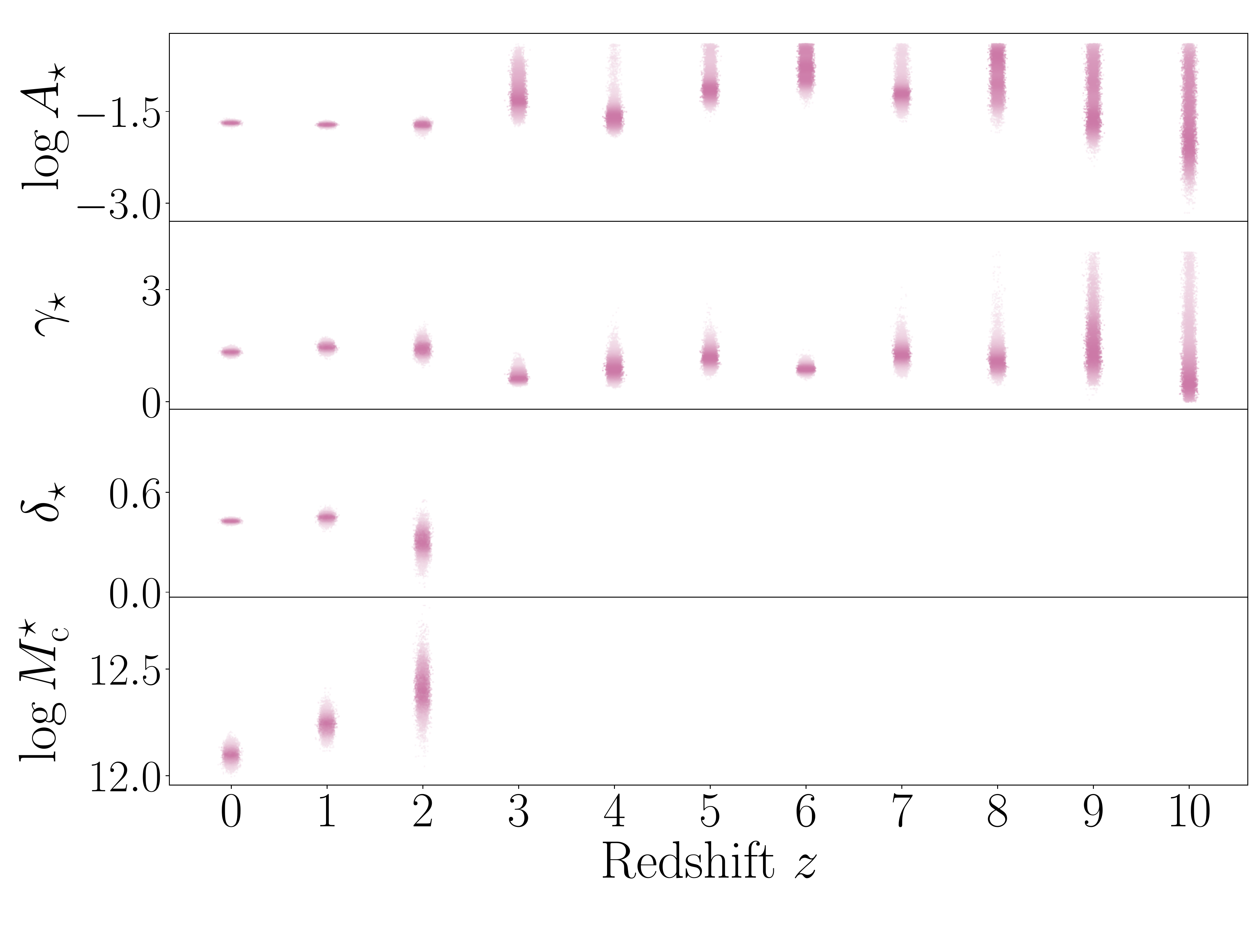}
         \caption{}
         \label{fig:mstar_parameter}
     \end{subfigure}
     \hfill
     \begin{subfigure}[b]{\columnwidth}
         \centering
         \includegraphics[width=\textwidth]{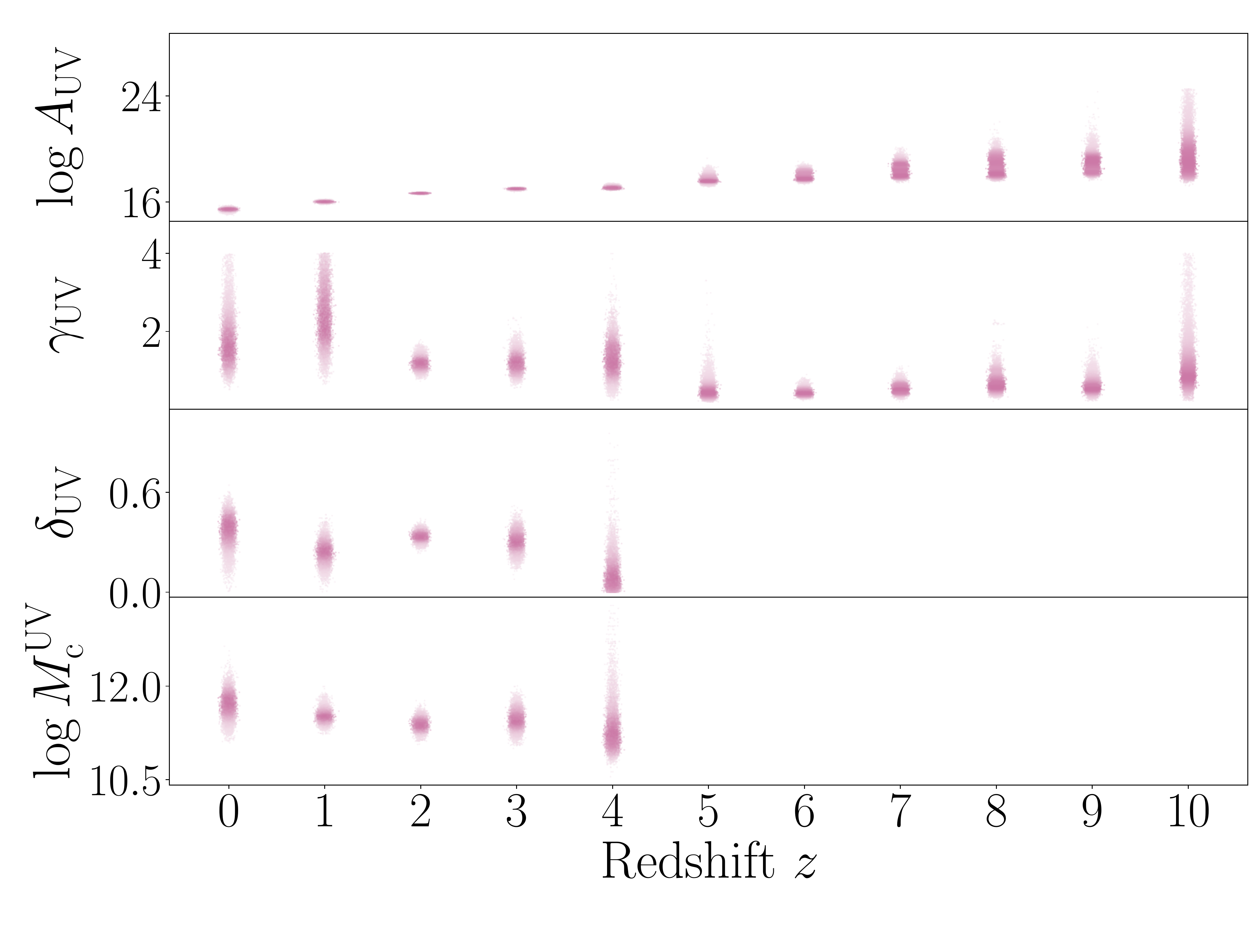}
         \caption{}
         \label{fig:Muv_parameter}
     \end{subfigure}
     
    \begin{subfigure}[b]{\columnwidth}
         \centering
         \includegraphics[width=\textwidth]{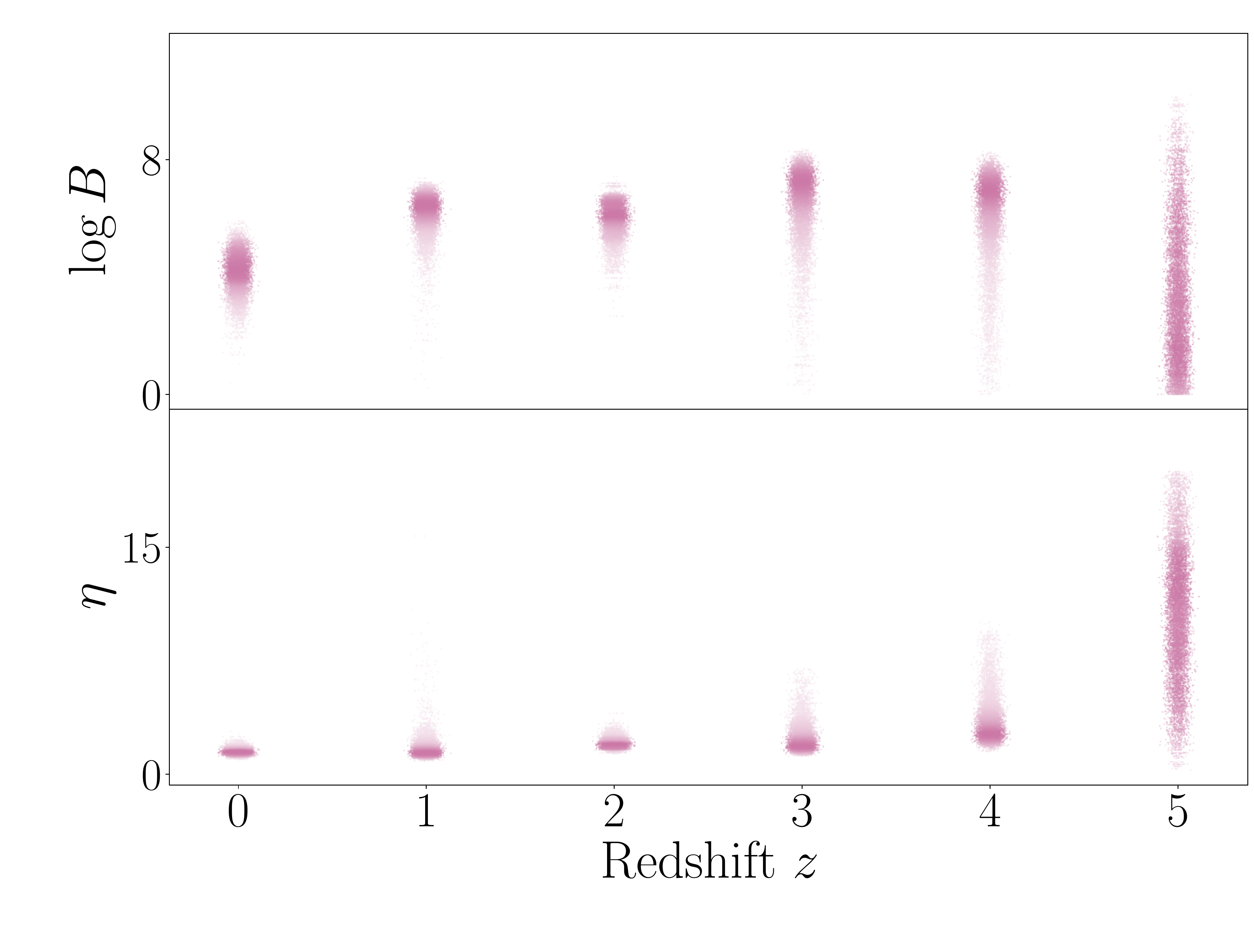}
         \caption{}
         \label{fig:mbh_parameter}
     \end{subfigure}
     \hfill
     \begin{subfigure}[b]{\columnwidth}
         \centering
         \includegraphics[width=\textwidth]{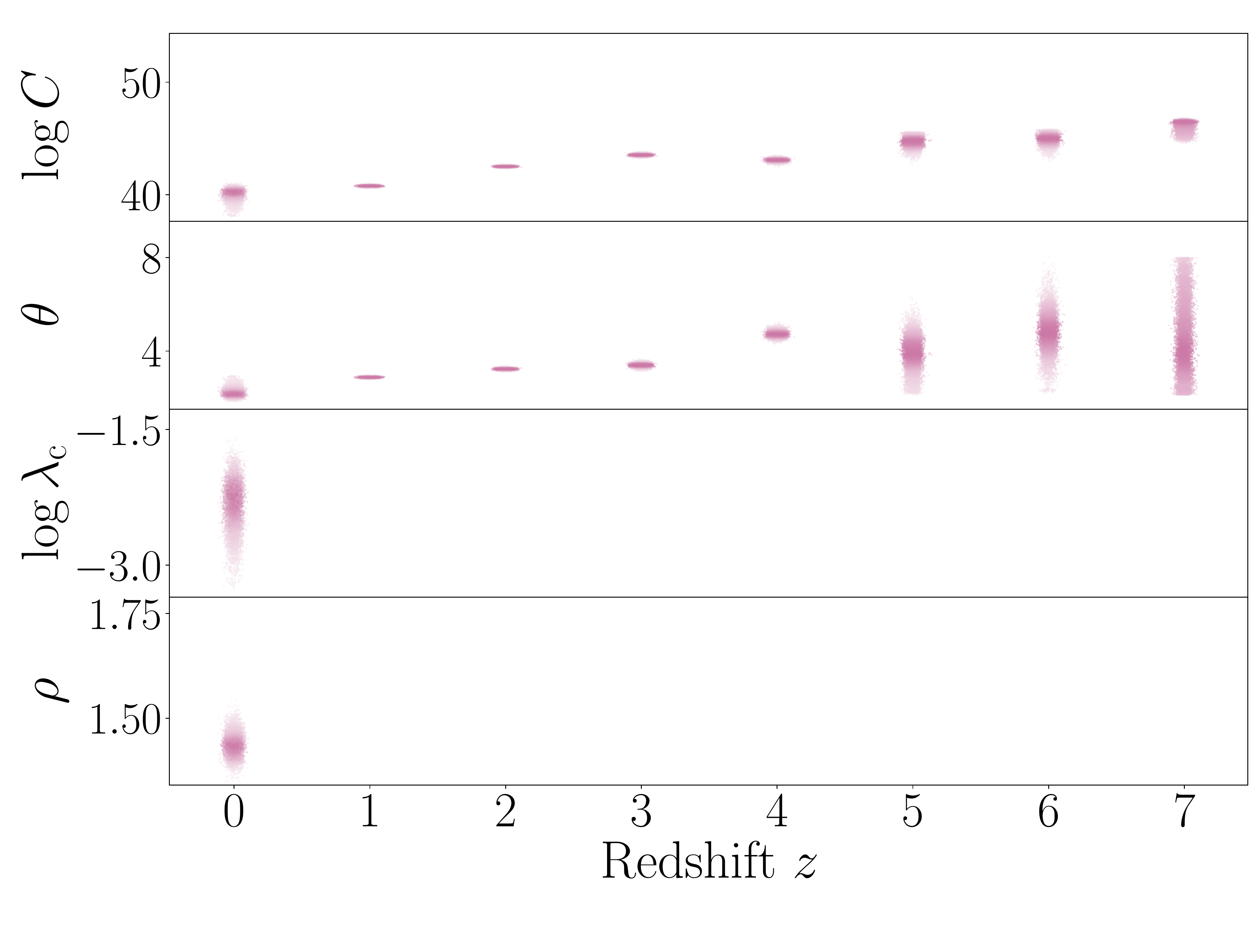}
         \caption{}
         \label{fig:Lbol_parameter}
     \end{subfigure}
     \caption{\textbf{Parameter sample across redshift:} A random sample of the model parameter drawn from the posterior distributions at every redshift. Increased colour saturation indicates a larger value for the probability density. (a) Stellar mass function parameter sample. The critical mass parameter $M_\mathrm{c}^\star$ and AGN feedback parameter $\delta$ are weakly constrained at $z>2$ and treated as nuisance parameters at higher redshift. (b) UV luminosity function parameter sample. The normalisation $A_\mathrm{UV}$ is given in erg s$^{-1}$ Hz$^{-1}$ $M_\odot^{-1}$. (c) Type 1 active black hole mass function parameter sample. (d) Quasar luminosity function parameter sample. The normalisation $C$ is given in erg s$^{-1}$ $M_\odot^{-1}$. }
     \label{fig:parameter_sample}
\end{figure*}
With the validity of the model established and potential caveats and pitfalls demonstrated, we can turn to redshift evolution of the model quantities. For each integer redshift bin, we can estimate the relationship between observable quantities and halo mass non-parametrically by matching the model to observations. This allows us to obtain a comprehensive picture of these relations across the entire observed redshift range.

 This section is split into two parts, first, we interpret the evolution of the modelled number density function and compare it to more comprehensive modelling approaches, while the second part is focused on extrapolating the observed trends to as-of-yet unobserved redshift.

\subsection{Comparison with observations and other models}
\label{subsec:redshiftevodisscussion}
\cref{fig:parameter_sample} shows $10^4$ samples of the parameter posterior distribution drawn at every redshift bin and for every quantity, with the level of colour saturation representing the posterior probability. The parameter distributions for the GSMF and UVLF (\cref{fig:mstar_parameter,fig:Muv_parameter}) are marginalised over the critical mass $M_\mathrm{c}$ and the AGN feedback parameter $\delta$ after $z=2$ and $z=4$, respectively.
\subsubsection{Galaxy evolution}
\begin{figure}
     \centering
     \begin{subfigure}[b]{\columnwidth}
         \centering
         \includegraphics[width=\textwidth]{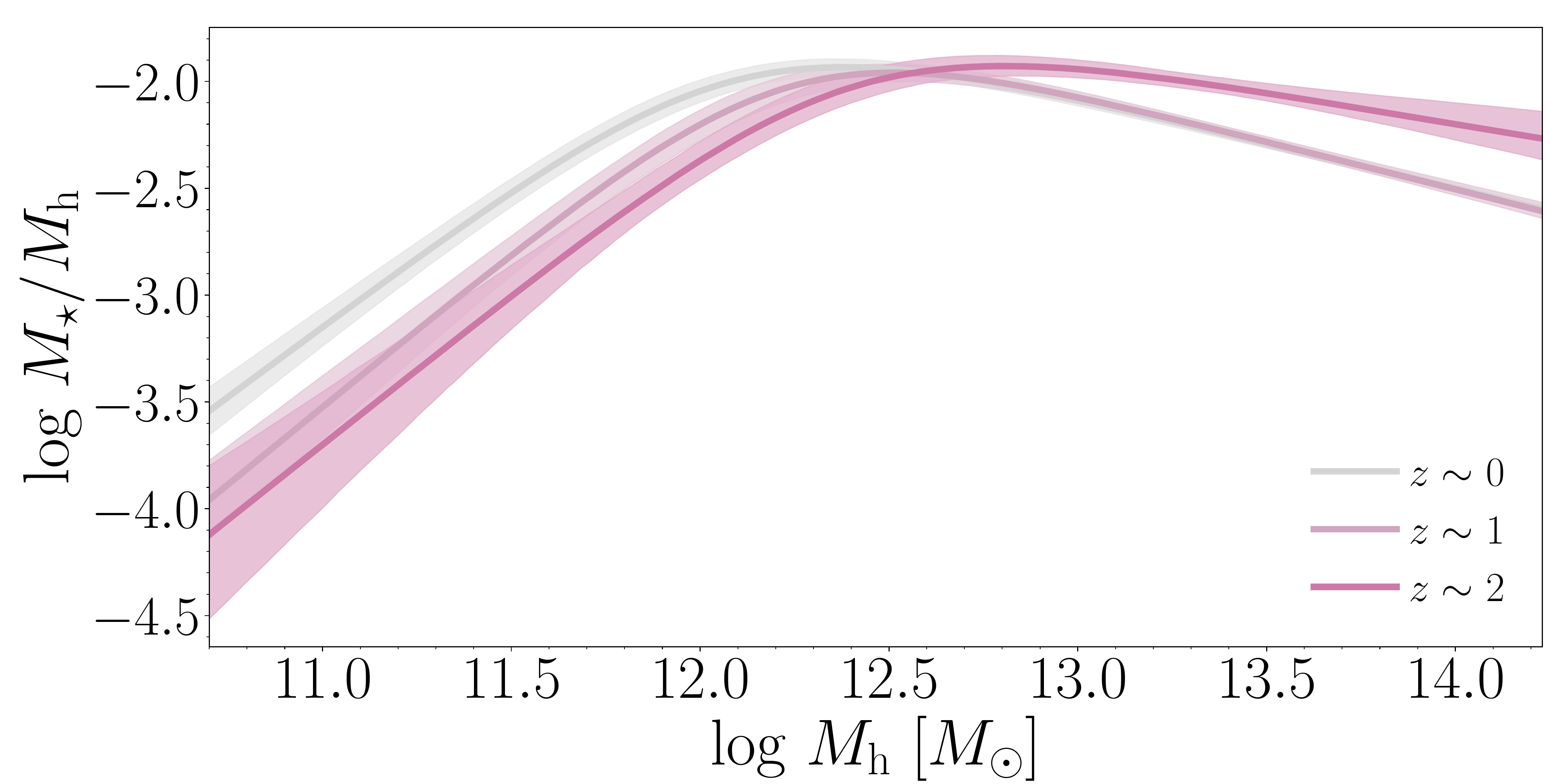}
         \caption{}
         \label{fig:mstar_qhmr_low_z}
     \end{subfigure}

     \begin{subfigure}[b]{\columnwidth}
         \centering
         \includegraphics[width=\textwidth]{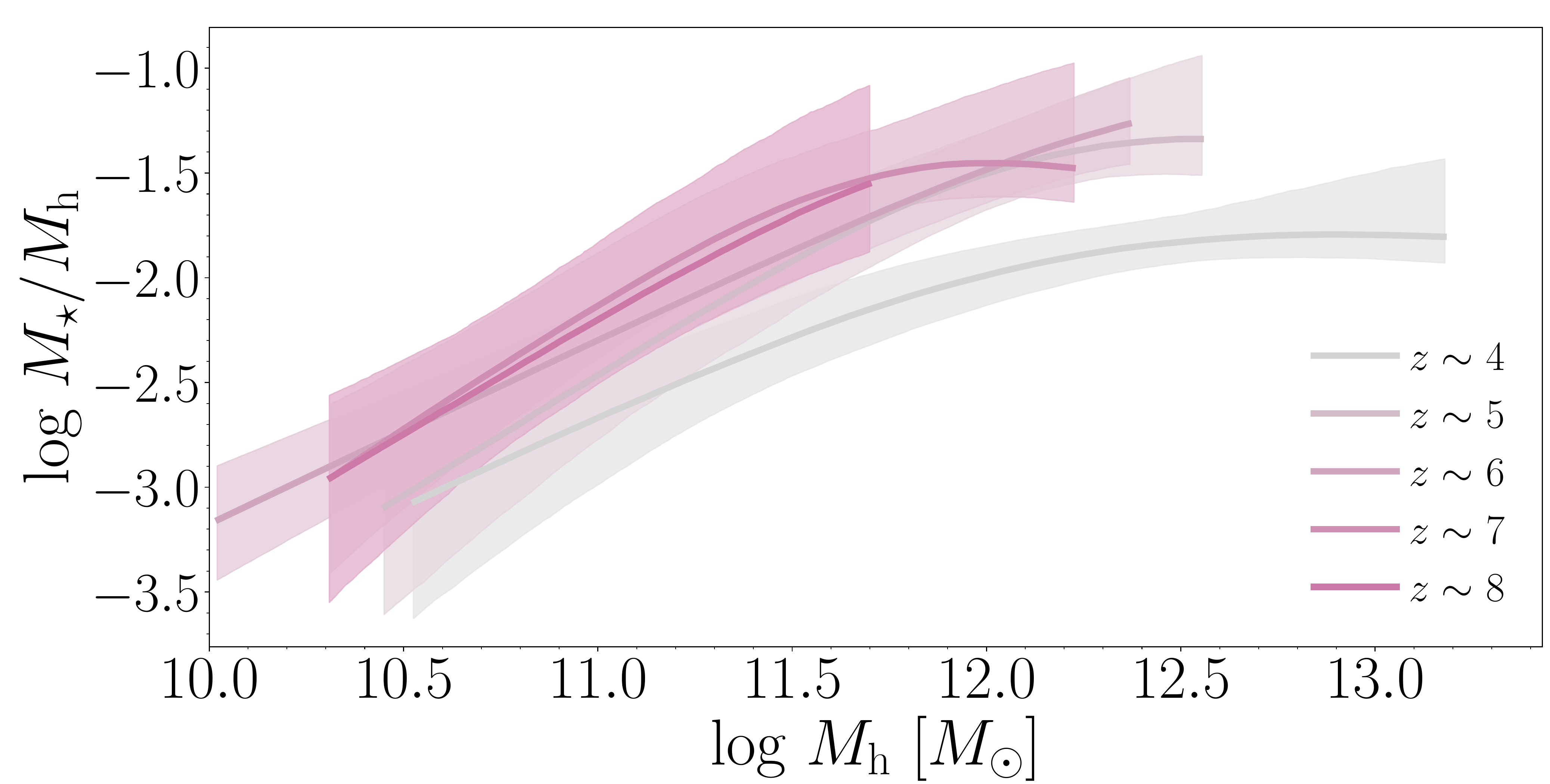}
         \caption{}
         \label{fig:mstar_qhmr_high_z}
     \end{subfigure}
     \caption{\textbf{Stellar mass -- halo mass ratio across redshift.} The SHMR is calculated to the parameter samples shown in \cref{fig:parameter_sample}. The pink line shows the model median while the shaded areas represent the 95\% credible regions. (a) Low redshift. (b) High redshift. The shown ranges in the figure coincide with the observed GSMFs. The high-mass slope is outside of the observed range}
     \label{fig:stellartohalomassratio}
\end{figure}
Our results show that the critical mass parameter ($M_\mathrm{c}^*$) and the normalisation parameter $A$ in the galaxy stellar mass function both increase with redshift. Specifically, $M_\mathrm{c}^*$ increases up to the highest redshift bin constrained in our study ($z=2$), while $A$ approaches values close to the cosmic baryon fraction at $z=6$. On the other hand, the feedback parameters ($\gamma$ and $\delta$) exhibit some variation over their respective redshift ranges but generally show little to no evolutionary trend.

The stellar mass to halo mass ratio (SHMR) provides a straightforward way to analyse the trends in galaxy evolution. \cref{fig:stellartohalomassratio} shows that at $z < 3$, the SHMR exhibits a constant slope and normalisation. Notably, the critical mass (location of the turnover point in the SHMR) is the only parameter that undergoes a significant change. At $z > 3$, the SHMR is only plotted over the ranges with observational data for the GSMF. This emphasises the lack of constraints on the AGN feedback-dominated high-mass slope at high redshift. In contrast, the stellar feedback-dominated low-mass slope remains consistent within the credible regions of the model. Interestingly, the critical mass appears to decrease with increasing redshift. While the normalisation is higher compared to low redshift, the uncertainty is too large to infer any trends.

When compared to the SHMR obtained by a more comprehensive empirical model such as \textsc{UniverseMachine} \citep{Behroozi2019}, we find that the evolution of the critical mass and low-mass slope are consistent with their results. On the other hand, in \textsc{UniverseMachine} the high-mass slope of the SHMR consistently increases with redshift, while they find a stronger evolution in the normalisation which consistently decreases with redshift. \textsc{UniverseMachine} has a larger focus on environmental effects and constructs the galaxy properties by populating dark matter halo in a probabilistic fashion, which leads to a more diverse galaxy population compared to our model. This likely contributes to this discrepancy. In particular, \citet{Behroozi2019} finds a redshift- and halo mass-dependent scatter in the stellar mass -- halo mass relation, with an average of $\approx 0.25$ dex, in line with observational findings \citep{Wechsler2018}.

In \cref{fig:mstar_scatter_ndf}, we illustrate the effects of the scatter assumption on our estimate of the GSMF, calculated as described in Appendix B. The slope at high masses is highly sensitive to the amount of scatter and, consequently, normalisation, whereas the slope at low masses is not. We find that the stellar feedback parameter $\gamma$ and turnover mass $M_\mathrm{c}$ change only weakly for a Gaussian scatter of 0.2 and 0.5 dex. On the other hand, the AGN feedback parameter $\delta$ increases by 10 -- 20\% for the 0.2 dex case and more than doubles for the 0.5 dex case. The stellar mass -- halo mass relations at low redshift are shown in \cref{fig:scatter_comparison_mstar}.

\begin{figure*}
     \centering
     \begin{subfigure}[b]{\columnwidth}
         \centering
         \includegraphics[width=\textwidth]{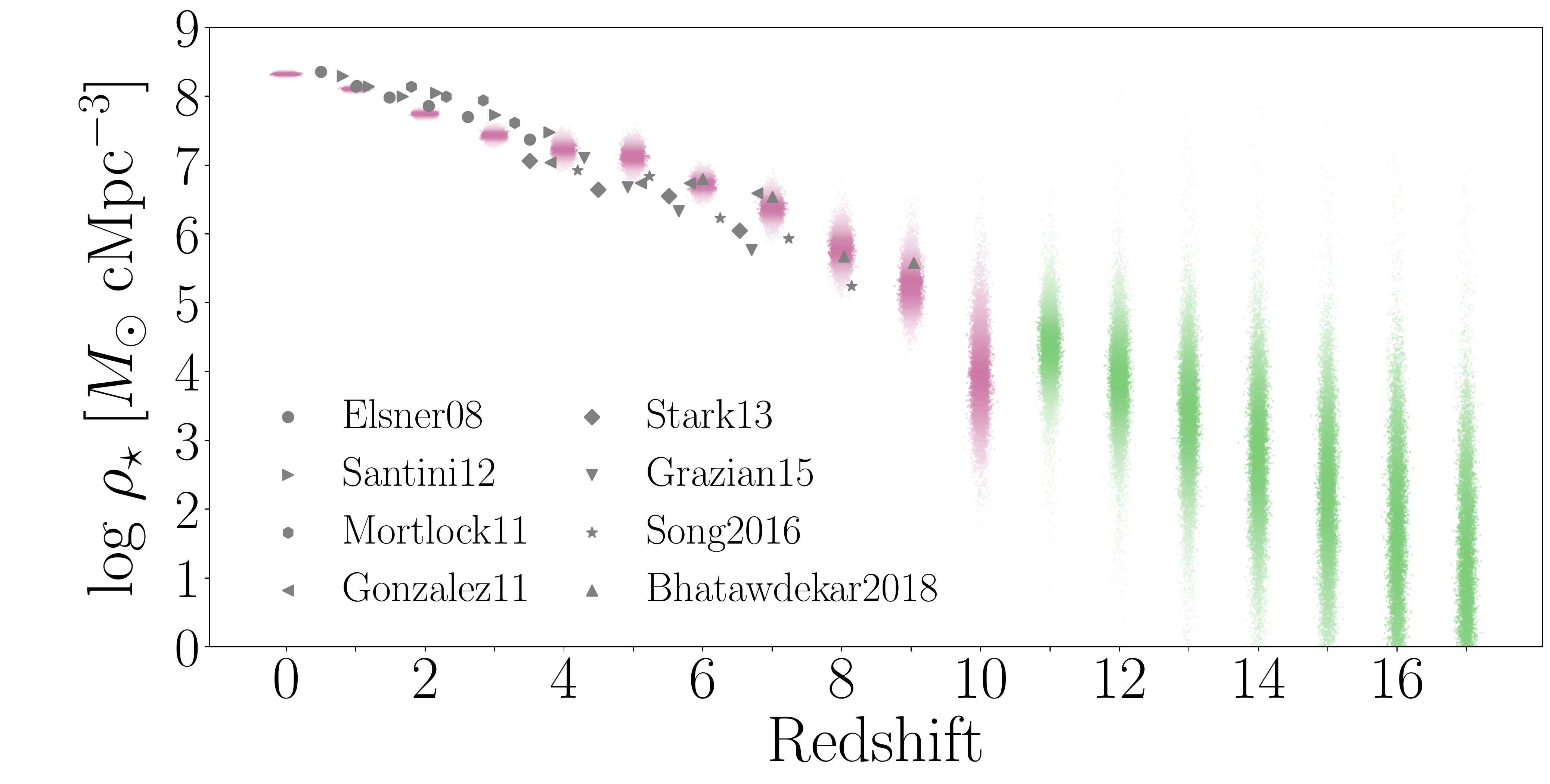}
         \caption{}
         \label{fig:mstar_density_evolution}
     \end{subfigure}
     \hfill
     \begin{subfigure}[b]{\columnwidth}
         \centering
         \includegraphics[width=\textwidth]{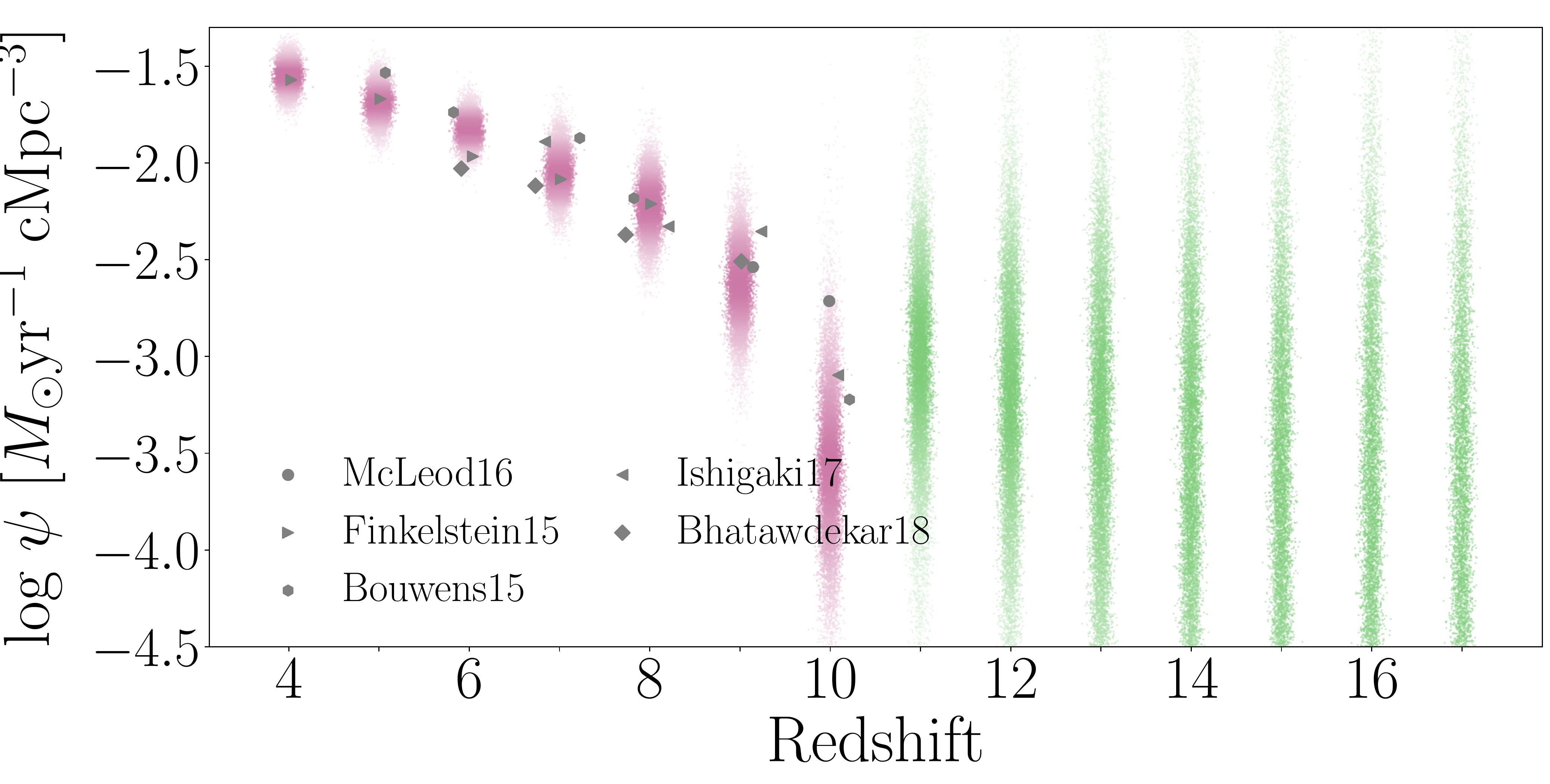}
         \caption{}
         \label{fig:Muv_density_evolution}
     \end{subfigure}
     
    \begin{subfigure}[b]{\columnwidth}
         \centering
         \includegraphics[width=\textwidth]{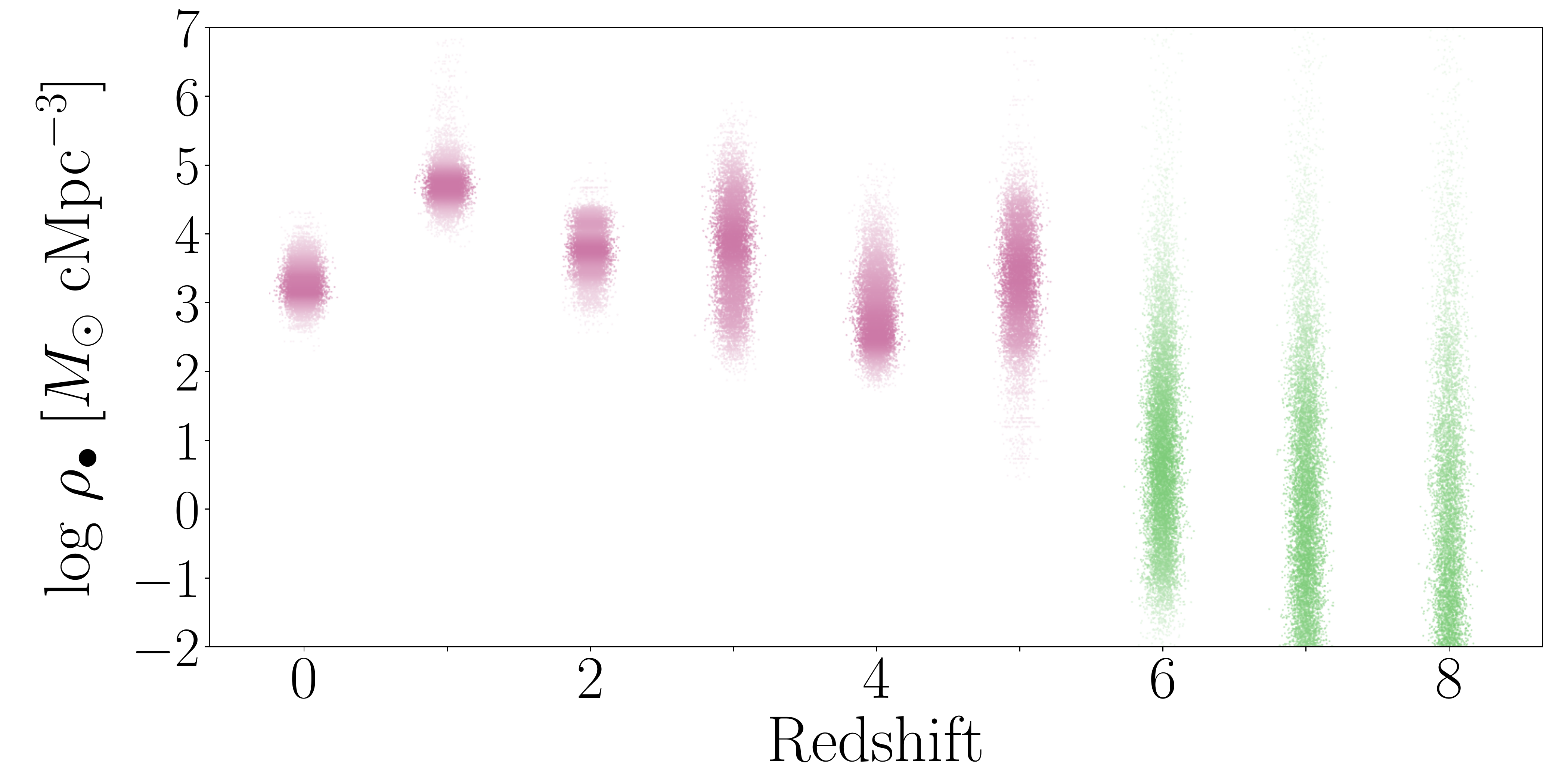}
         \caption{}
         \label{fig:mbh_density_evolution}
     \end{subfigure}
     \hfill
     \begin{subfigure}[b]{\columnwidth}
         \centering
         \includegraphics[width=\textwidth]{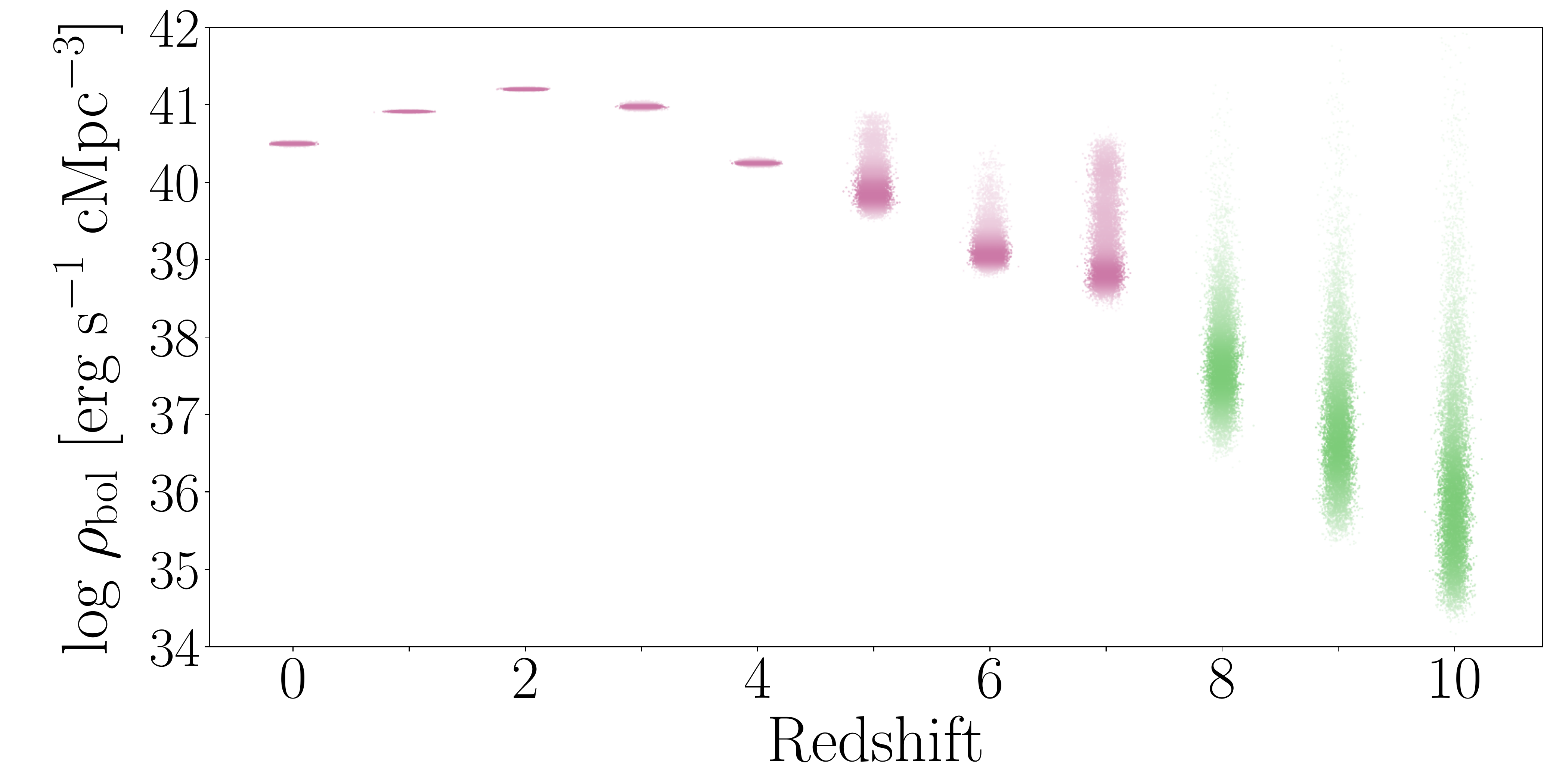}
         \caption{}
         \label{fig:Lbol_density_evolution}
     \end{subfigure}
     \caption{\textbf{Redshift evolution of the integrated densities:} Integrated number density function calculated for a sample of the model parameter drawn from the posterior. Increased colour saturation indicates a larger value for the probability density. Distributions in pink have been constrained by data, while green distributions are linearly extrapolated. For \cref{fig:mstar_density_evolution} and \cref{fig:Muv_density_evolution}, observations collected by \citet{Bhatawdekar2018} are shown in grey. (a) Stellar mass density. (b) SFR density. (c) Type 1 active black hole mass density. (d) AGN bolometric luminosity density.}
     \label{fig:quantity_density_sample}
\end{figure*}

This behaviour is consistent with the discrepancies observed between our results and those of \textsc{UniverseMachine}. This also explains why we find a consistently higher normalisation (which even approaches the cosmic baryon fraction at high redshift) in the stellar mass -- halo mass relation. Compared to \citet{Behroozi2019}, we thus predict more massive galaxies (in terms of stellar mass) for massive haloes above the turnover mass, especially at low redshift. \citet{Behroozi2019} predict this turnover mass to decrease at high redshift, which is however not constrained by data and our model. Nonetheless, in our model framework, we can reasonably reproduce the GSMF with a stellar mass -- halo mass relation that only evolves in the critical mass and normalisation while feedback slopes are fixed. Further, we find the total stellar mass density, defined as
\begin{equation}
        \rho_{M_\star} = \int_{\log M_\star^\mathrm{min}}^{\log M_\star^\mathrm{max}} M_\star \diff {n}{\log M_\star} \dif \log M_\star,
        \label{eq:stellarmassdensity}
\end{equation}
where we employ $M_\star^\mathrm{min} = 10^8 M_\odot$ and  $M_\star^\mathrm{max} = 10^{13} M_\odot$ for comparability with earlier studies, to decrease with redshift in a log-linear fashion (\cref{fig:mstar_density_evolution}). This is consistent in slope and normalisation with observations \citep{Bhatawdekar2018}.

In contrast, the UV luminosity function (\cref{fig:Muv_parameter}) reveals a distinct pattern. The posterior distribution suggests that the critical mass remains constant, while the normalisation consistently increases with increasing redshift. In contrast to the relatively constant AGN feedback factor, the stellar feedback parameter exhibits a trend with redshift. It assumes larger values at $z=0$ and $z=1$ compared to higher redshifts, after which it decreases to a more consistent value.

\begin{figure}
    \centering
    \includegraphics[width=\columnwidth]{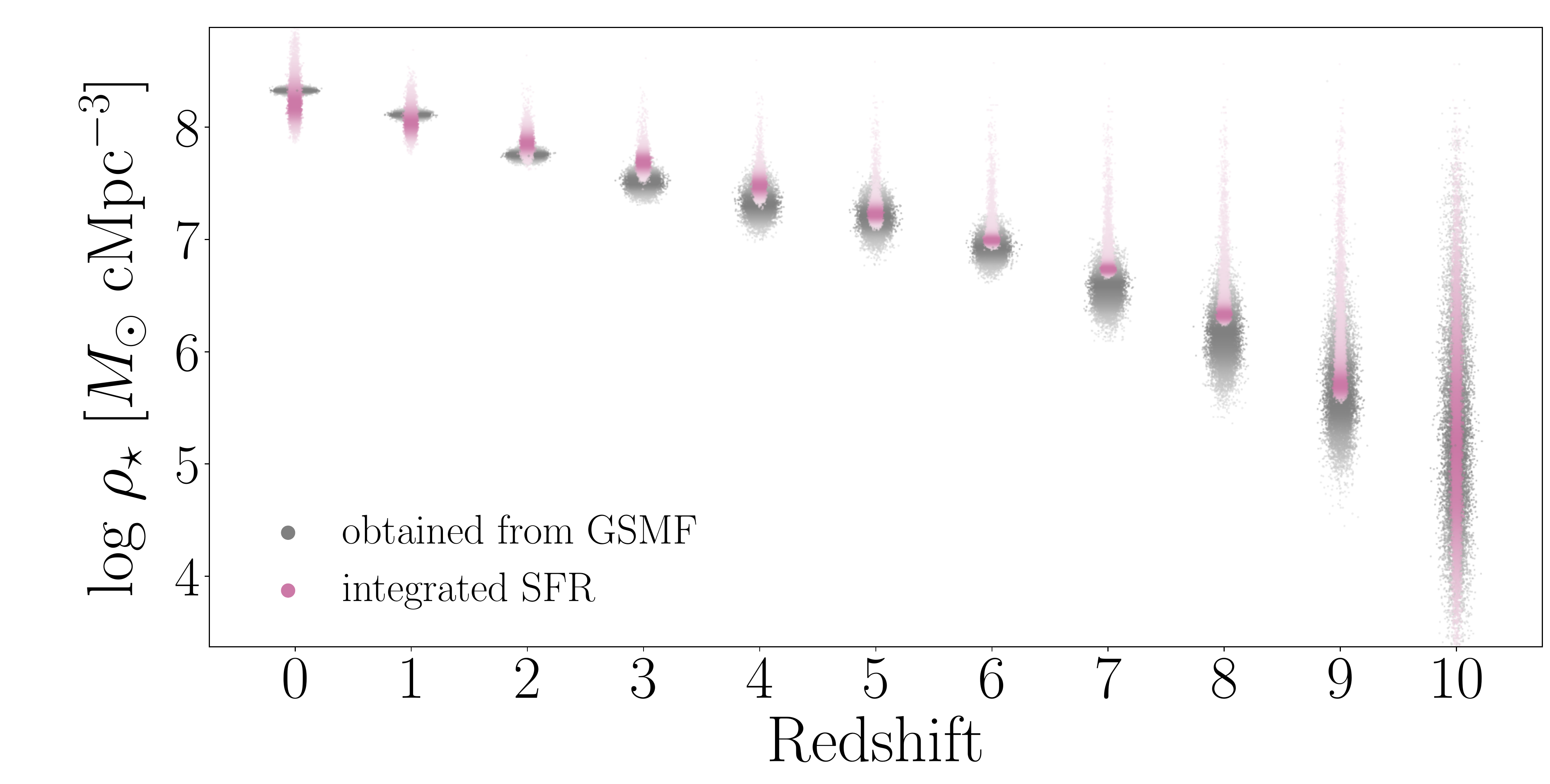}
    \caption{\textbf{Evolution of the stellar mass density.} Shown are samples of the stellar mass density predicted by the model, calculated by integrating the GSMF at every redshift (grey) and by integrating the model SFR density over cosmic time (pink).}
    \label{fig:mstar_Muv_SMD_evolution}
\end{figure}

This behaviour at low redshift is likely caused by the comparably large spread in different estimates of the UVLF rather than physical effects. This is due to the lag of large UV surveys at low redshift, which manifests in the large uncertainties of the parameter estimates at low redshift. The estimates are much more robust for $z \geq 2$ when the emitted UV is shifted to the rest-frame optical bands. We show the integrated SFR for $z \geq 4$ in \cref{fig:Muv_density_evolution}. \footnote{We estimate the SFR from the UV luminosity using $\psi = \mathcal{K}_\mathrm{UV} L_\mathrm{UV} $ with $\mathcal{K}_\mathrm{UV} = 1.4 \cdot 10^{-28} \frac{\mathrm{[M_\odot yr^{-1}]}}{\mathrm{[erg s^{-1} Hz^{-1}]}}$ for a \citet{Chabrier2003} IMF, and integrated the SFR from $\mathcal{M}_\mathrm{UV}^\mathrm{min} = -17$ to $\mathcal{M}_\mathrm{UV}^\mathrm{max} = -25$ for comparability with previous studies.} They are consistent with observational estimates for high redshift surveys \citep{Oesch2018, Bhatawdekar2018} and show a similar trend to \textsc{UniverseMachine}, although their estimated SFR density drops more slowly with redshift. For consistency, we compare the stellar mass density calculated from the GSMF to the one obtained by integrating the SFR density over cosmic time. To include all relevant contributions, we integrate the model GSMF and UVLF over the same range defined by through the halo mass, i.e. we integrate from compare $q^\mathrm{min} = {q(M_\mathrm{h}^\mathrm{min})}$ to $q^\mathrm{max}= {q(M_\mathrm{h}^\mathrm{max})}$
where we impose $M_\mathrm{h}^\mathrm{min} = 10^3 M_\odot$ and $M_\mathrm{h}^\mathrm{max} = 10^{21} M_\odot$. The stellar mass density can be obtained from the SFR density using
\begin{align}
        \rho_\star^\mathrm{UV} (z) & = \left(1-R\right) \int_0^{t(z)} \psi \dif t^\prime \nonumber \\
                                & = \left(1-R\right) \int_{t(z=10)}^{t(z)} \psi \dif t^\prime
                                + \rho_\star(z=10),
\end{align}
where $R=0.41$ is the return fraction for a \citet{Chabrier2003} IMF and $\rho_\star(z=10)$ is the stellar mass density obtained from integrating the GSMF at $z=10$. \cref{fig:mstar_Muv_SMD_evolution} shows that both methods produce similar results for the stellar mass density. It is worth noting that this consistent result is not necessarily expected. \citet{Madau2014}, for example, shows that the integrated SFR slightly overestimates the stellar mass density compared with direct measurements, particularly at $z<2$. Due to the relatively poor constraints on the UV luminosity function in this regime, we are however unable to find a statistically significant difference in our model.

\subsubsection{AGN evolution}
\begin{figure}
    \centering
    \includegraphics[width=\columnwidth]{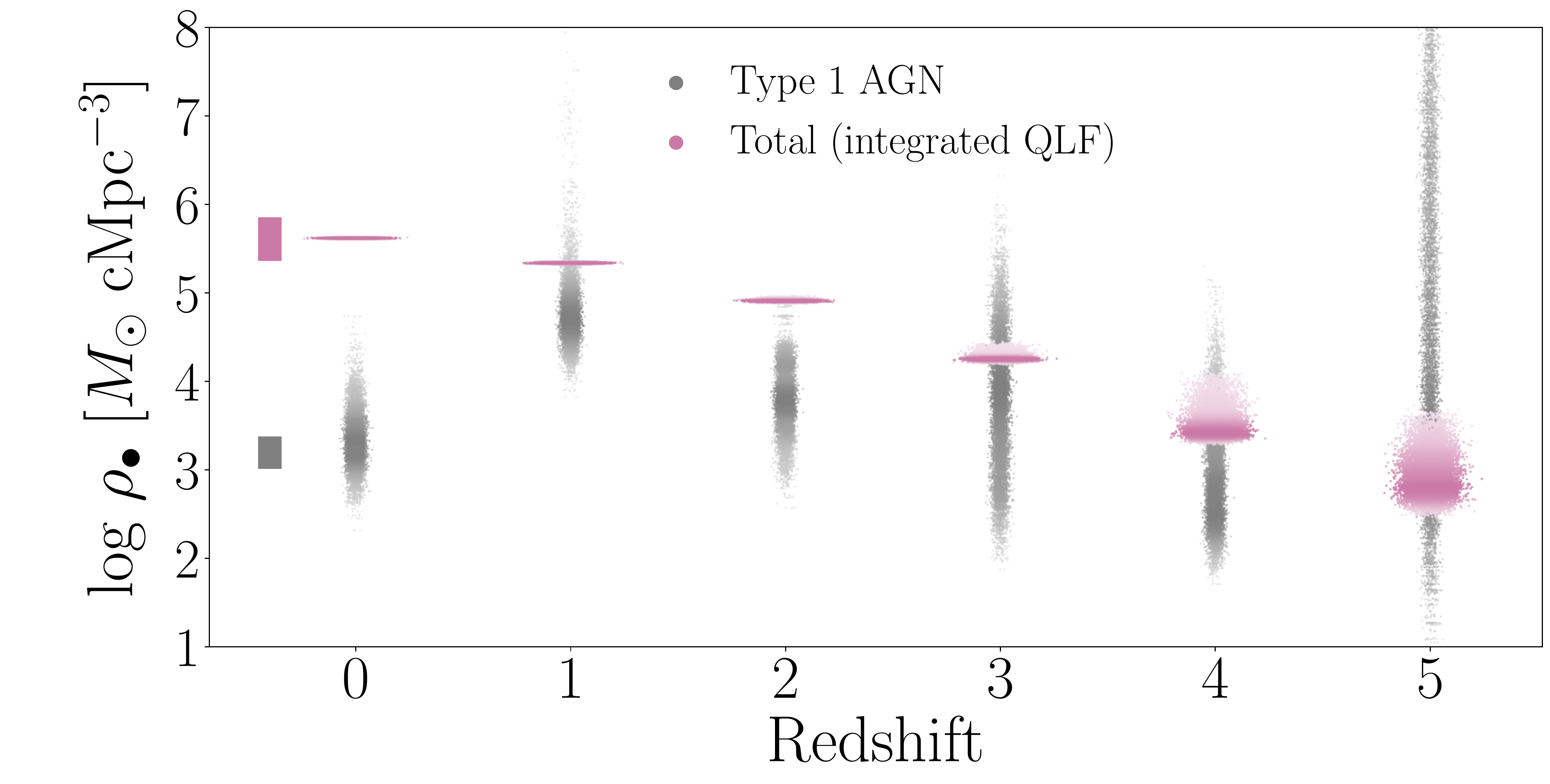}
    \caption{\textbf{Evolution of the black mass density.} Shown are samples of the Type 1 AGN mass density predicted by the model, same as in \cref{fig:mbh_density_evolution} (grey), as well as the estimated total black hole mass density calculated by integrating the model QLF over cosmic time (pink). Shown on the left are the observational estimates for the Type 1 AGN mass density \citep[grey,][]{Mullaney2012} and total black hole mass density \citep[pink,][]{Shankar2009a} at $z=0$.}
    \label{fig:mbh_Lbol_BHMD_evolution}
\end{figure}
The parameter distribution of the active Type 1 AGN mass function (\cref{fig:mbh_parameter}) displays a consistent trend of increasing normalisation and slope as redshift increases. This implies that, according to the model, at higher redshifts, supermassive black holes are more massive for a given halo mass and exhibit a faster increase in mass with halo mass. This is consistent with cosmic downsizing as discussed earlier and found in previous studies on the active BHMF \citep{Kelly2013,Schulze2015}. The comprehensive halo -- galaxy -- SMBH evolution model \textsc{Trinity} \citep{Zhang2021} find similar results, in the sense that more massive black holes become less active earlier in cosmic time compared to smaller black holes. 

The evolution of the parameter distribution for the QLF (as shown in \cref{fig:Lbol_parameter}) is in line with this finding, where the normalisation and slope both increase with redshift. This means that for a given halo mass, the model predicts AGN to become more luminous with increasing redshift, assuming a fixed Eddington ratio distribution function.

 Put together, this means the model predicts that at higher redshift massive and luminous AGN will be hosted in lower mass compared to AGN at low redshift, suggesting that massive black holes grow early. This holds at least under the assumption of an unevolving ERDF, which can reproduce the observational data. We note that the normalisation parameter and slope parameter are very strongly correlated since they have similar effects on the estimated number density (see \cref{fig:corner_plots}).  Connected with the fact that the data only constrains the power law section of the BHMF (\cref{subsubsec:ABHMFdata}, this leads to a strong degeneracy between the two parameters. \textsc{Trinity} in comparison suggests that the ERDF evolves, with the average Eddington ratio increasing with redshift.

Integrated quantities like the mass density of active SMBH and the AGN luminosity density peak at z~2-3, consistent with the peak of SMBH growth. \footnote{We integrate the BHMF from $M_\bullet^\mathrm{min} = 10^5 M_\odot$ to $M_\bullet^\mathrm{min} = 10^{15} M_\odot$, and the QLF from $L_\mathrm{bol}^\mathrm{min} = 10^{38} \mathrm{erg s}^{-1}$ to $L_\mathrm{bol}^\mathrm{min} = 10^{55} \mathrm{erg s}^{-1}$.} In comparison, \textsc{Trinity} suggests that total SMBH mass density decreases with increasing redshift and has no peak. This reinforces the idea that the active fraction evolves with redshift (and/or halo mass), meaning the simple scaling done in \cref{subsec:mbhmstarrel} is therefore only a crude approximation and even if done, scaling factors likely different in different redshift bins.

We also calculate the integrated total black hole mass density by connecting bolometric luminosity to the mass accretion rate of the SMBH $L_\mathrm{bol} = \epsilon \dot{M}_\bullet c^2$, assuming growth to be dominated by gas accretion \citep{Soltan1982}. Choosing a constant efficiency factor $\epsilon = 0.065$, the assumption used by \citet{Shankar2009a}, produces a local density consistent with their original result. \cref{fig:mbh_Lbol_BHMD_evolution} shows the evolution of the mass density for the total black hole population estimated in this way, and for the Type 1 AGN population estimated earlier. At $z=0$, the mass density of both populations differs by a factor of 10-100, consistent with our estimate in \cref{subsec:mbhmstarrel}.

 At higher redshift, the two quantities become closer in magnitude. For $z \sim 1-3$ this is likely connected to increased AGN activity as evident from the quasar luminosity function, while for $z>3$ this is more likely an artefact of limited data availability.

\subsection{Extrapolating the population statistics to higher redshift}
\begin{table*}
\centering
\caption{Upper limits on the expected number of galaxies for JWST surveys detected in the rest-frame UV.}
\begin{tabular}{lrr|rrrrr}
\toprule
      Survey & Area (arcmin$^2$) & $5\sigma$ Depth & $z=7$ & $z=9$ & $z=11$ & $z=13$ & $z=15$ \\
\midrule
       CEERS &               100 &            29.2 &    54 &     3 &      3 &      2 &      2 \\
  Cosmos-Web &              1929 &            28.2 &   224 &    10 &      8 &      6 & 8\footnotemark{}\\
  JADES-Deep &                46 &            30.7 &   205 &    20 &     17 &     12 &     11 \\
JADES-Medium &               190 &            29.8 &   247 &    18 &     16 &     14 &     11 \\
      PRIMER &               378 &            29.5 &   319 &    20 &     18 &     15 &     13 \\
\bottomrule
\end{tabular}
\tablefoot{The expected number of objects is calculated by integrating the UV luminosity function up to the $5\sigma$ depth of the surveys. The redshift bins encompass $[z-0.5, z+0.5]$. Shown are the upper limits, given by the 95th percentile of the distribution obtained from sampling the model UVLF at different redshifts. $5\sigma$ depth and survey areas taken from \citet{Casey2022b}.}
\label{tab:JWST_surveys}
\end{table*}
\footnotetext{An increase in the upper limit of expected galaxies is caused by increased uncertainty in the UVLFs with increasing redshift and is associated with a wider spread of the posterior distribution rather than a shift of the distribution towards larger values.}

\begin{table*}
\centering
\caption{Upper limits on the expected number of galaxies for \textit{Euclid} deep field surveys detected in the rest-frame UV.}
\begin{tabular}{lrr|rrrrr}
\toprule
               Survey & Area (arcmin$^2$) & $5\sigma$ Depth & $z=5$ & $z=6$ & $z=7$ & $z=8$ & $z=11$ \\
\midrule
\textit{Euclid} DF North &             72000 &            26.4 &  5286 &   776 &   325 &   116 &      5 \\
\textit{Euclid} DF South &             72000 &            26.4 &  5286 &   776 &   325 &   116 &      5 \\
     \textit{Euclid} DF Fornax &             36000 &            26.4 &  2647 &   399 &   160 &    53 &      3 \\
\bottomrule
\end{tabular}
\tablefoot{Same as \cref{tab:JWST_surveys}. The $5\sigma$ depth taken from \citet{vanMierlo2021}.}
\label{tab:Euclid_surveys}
\end{table*}

The model's simplicity allows us to extend the results obtained in the previous section to higher, unobserved redshift bins to make qualitatively the expected number densities. We will do this
through two approaches. In the first, we keep the quantity -- halo mass relations fixed to the parameter distributions found at the last observed redshift (UVLF/GSMF at z=10, Type 1 AGN BHMF at z=5, QLF at z=7) and only evolve the halo mass function (HMF). In the second approach, we linearly extrapolate the quantity -- halo mass relations to higher redshifts. 

The extrapolation is done by drawing parameter samples from different redshift bins ($z=5-10$ for the GSMF/UVLF, $z=1-5$ for the Type 1 AGN BHMF and $z=2-7$ for the QLF), and calculating the linear trend using multivariate regression. The extrapolated number density functions are shown in \cref{fig:mstar_ndf_predictions} (GSMF), \cref{fig:Muv_ndf_predictions} (UVLF), \cref{fig:mbh_ndf_predictions} (BHMF) and \cref{fig:Lbol_ndf_predictions} (QLF), with the bands covering the 68\% credible regions. For all quantities, the extrapolation method leads to larger expected number densities compared with purely evolving the HMF. This is consistent with the results found in the previous subsection, which suggests the quantity -- halo mass relations tend to increase in normalisation with redshift. Despite the qualitative nature of these predictions, we find that the extrapolation of our model is well in agreement with estimates of the high-z UVLF from the JWST Early Data Release by \citet{Harikane2022} and \citet{Donnan2022}.

Since the original submission of this work, a wealth of new high-redshift data has become available, significantly advancing our understanding of early galaxy and black hole growth. JWST observations \citep{Harikane2023, Maiolino2024, Greene2024, Kokorev2024, Akins2024} have provided new constraints on the AGN luminosity function at $z=7-12$, as well as the relationship between stellar mass 
and black hole mass \citep[e.g.][]{Harikane2023, Kocevski2023, Furtak2023, Bogdan2023, Kokorev2024, Juodzbalis2024, Maiolino2024, Kovacs2024}. Many of these observations show unprecedentedly high black hole mass -- stellar mass ratios of  
>30\% \citep{Bogdan2023, Juodzbalis2024, Kovacs2024, Kokorev2024} that deviate 
from local relations at the > 3$\sigma$ level (e.g. \citealt{Pacucci2023}, see 
however \citealt{Li2024}) leading to intense discussions regarding the seeding and growth mechanisms required to grow such obese black holes. A number of works crucially warn that these black hole masses could be over-estimated by as much as an order of magnitude since the systems might be dominated by outflows 
\citep{King2024} or the assumption of the standard Sunyaev-Shakura accretion disk \citep{Lupi2024}. Reconciling our model with these new findings presents a challenge. Even when using our locally calibrated (z=0) black hole mass function, we would require a significantly lower duty cycle (<0.1\%) to match the observed high-redshift black hole masses. This suggests that a simple halo mass- and redshift-independent duty cycle is insufficient to capture the complex co-evolution of black holes and galaxies at high redshift. Given the theoretical options currently being explored in the data (overly efficient mergers of Black holes, black holes growing before/more efficiently than their hosts, these being the initial phases in the growth of heavy seeds, stellar mass being under-estimated due to dust/low surface brightness, baryons existing in the right amount, but unable to form stars), such an exploration goes beyond the scope of this work.

We calculate the stellar mass density and other integrated quantities from the linearly extrapolated parameter samples (depicted in green in \cref{fig:quantity_density_sample}; see \cref{subsec:redshiftevodisscussion}). These quantities exhibit a continuing decline with increasing redshift. This decrease occurs despite the inferred increases in the relationship between galaxy and halo properties, which would lead to a higher anticipated number density. This suggests that the decrease in the integrated densities is primarily driven by the evolution of the halo mass function, rather than changes in the quantity -- halo mass relations.

Finally, in \cref{tab:JWST_surveys} and \cref{tab:Euclid_surveys} we present an upper limit on the expected number of galaxies detected in the rest-frame UV for the \textit{Euclid} deep field and JWST CEERS, Cosmos-Web, JADES and PRIMER surveys, based on our modelled UV luminosity function. We calculate the expected number of galaxies by integrating the UV luminosity function up to the reported $5\sigma$ depths \citep{vanMierlo2021, Casey2022b} and for redshift bins $\Delta z = 1$. The upper limits are given by the 95th percentile of the expected number of objects calculated from the model UVLFs shown in \cref{fig:Muv_ndf_intervals} and \cref{fig:Muv_ndf_predictions}. Based on our estimation,\textit{Euclid} will be able to detect thousands of galaxies at $z \sim 5$, and some of the brightest galaxies up to $z \sim 11$. Across the different JWST surveys, dozens to hundreds of galaxies can be expected at $z>10$ including fainter galaxies representative down to $\mathcal{M}_\mathrm{UV} \approx 20$ at $z=15$. This will be further enhanced by JWST lensing surveys like GLASS \citep{Roberts-Borsani2022}, which we have not considered in our calculations.

\section{Conclusion}
\label{sec:conclusion}
We present an analytical model that explores the co-evolution of galaxies, SMBHs, and dark matter halos. Our model links the evolution of the halo mass function with prescriptions for baryonic physics. This analytical approach enables a computationally efficient full Bayesian treatment. Remarkably, simple prescriptions capture the observed galaxy stellar mass function, galaxy UV luminosity function, active black hole mass function, and quasar bolometric luminosity function up to redshift z = 10. Our model establishes connections between different observable properties and provides qualitative predictions for quantities at poorly constrained or unobserved redshifts.

The strength of this modelling approach lies in which model parameters can be adjusted and physically interpreted. We show that by calibrating our baryonic parametrisations to the observed number densities of the different physical quantities, we were able to qualitatively reproduce the observed relations between these quantities. In particular, we reproduced the slope of the UV luminosity -- stellar mass relation, as well as the SMBH mass -- galaxy stellar mass relation. Both of these relations have a systematic offset compared with observations, which can relate to a model assumption. In particular, the former is likely caused by the assumption that observable properties and halo masses are connected uniquely. The latter can be explained by the fact that we did not explicitly model the duty cycles and active fraction of black holes. We discuss a future avenue to include the effect of scatter in \cref{ApB:scatter}. The model was further able to reproduce the expected black hole mass distribution of a sample of type 1 AGN, by only calibrating the model quasar luminosity function to observations.

Using the model, we disentangled the effects of dark matter structure evolution and baryonic physics on the evolution of the observed galaxy properties and, in particular, we were able to study the evolution of the relation between dark matter halo mass and the baryonic properties. Key results are as follows:
\begin{enumerate}
        \item The stellar mass -- halo mass relation, within uncertainties, exhibits an unchanging low-mass slope up to redshift $z=10$ and an unchanging high-mass slope up to at least $z=2$. The prominent evolution occurs in the overall normalisation, which increases with increasing redshift. Additionally, the feedback turnover mass increases from approximately $10^{12} M_sun$ at $z=0$ to approximately $10^{12.4} M_sun$ at $z=2$.
        \item At the faint end, the UV luminosity -- halo mass relation maintains a consistent slope between $z$ = 2-10, and at the bright end, it does so at least for $z$ = 2-4. The overall normalisation rises with redshift, whereas the feedback turnover mass remains constant at $\sim 10^{11.8} M_\odot$ between $z$ = 2-4. Drawing solid conclusions from the data at $z$ < 2 is challenging due to data-related uncertainties.
        \item The observed quasar luminosity function can be reproduced up to $z=7$ with a redshift- and halo mass-independent Eddington rate distribution function.
\end{enumerate}
The model can reproduce the observed stellar mass density for $z=1-9$ and observed SFR density $z=4-10$, and it produces a self-consistent evolution of the stellar mass density when calculated directly and inferred from the SFR. 

The model shows that the stellar mass density, SFR density, black hole mass density, and bolometric luminosity density are all decreasing for $z>3$ in a near log-linear fashion. We used the results obtained from the calibration of baryonic parametrisations to make qualitative predictions on the number densities at redshifts beyond those used for calibration, by linearly extrapolating the calibrated parametrisations. To assess the reasonableness of our predictions, we compared them with the Early Science Release of the UV luminosity function of  JWST. Furthermore, using these extrapolations, we present upper limits on the expected number of objects for scheduled JWST (\cref{tab:JWST_surveys})  and  \textit{Euclid}  (\cref{tab:Euclid_surveys}) surveys.

In summary, our empirical model accurately captures the observed changes in galaxy and SMBH characteristics. It can make both qualitative and quantitative predictions about how these properties are interrelated and change over time, even beyond the data used to constrain the model. Compared to more comprehensive and computationally demanding models, conceptual and empirical models offer a rapid, straightforward, understandable, and adaptable framework for studying galaxy evolution. Therefore, these models can complement each other.

\section*{Data availability}
The code, data used for calibration and posterior parameter distributions obtained from the MCMC sampling can be accessed under \href{https://doi.org/10.5281/zenodo.7552484}{https://doi.org/10.5281/zenodo.7552484}.

\begin{acknowledgements}
CB acknowledges generous support from the young Academy Groningen through the award of an interdisciplinary PhD fellowship and thanks Irene Tieleman for her support, as well as Piero Madau and Mark Dickinson for generously providing data. MT and PD acknowledge support from the NWO grant 0.16.VIDI.189.162 (``ODIN''). PD acknowledges support from the University of Groningen's CO-FUND Rosalind Franklin Program.
\end{acknowledgements}

\bibliographystyle{aa} 
\bibliography{draft} 


\begin{appendix}
\section{The halo mass function}
\label{ApA:HMF}
Compared to the evolution of galaxies and baryonic matter in general, the formation and evolution of dark matter halos is a well-understood process, due to the comparably simple physics involved in the process. The halo mass function describes the distribution of dark matter halo masses across cosmic history and redshift $z$. If $\diff {n}{\log M_\mathrm{h}}$ denotes the number density $\dif n$ of halos (per comoving Mpc) per infinitesimal mass bin $\dif \log M_\mathrm{h}$, this quantity can be expressed as
\begin{equation}
    \phi (M_\mathrm{h}) = \diff{n}{\log M_\mathrm{h}} (M_\mathrm{h}, z) = \frac{\overline{\rho}}{M_\mathrm{h}} f\left(\nu (M_\mathrm{h}, z)\right) \left|\diff{\log \nu(M_\mathrm{h}, z) }{\log M_\mathrm{h}}\right|.\end{equation}
Integration over $M_\mathrm{h}$ yields the total number density of halos in the mass range $[M_1, M_2]$ at redshift z,
\begin{equation}
    n(z)\bigg|_{M_1}^{M_2} = \int_{M_1}^{M_2} \frac{\overline{\rho}}{M_\mathrm{h}^2} f\left(\nu (M_\mathrm{h}, z)\right) \left|\diff{\log \nu(M_\mathrm{h}, z) }{\log M_\mathrm{h}}\right| \dif M_\mathrm{h}.
\end{equation}
Here, $\overline{\rho}$ is the comoving mean matter density and $\nu$ is defined by $ \nu(M_\mathrm{h},z) = \frac{\delta^2_\mathrm{c}(z)}{\sigma^2(M_\mathrm{h},z)}$, where $\delta_\mathrm{c}(z)$ critical overdensity needed for the collapse of a halo and $\sigma^2(M_\mathrm{h},z)$ is the mass variance of the smoothed overdensity field (see e.g. \citet{Mo2010} for more details). Finally, the function $f(\nu)$ is called the multiplicity function and is given by
\begin{equation}
    f(\nu) = C \left( 1 + \frac{1}{\nu'^p} \right) \left(\frac{\nu'}{2\pi} \right)^{\nicefrac{1}{2}} e^{\nicefrac{-\nu'}{2}},
    \label{eq:multiplicityfunc}
\end{equation}
where $\nu' = a \nu$. (It is common in the literature to define a function $\widetilde{f}(\nu)$ instead, with $f(\nu)= \nu  \widetilde{f}(\nu)$.) The parameters $(a, p, C)$ define the high-mass cutoff, the shape at lower masses, and the normalisation of the curve, respectively \citep{Sheth2001, Despali2015}. These parameters can be estimated in various ways from theoretical considerations \citep{Press1974, Sheth2001} or numerical simulations \citep{Tinker2008, Despali2015}. Though there has been much debate \citep{Tinker2008, Courtin2010}, it is assumed that the functional form and parameters $(a, p, A)$ of the multiplicity function are to a good approximation universal, meaning they are independent of redshift and specific cosmology \citep{Despali2015}. Discussions are still ongoing on how much the idea of universality can be extended, and to what degree universality holds \citep{Bocquet2015, Bocquet2020,Diemer2020}.  Nonetheless, mass functions that are well described by \cref{eq:multiplicityfunc}, and even the simple analytical expression such as the Press-Schechter \citep{Press1974} and extended Press-Schechter \citep{Bond1991, Sheth2001} formalisms seem to be good first-order approximations, especially at low to medium redshift. For our quantitative analysis, we have used the extended Press-Schechter HMF for ellipsoidal collapse given by \citet{Sheth2001}. They derive an HMF of the form given by \cref{eq:multiplicityfunc} with the parameter set $(a,p,C) = (0.84, 0.3, 0.644)$. For many $\Lambda$CDM simulations, the extended Press-Schechter halo mass function yields a good approximation of the HMFs derived from numerical simulations up to $z \approx 15-20$.

\section{The effect of scatter}
\label{ApB:scatter}
The focus of this work is on reproducing average relations between physical quantities, using the simplifying assumption that the observables and halo masses are linked by a one-to-one relation given by \cref{eq:qhmrel}. In reality, however, environmental effects, and the inherent stochastic nature of baryonic processes lead to a scatter in this relation, which means each halo mass will have an associated distribution for every observable quantity. As described in \cref{subsubsec:QLF} for the Eddington ratios, large scatter in primary quantities can have striking effects on the number statistics and need to be included for the model to be able to match observations.

To include these effects systematically, it helps to recast the model as we have described in \cref{subsec:modelbasics} in the language of probability distributions. The HMF, if normalised to unity (where we can introduce a low-mass cutoff in case the integral diverges), constitutes the probability density function (PDF) of halo masses in a given cosmic volume, i.e. 
\begin{equation}
    f_{M_\mathrm{h}} (m_\mathrm{h}) = \frac{\phi(m_\mathrm{h})}{N},
   \quad \text{where} \quad 
    N= \int_{m_\mathrm{min}}^\infty \phi(m_\mathrm{h}) \dif m_\mathrm{h}
\end{equation}
is the total number of halos in the volume. The PDF $f_Q (q)$ for the observable $q$ is then obtained by a simple change of variables using \cref{eq:qhmrel}. To include scatter, rather than performing a change of variables we define a joint probability distribution,
\begin{equation}
    f_{Q, M_\mathrm{h}} (q, m_\mathrm{h}) = f_{Q|M_\mathrm{h}} \left(q|M_\mathrm{h} = m_\mathrm{h}\right) \cdot f_{M_\mathrm{h}} (m_\mathrm{h}),
    \label{eq:jointpdf}
\end{equation}
where we now treat $Q$ and $M_\mathrm{h}$ as random variables. The conditional probability $f_{Q|M_\mathrm{h}} (q|M_\mathrm{h} = m_\mathrm{h})$ of $q$ with respect to $m_\mathrm{h}$ describes the distribution of the quantity $Q$ at a fixed halo mass $M_\mathrm{h} = m_\mathrm{h}$, i.e. the scatter in the relation we want to include. The marginal distribution for the $Q$ is given by 
\begin{align}
    f_Q (q) &= \int_{m_\mathrm{min}}^\infty f_{Q, M_\mathrm{h}} (q, m_\mathrm{h}) \dif m_\mathrm{h} \nonumber\\
            & =\int_{m_\mathrm{min}}^\infty f_{Q|M_\mathrm{h}} (q|M_\mathrm{h} = m_\mathrm{h}) \cdot f_{M_\mathrm{h}} (m_\mathrm{h}) \dif m_\mathrm{h},
            \label{eq:marginalpdf}
\end{align}
and the number density of the observable is $\phi(q) = N \cdot f_Q (q)$.

As an example, a scatter-free relation between observable and halo mass as described by \cref{eq:qhmrel} would be described by a conditional probability of the form
\begin{equation}
    f_{Q|M_\mathrm{h}} (q|M_\mathrm{h} = m_\mathrm{h}) = \delta \left(q - \mathcal{Q}(m_\mathrm{h})\right),
    \label{eq:deltacondpdf}
\end{equation}
with $\delta$ being the Dirac delta distribution. The marginal probability for $Q$ in this case is given by
\begin{align}
        f_Q (q) 
        & = \int_{m_\mathrm{min}}^\infty f_{Q, M_\mathrm{h}} (q, m_\mathrm{h}) \dif m_\mathrm{h}              \nonumber\\
        & = \int_{m_\mathrm{min}}^\infty \delta\left(q - \mathcal{Q}(m_\mathrm{h})\right) \cdot               f_{M_\mathrm{h}} (m_\mathrm{h}) \dif m_\mathrm{h} \nonumber\\
        & = \int_{m_\mathrm{min}}^\infty 
            \frac{\delta(m_\mathrm{h}-m_\mathrm{h}^*)}
                 {|\mathcal{Q}^\prime (m_\mathrm{h}^*)|}
            \cdot f_{M_\mathrm{h}} (m_\mathrm{h}) \dif m_\mathrm{h} \nonumber \\
            & = \frac{f_{M_\mathrm{h}} (m_\mathrm{h}^*)}{|\mathcal{Q}^\prime (m_\mathrm{h}^*)|},
\end{align}
which recovers \cref{eq:qNDF} obtained in the original approach. Here, $m_\mathrm{h}^*$ is 
the halo mass that solves \cref{eq:qhmrel} for a given $q$ and we make use of the function composition 
property of the delta distribution. To include scatter, we could e.g. assume the marginal distribution to 
be a Gaussian with a central value given by \cref{eq:qhmrel} and a halo mass-independent variance $\sigma^2$,
\begin{equation}
    f_{Q|M_\mathrm{h}} (q|M_\mathrm{h} = m_\mathrm{h}) = \mathcal{N}\left(\mathcal{Q}(m_\mathrm{h}), \sigma^2\right).
    \label{eq:gaussiancondpdf}
\end{equation}
This generalised approach has several advantages besides the ability to include scatter. For one, there is no need for the function $\mathcal{Q}$ to be invertible anymore, since multiple halo masses can be assigned to the same observable value. Further, it is easy to calculate higher moments of the various quantities, enabling the study of the scatter and skewness of the distributions rather than just mean relations. To study the interrelation of observables, one can construct the probability distributions of one quantity concerning another, i.e.
\begin{equation}
    p(q_1|q_2) = \int_{m_\mathrm{min}}^\infty f_{Q_1|M_\mathrm{h}} (q_1|m_\mathrm{h}) \cdot f_{M_\mathrm{h}|Q_2} (m_\mathrm{h}|q_2) \dif m_\mathrm{h},
    \label{eq:q1q2wscatter}
\end{equation}
where $f_{M_\mathrm{h}|Q_2} (m_\mathrm{h}|q_2)$ is given by
\begin{equation}
    f_{M_\mathrm{h}|Q_2} (m_\mathrm{h}|q_2) = \frac{f_{Q_2, M_\mathrm{h}} (q_2, m_\mathrm{h})}
                                                     {f_{Q_2} (q_2)}.
    \label{eq:condpdfinverse}
\end{equation}
\begin{figure}
    \centering
    \includegraphics[width=\columnwidth]{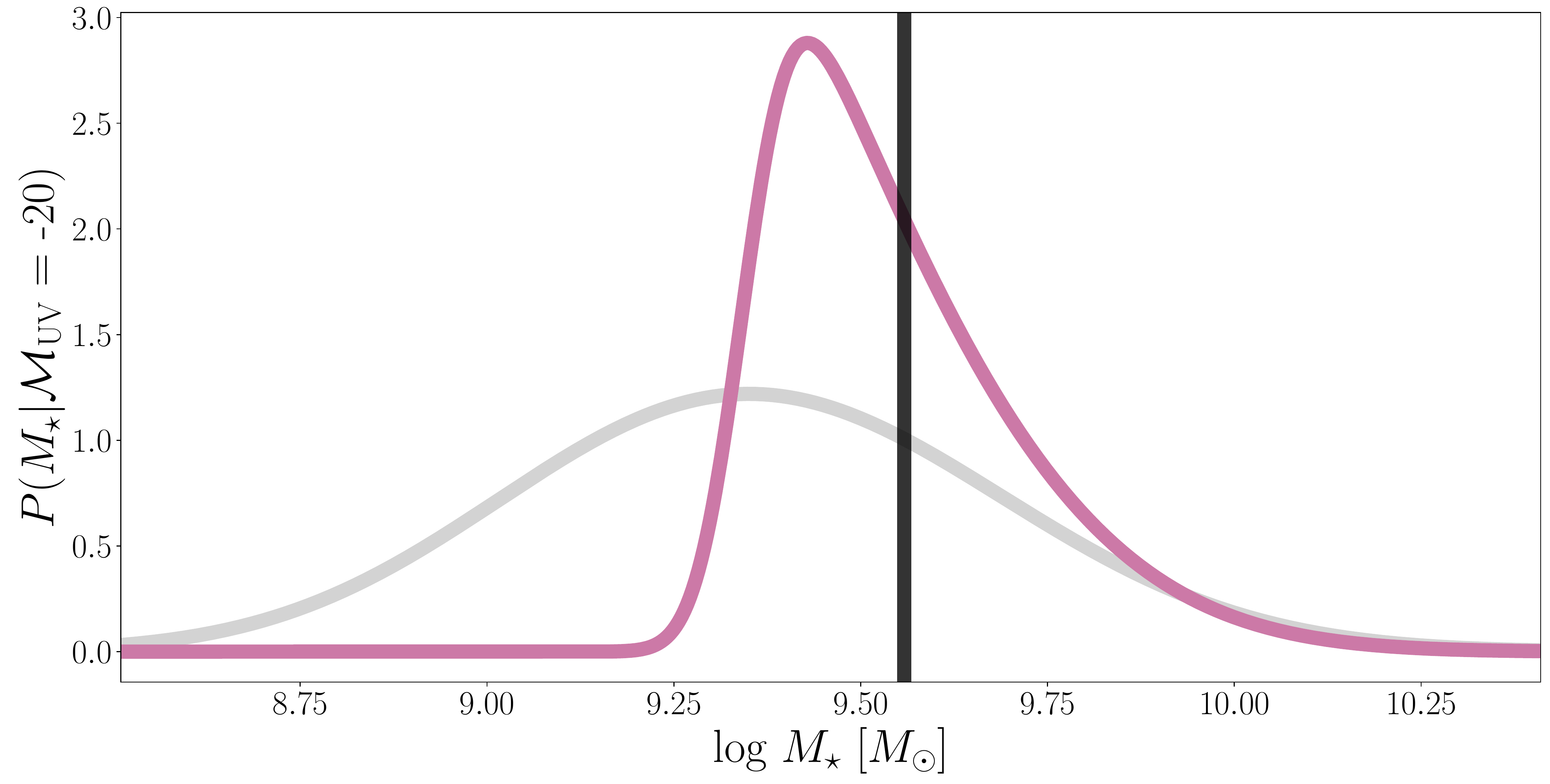}
    \caption{\textbf{Influence of scatter on the stellar mass -- UV luminosity relation.} Shown are the 
    conditional probability distributions of the galaxy stellar mass for a fixed UV magnitude $\mathcal{M}_\mathrm{UV} = -20$ when including scatter in the stellar mass -- halo mass and UV luminosity -- halo mass relations, calculated using \cref{eq:q1q2wscatter} for a fixed set of model parameter. The grey line shows the probability distribution when assuming $\log M_\star$ and $\log L_\mathrm{UV}$ are distributed according to a Gaussian distribution with $\sigma_{M_\star} = 0.05$ and $\sigma_{L_\mathrm{UV}} = 0.25$. The pink line assumes a skew-normal distribution (a generalisation of the normal distribution that allows for non-zero skewness) for $L_\mathrm{UV}$ with a skewness parameter $\alpha=-40$. All distributions have a median given by \cref{eq:qhmrel}. The black vertical line shows the value of $M_\star$ calculated without assuming scatter. The scatter influences the location and shape of the distributions, the skewed case produces a distribution that more closely matches the observations shown in \cref{fig:Muv_mstar_relation}.}
    \label{fig:mstar_Muv_scatter_distribution}
\end{figure}
\begin{figure}
    \centering
    \includegraphics[width=\columnwidth]{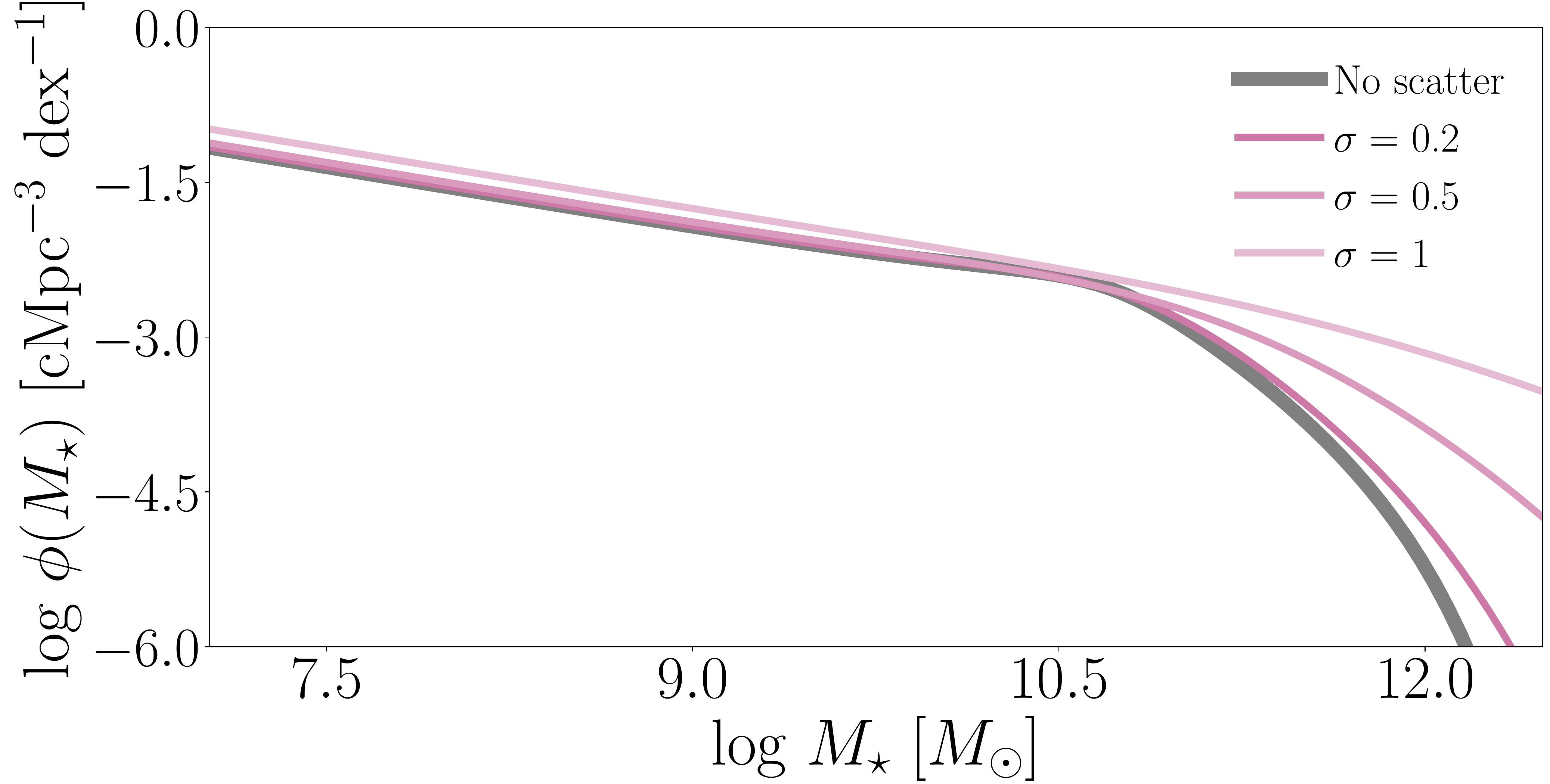}
    \caption{\textbf{Influence of scatter on the GSMF.} Shown are the modelled stellar mass functions for a fixed set of model parameters at $z=0$. The grey line shows the GSMF calculated assuming a one-to-one relationship between the quantities and halo masses, as done throughout this paper. The pink lines show the GSMF calculated using the same parameter but adding a Gaussian scatter to the stellar mass -- halo mass relation with different variances. The high-mass slope is strongly sensitive to the amount of scatter, while the normalisation is weakly affected and the low-mass slope unaffected. }
    \label{fig:mstar_scatter_ndf}
\end{figure}
\begin{figure}
    \centering
    \includegraphics[width=\columnwidth]{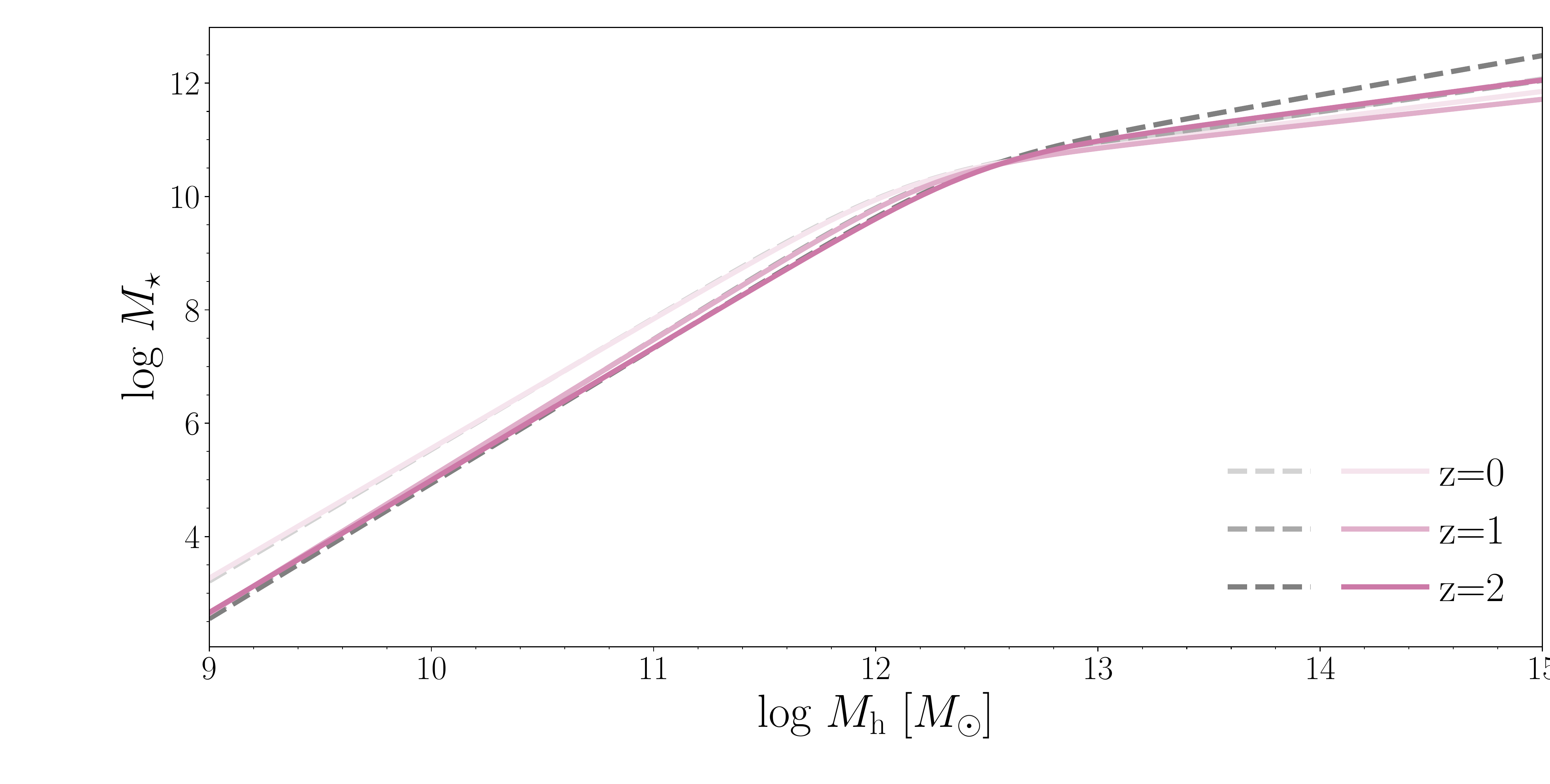}
    \caption{\textbf{Influence of scatter on the stellar mass -- halo mass relation.} Shown are the stellar mass -- halo mass relations at redshift $z = 0$ - 2, with (grey, dashed lines) and without (pink, solid lines) scatter. The scatter is assumed to be Gaussian with a variance of 0.2 dex. The low-mass end, normalisation and turnover mass are barely affected by the scatter. The slope of the high-mass end is steeper for the case with scatter, corresponding to an increase of the parameter $\delta_\star$ by about 10-20\%. The same result holds for a larger scatter of 0.5 dex, with the parameter $\delta_\star$ increasing by about a factor of 2.}
    \label{fig:scatter_comparison_mstar}
\end{figure}

\section{Additional figures and tables}
\label{ApC:parametertable}
\begin{table*}
  \centering
  \caption{Summary of modelled population statistics, their parameter, and asymptotic behaviour.}
  \scalebox{0.80}{
  \begin{tabular}{lllcc}
    Number Density Function & Parameter & Interpretation & Low-Mass Slope & High-Mass Slope\\
    
    \midrule\midrule[.1em]
    \multirow{3}{3.5cm}{Galaxy Stellar Mass Function} 
      & $A_\star$
      & Average stellar mass -- halo mass ratio at $M_\mathrm{h}=M_\mathrm{c}^*$. 
      & \multirow{3}{*}{$\nicefrac{\alpha_\mathrm{HMF}}{(1+\gamma_\star)}$} 
      & \multirow{3}{*}{exponential} 
      \\
      & $\gamma_\star$
      & Strength of stellar feedback.
      \\
      & $\delta_\star$
      & Strength of AGN feedback.
      \\
      & $M_\mathrm{c}^*$
      & Turnover (halo) mass for stellar and AGN feedback-dominated regime.
      \\
    
    \midrule[.1em] 
    \multirow{3}{3.5cm}{Galaxy UV Luminosity Function} 
      & $A_\mathrm{UV}$
      & Average UV luminosity -- halo mass ratio at $M_\mathrm{h}=M_\mathrm{c}$. 
      & \multirow{3}{*}{$\nicefrac{\alpha_\mathrm{HMF}}{(1+\gamma_\mathrm{UV})}$} 
      & \multirow{3}{*}{exponential} 
      \\
      & $\gamma_\mathrm{UV}$
      & Strength of stellar feedback.
      \\
      & $\delta_\mathrm{UV}$
      & Strength of AGN feedback.
      \\
      & $M_\mathrm{c}^\mathrm{UV}$
      & Turnover (halo) mass for stellar and AGN feedback-dominated regime
      \\
      
    \midrule[.1em] 
    \multirow{3}{3.5cm}{Black Hole Mass Function} 
      & $B$
      & Average mass of SMBH at $M_\mathrm{h}=M_\mathrm{c}^\bullet$. 
      & \multirow{2}{*}{$\nicefrac{\alpha_\mathrm{HMF}}{\eta}$} 
      & \multirow{2}{*}{exponential} 
      \\
      & $\eta$
      & Slope of BH mass growth with halo mass for acreeting SMBHs.
      \\
      & $M_\mathrm{c}^\bullet$
      & Critical (halo) mass for SMBH mass model.
      \\
      
    \midrule[.1em] 
    \multirow{4}{3.5cm}{Quasar Luminosity Function} 
      & $C$
      & Average bolometric AGN luminosity at $M_\mathrm{h}=M_\mathrm{c}^\mathrm{bol}$ and $\lambda=1$.
      & \multirow{4}{*}{$\nicefrac{\alpha_\mathrm{HMF}}{\theta}$} 
      & \multirow{4}{*}{$-\rho$} 
      \\
      & $\theta$
      & Slope of luminosity increase with halo mass for acreeting SMBHs.
      \\
      & $\lambda_\mathrm{c}$
      & Critical Eddington ratio for power law drop-off of ERDF.
      \\
      & $\rho$
      & Slope of ERDF drop-off.
      \\
      & $M_\mathrm{c}^\mathrm{bol}$
      & Critical (halo) mass for AGN luminosity model.
      \\
      
  \end{tabular}
  }
  \label{tab:summary}
\end{table*}

\begin{table*}
    \centering
    \caption{Reference list for the galaxy property-related parameter.}
    \begin{tabular}{rrrrrrrrr}
        \toprule
         $z$ & $\log M_\mathrm{c}^\star$ &          $\log A_\star$ &         $\gamma_\star$ &         $\delta_\star$ & $\log M_\mathrm{c}^\mathrm{UV}$ &    $\log A_\mathrm{UV}$ &   $\gamma_\mathrm{UV}$ &   $\delta_\mathrm{UV}$ \\
        \midrule
           0 &    $12.1_{-0.05}^{+0.06}$ & $-1.69_{-0.04}^{+0.04}$ &   $1.32_{-0.1}^{+0.1}$ & $0.43_{-0.02}^{+0.02}$ &         $11.66_{-0.35}^{+0.39}$ & $15.44_{-0.21}^{+0.18}$ & $1.67_{-0.76}^{+1.57}$ & $0.36_{-0.19}^{+0.14}$ \\
           1 &   $12.24_{-0.07}^{+0.07}$ & $-1.72_{-0.04}^{+0.04}$ & $1.45_{-0.15}^{+0.16}$ & $0.44_{-0.04}^{+0.03}$ &          $11.5_{-0.15}^{+0.19}$ & $16.01_{-0.12}^{+0.11}$ & $2.34_{-1.18}^{+1.46}$ & $0.23_{-0.12}^{+0.12}$ \\
           2 &    $12.4_{-0.17}^{+0.17}$ & $-1.72_{-0.09}^{+0.07}$ &  $1.4_{-0.25}^{+0.34}$ &  $0.3_{-0.13}^{+0.12}$ &         $11.39_{-0.17}^{+0.17}$ & $16.65_{-0.06}^{+0.05}$ & $1.17_{-0.25}^{+0.32}$ & $0.33_{-0.05}^{+0.05}$ \\
           3 &                           & $-1.27_{-0.35}^{+0.63}$ & $0.61_{-0.13}^{+0.36}$ &                        &          $11.44_{-0.24}^{+0.3}$ & $16.99_{-0.11}^{+0.08}$ & $1.19_{-0.39}^{+0.49}$ &  $0.3_{-0.11}^{+0.11}$ \\
           4 &                           & $-1.61_{-0.22}^{+0.72}$ &  $0.89_{-0.4}^{+0.59}$ &                        &         $11.24_{-0.34}^{+0.91}$ & $17.08_{-0.14}^{+0.22}$ & $1.19_{-0.67}^{+0.94}$ & $0.08_{-0.08}^{+0.26}$ \\
           5 &                           &  $-1.11_{-0.3}^{+0.64}$ & $1.16_{-0.33}^{+0.56}$ &                        &                                 &  $17.6_{-0.31}^{+0.66}$ & $0.44_{-0.18}^{+0.59}$ &                        \\
           6 &                           &  $-0.8_{-0.39}^{+0.39}$ & $0.88_{-0.16}^{+0.21}$ &                        &                                 &  $17.9_{-0.36}^{+0.64}$ &  $0.42_{-0.11}^{+0.2}$ &                        \\
           7 &                           &   $-1.2_{-0.3}^{+0.69}$ &  $1.25_{-0.4}^{+0.62}$ &                        &                                 &  $18.31_{-0.6}^{+1.03}$ & $0.53_{-0.17}^{+0.27}$ &                        \\
           8 &                           &  $-0.92_{-0.6}^{+0.49}$ & $1.12_{-0.43}^{+0.84}$ &                        &                                 & $18.52_{-0.79}^{+1.51}$ & $0.64_{-0.24}^{+0.52}$ &                        \\
           9 &                           &  $-1.34_{-0.65}^{+0.9}$ & $1.52_{-0.89}^{+1.99}$ &                        &                                 &  $18.76_{-0.8}^{+1.66}$ & $0.61_{-0.25}^{+0.52}$ &                        \\
          10 &                           &  $-1.63_{-1.0}^{+1.15}$ & $1.02_{-0.89}^{+2.64}$ &                        &                                 & $19.25_{-1.49}^{+4.12}$ &   $1.1_{-0.61}^{+2.1}$ &                        \\
        \bottomrule
    \end{tabular}
    \tablefoot{Given are the median parameter values and uncertainties that cover the 95\% credible interval. Note that the median values do not necessarily correspond to the most likely parameter (as defined by the MAP estimator). For calculations, the provided MCMC chains should be used. For $z>3$ ($M_\star$) and $z>5$ ($L_\mathrm{UV}$), we marginalise over $M_\mathrm{c}$ and $\delta$ in our analysis, so that no values are given.}
    \label{tab:gal_parameter}
\end{table*}

\begin{table*}
    \centering
    \caption{Reference list for the SMBH property-related parameter.}
    \begin{tabular}{rrrrrrr}
        \toprule
         $z$ &               $\log B$ &                 $\eta$ &                $\log C$ &               $\theta$ & $ \log \lambda_\mathrm{c}$ &                 $\rho$ \\
        \midrule
           0 & $4.17_{-1.48}^{+1.13}$ &  $1.5_{-0.35}^{+0.48}$ & $40.08_{-1.13}^{+0.62}$ & $2.23_{-0.25}^{+0.42}$ &    $-2.35_{-0.55}^{+0.43}$ & $1.44_{-0.05}^{+0.05}$ \\
           1 & $6.23_{-1.85}^{+0.76}$ & $1.56_{-0.47}^{+1.53}$ & $40.77_{-0.12}^{+0.12}$ & $2.88_{-0.06}^{+0.05}$ &                            &                        \\
           2 &  $6.03_{-1.5}^{+0.76}$ & $2.01_{-0.41}^{+0.81}$ &  $42.5_{-0.11}^{+0.11}$ & $3.24_{-0.07}^{+0.07}$ &                            &                        \\
           3 & $6.73_{-3.93}^{+1.19}$ & $2.21_{-0.78}^{+2.73}$ & $43.52_{-0.17}^{+0.17}$ &  $3.4_{-0.14}^{+0.14}$ &                            &                        \\
           4 & $6.35_{-4.79}^{+1.42}$ & $3.22_{-1.27}^{+4.62}$ & $43.06_{-0.27}^{+0.25}$ & $4.73_{-0.23}^{+0.25}$ &                            &                        \\
           5 & $2.78_{-2.67}^{+5.56}$ & $10.95_{-8.4}^{+7.74}$ & $44.72_{-0.97}^{+0.78}$ &  $3.99_{-1.57}^{+1.2}$ &                            &                        \\
           6 &                        &                        &  $44.97_{-0.98}^{+0.7}$ & $4.83_{-1.83}^{+1.57}$ &                            &                        \\
           7 &                        &                        & $46.23_{-1.29}^{+0.39}$ & $4.49_{-2.28}^{+3.29}$ &                            &                        \\
        \bottomrule
    \end{tabular}
    \tablefoot{Same as \cref{tab:gal_parameter}, but the the SMBH parameter. The black hole mass function-related parameters ($B$, $\eta$ are estimated up to $z=5$. For the quasar luminosity function, the Eddington ratio-related parameters are fixed to the MAP estimator at $z=0$.}
    \label{tab:BH_parameter}
\end{table*}

\begin{figure*}
     \centering
     \includegraphics[width=\textwidth]{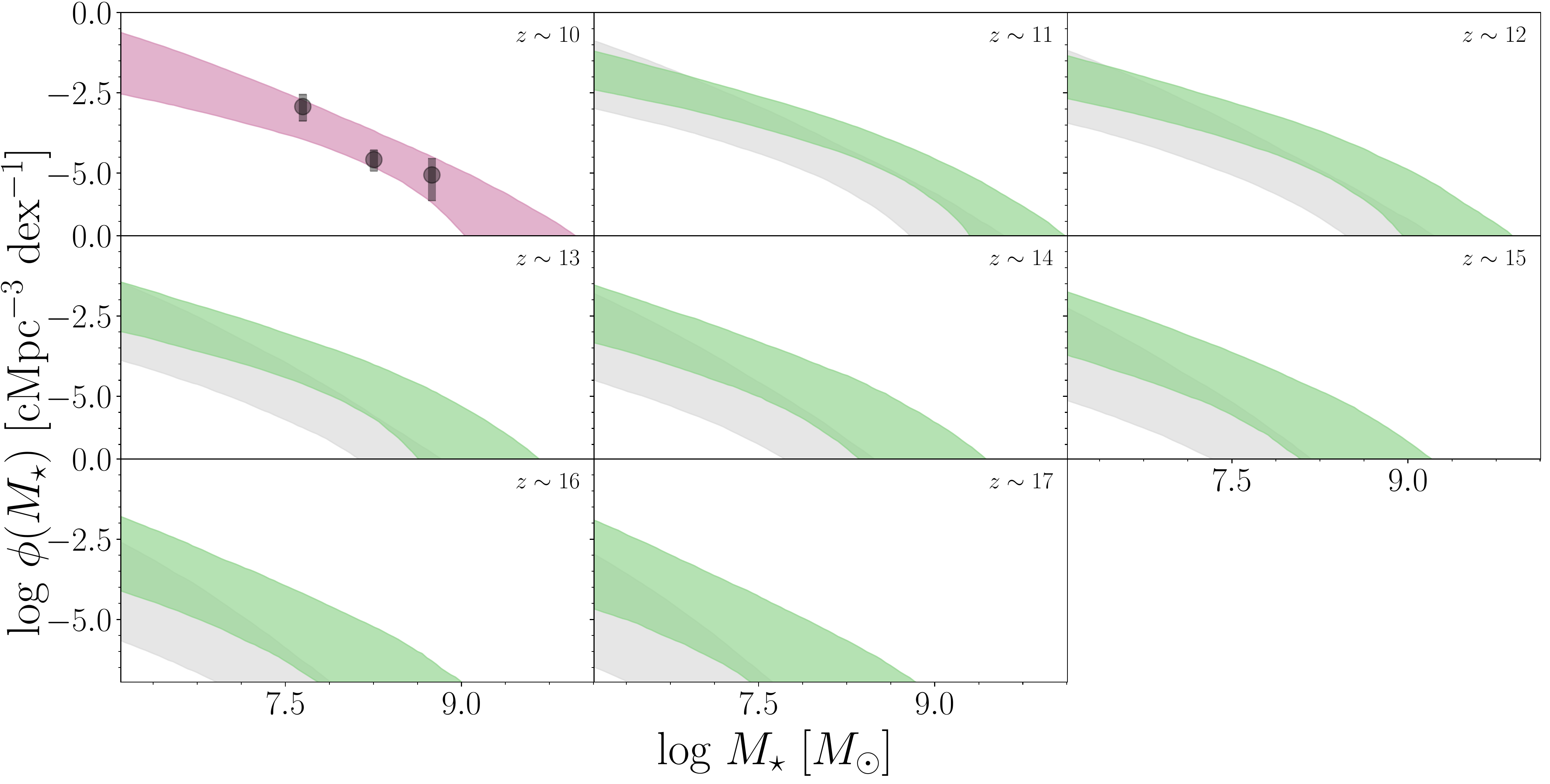}
     \caption{\textbf{Extrapolated galaxy stellar mass function.} The predicted GSMF is calculated by extrapolating the stellar mass -- halo mass relation to the redshift in question and evolving the HMF (green), and by fixing the stellar mass -- halo mass relation to the distribution at $z=10$ and only evolving the HMF(grey). The pink band is the last GSMF constrained by data. The shown bands correspond to the 68\% credible regions.}
     \label{fig:mstar_ndf_predictions}
\end{figure*}
\begin{figure*}
     \centering
     \includegraphics[width=\textwidth]{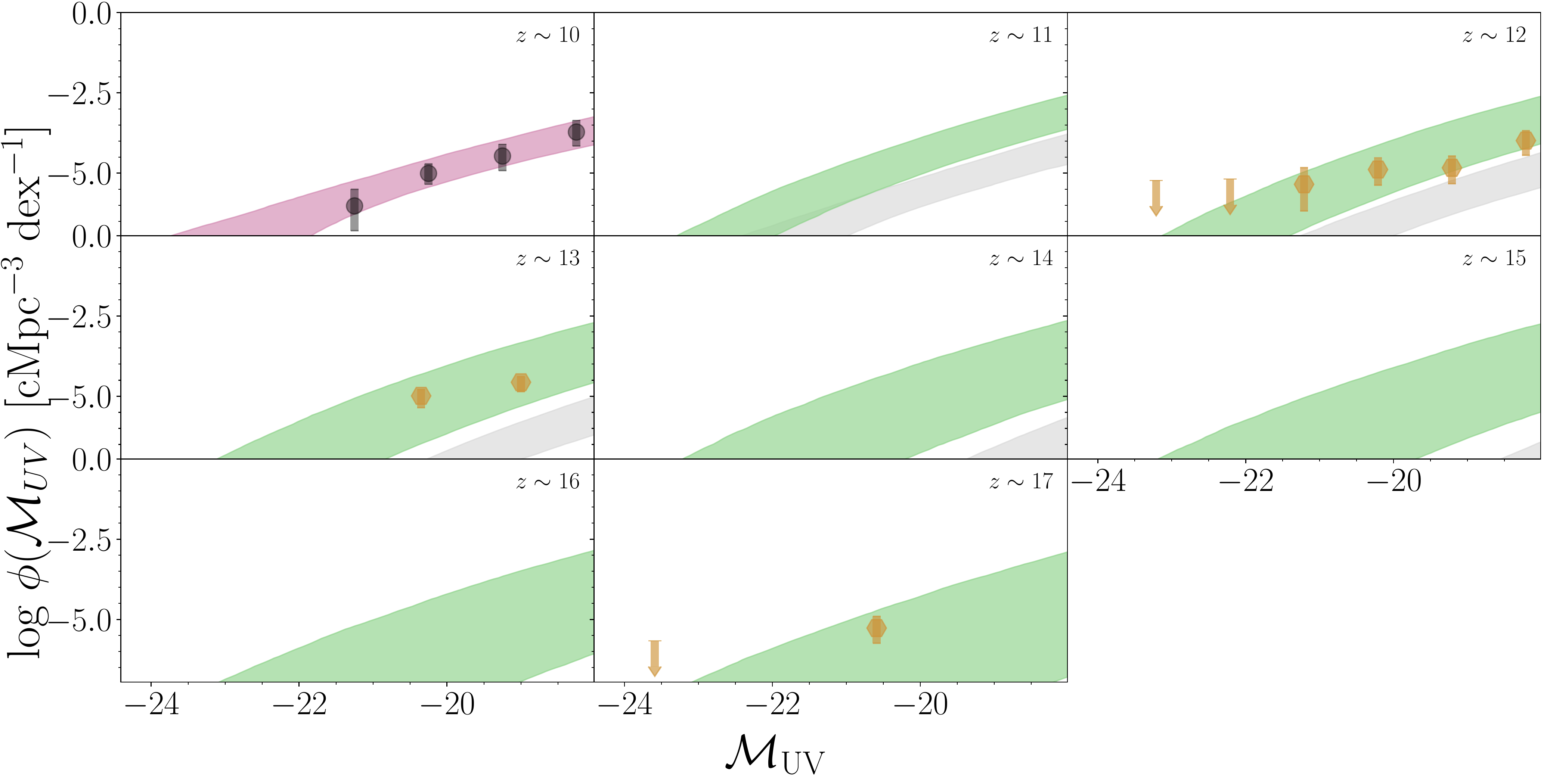}
      \caption{\textbf{Extrapolated galaxy UV luminosity functions:} Similar to \cref{fig:mstar_ndf_predictions}. The orange data points correspond to the estimates obtained from the JWST early data release estimated by \citet{Harikane2022} and \citet{Donnan2022}.}
     \label{fig:Muv_ndf_predictions}
\end{figure*}
\begin{figure*}
     \centering
     \includegraphics[width=\textwidth]{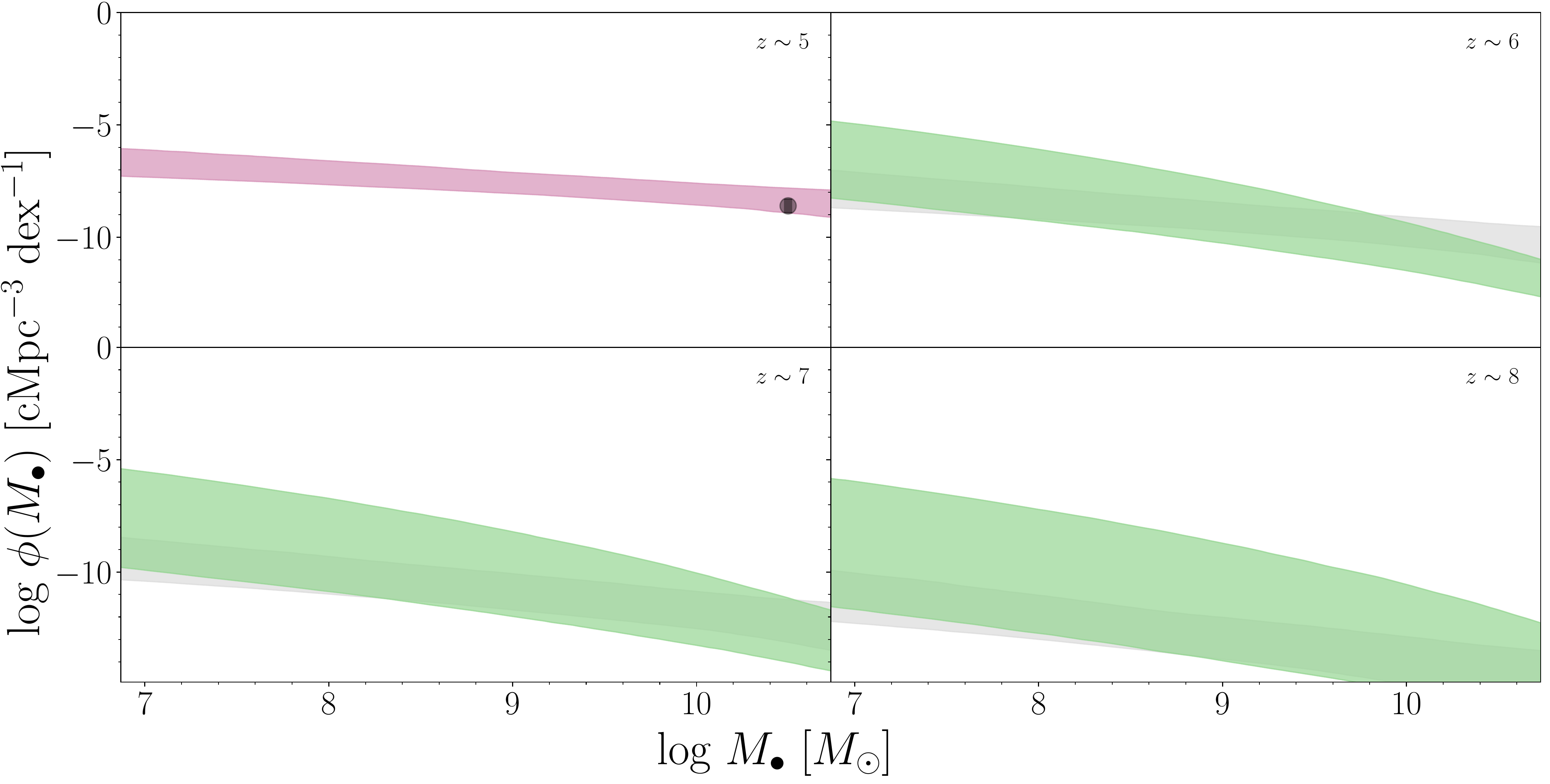}
      \caption{\textbf{Extrapolated type 1 active black hole mass functions:} Similar to \cref{fig:mstar_ndf_predictions}.}
     \label{fig:mbh_ndf_predictions}
\end{figure*}
\begin{figure*}
     \centering
     \includegraphics[width=\textwidth]{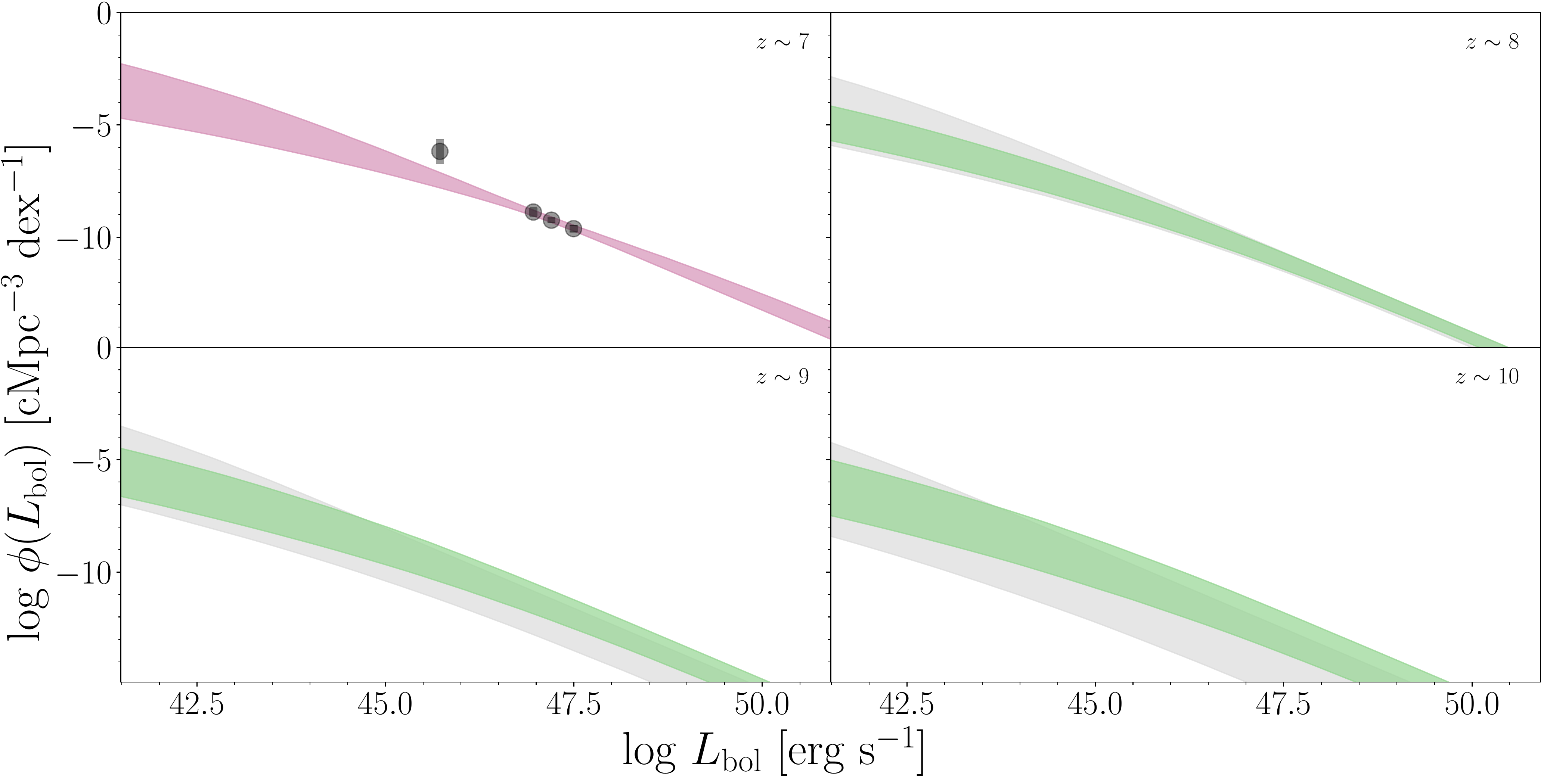}
      \caption{\textbf{Extrapolated quasar luminosity functions:}  Similar to \cref{fig:mstar_ndf_predictions}.}
     \label{fig:Lbol_ndf_predictions}
\end{figure*}

\begin{figure*}[ht]
     \centering
     \begin{subfigure}[b]{\columnwidth}
         \centering
         \includegraphics[width=\textwidth]{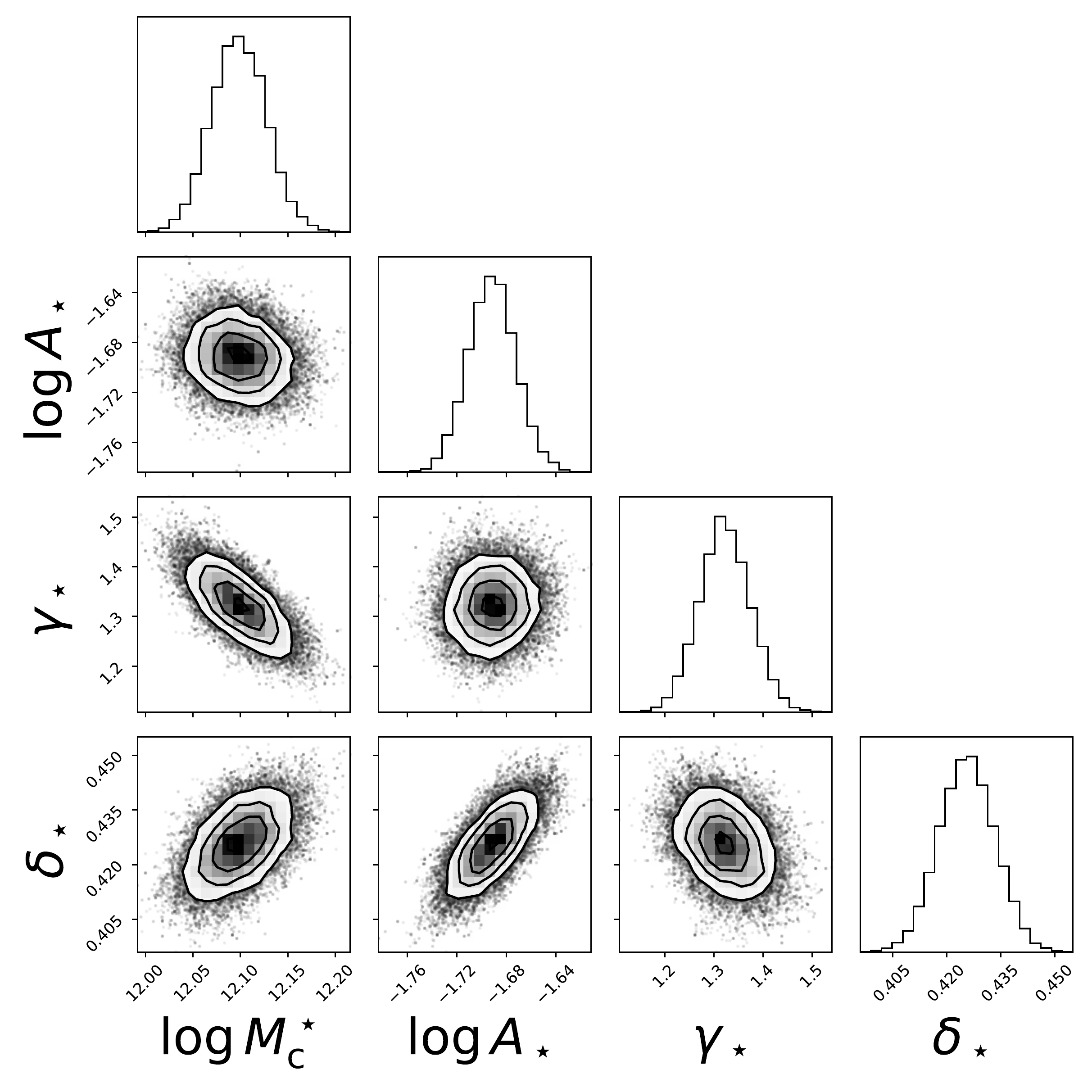}
         \caption{Stellar mass.}
         \label{fig:mstar_corner}
     \end{subfigure}
     \hfill
     \begin{subfigure}[b]{\columnwidth}
         \centering
         \includegraphics[width=\textwidth]{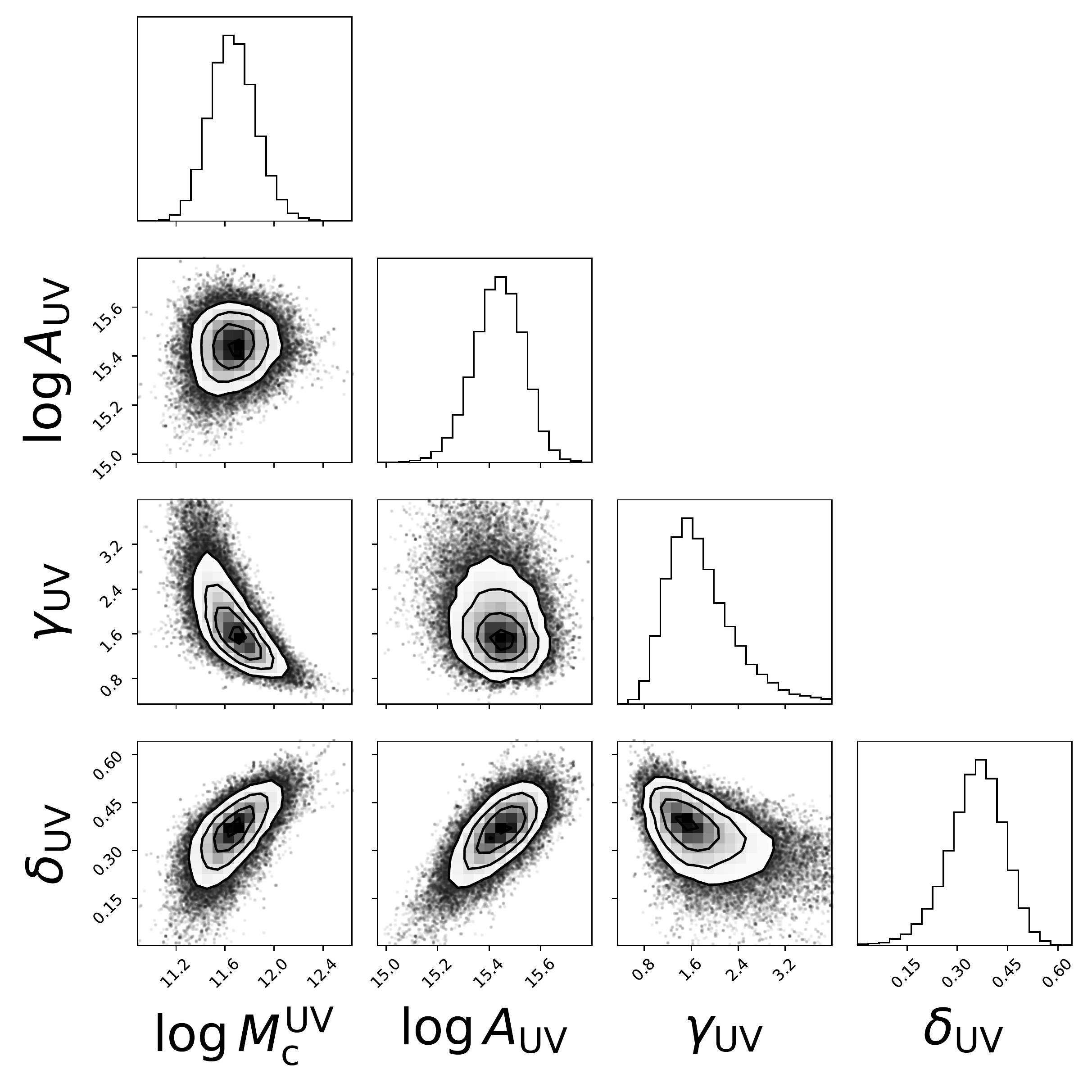}
         \caption{UV luminosity.}
         \label{fig:Muv_corner}
     \end{subfigure}
     
    \begin{subfigure}[b]{\columnwidth}
         \centering
         \includegraphics[width=\textwidth]{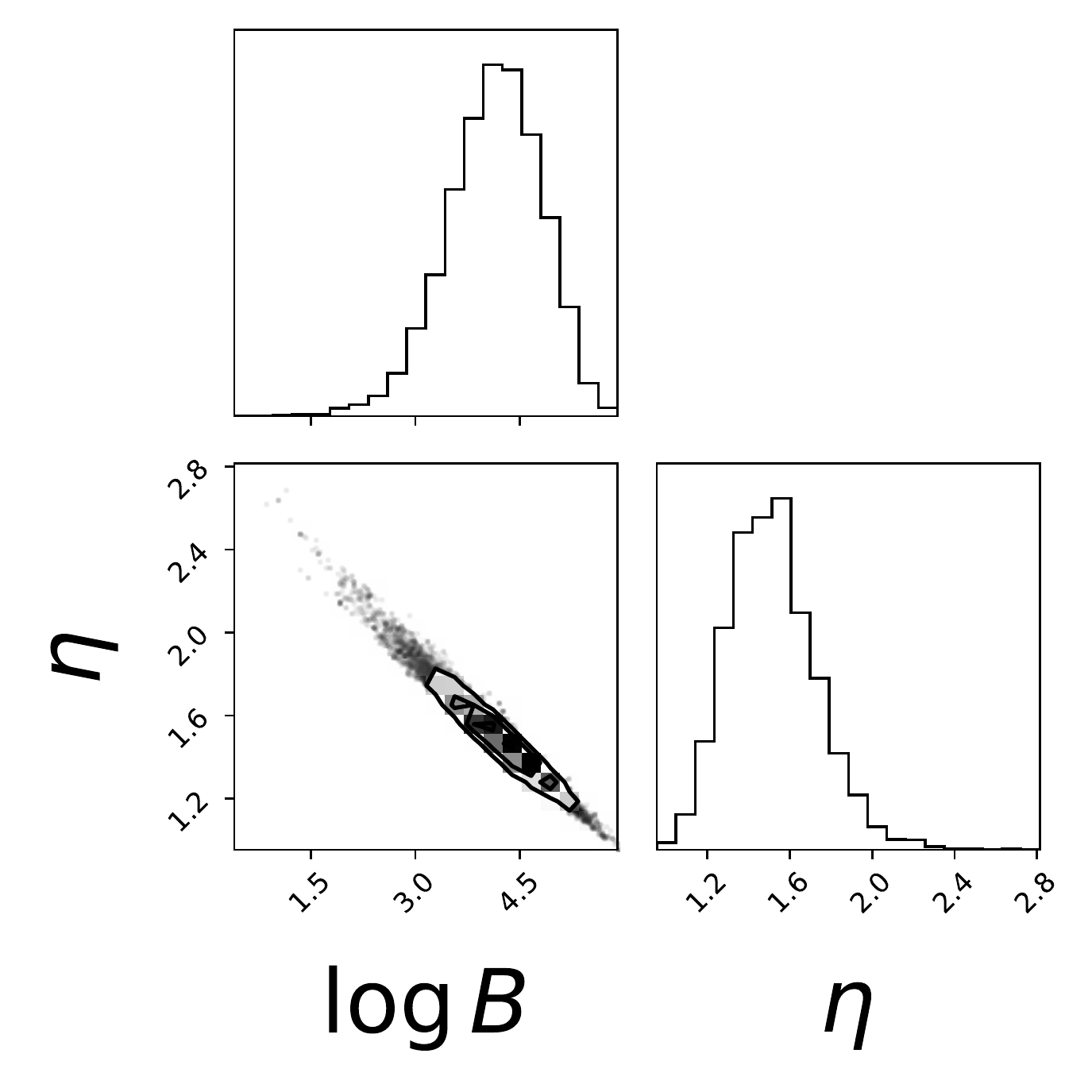}
         \caption{Type 1 active black hole mass.}
         \label{fig:mbh_corner}
     \end{subfigure}
     \hfill
     \begin{subfigure}[b]{\columnwidth}
         \centering
         \includegraphics[width=\textwidth]{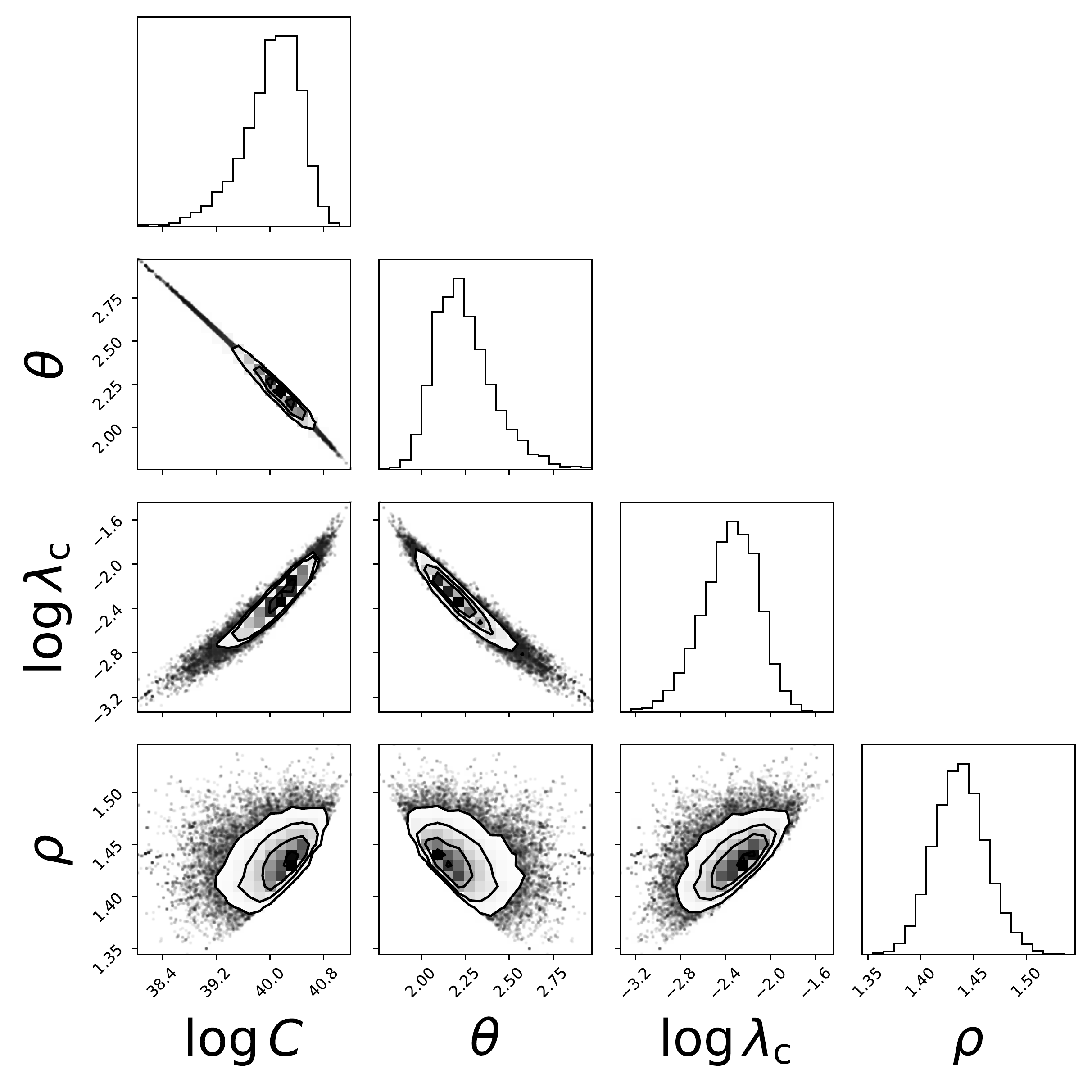}
         \caption{AGN bolometric luminosity.}
         \label{fig:Lbol_corner}
     \end{subfigure}
     \caption{\textbf{Correlation plots for the different models at} $z=0$ \textbf{:} Correlations between different different parameters obtained from the MCMC chains at z=0. For the stellar mass and UV luminosity, the parameter can be considered largely independent, the strongest (anti-) correlation can be found between $\gamma$ and $M_\mathrm{c}$. For the black hole models, there is a strong degeneracy between normalisation and power law slope.}
     \label{fig:corner_plots}
\end{figure*}
\end{appendix}

\end{document}